\documentclass[twocolumn]{aastex63}

\usepackage{ulem,xcolor}
\usepackage{epsf}
\usepackage{graphicx}
\usepackage{natbib}
\usepackage{url}
\usepackage{mathrsfs}
\usepackage{amsmath}

\newcommand{\Mstar}{\mbox{\,$M_{\star}$}}
\newcommand{\Msun}{\mbox{\,$M_{\odot}$}}

\newcommand{\Zsun}{\mbox{\,$Z_{\odot}$}}
\newcommand{\Avmw}{$A_{V, \rm MW}$} 
\newcommand{\dAv}{d$A_{V}$}
\newcommand{\dAvy}{d$A_{V_{\rm y}}$}
\newcommand{\sfravg}[2]{$\langle\rm SFR\rangle_{#1\text{\rm #2}}$}
\newcommand{\sfravgcol}[3]{$\langle\rm SFR\rangle^{\text{\rm #3}}_{#1\text{\rm #2}}$}
\newcommand{\asymerr}[3]{$#1^{+#2}_{-#3}$}
\defcitealias{Lee2009}{L09}
\defcitealias{Karachentsev2013}{K13}
\defcitealias{Gilbert2025}{Paper I}

\graphicspath{{./}}

\shorttitle{LUVIT II. Impact of UV CMDs on deriving recent SFHs}
\shortauthors{Choi et al.}

\begin{document}
\bibliographystyle{aasjournal}

\title{The Local Ultraviolet to Infrared Treasury II. Refining Star Formation Histories of Ten Metal-Poor Dwarf Galaxies with Simultaneous UV-Optical Two-CMD Fitting}

\correspondingauthor{Yumi Choi}
\email{yumi.choi@noirlab.edu}

\author[0000-0003-1680-1884]{Yumi Choi}
\affiliation{NSF National Optical-Infrared Astronomy Research Laboratory, 950 North Cherry Avenue, Tucson, AZ 85719, USA}

\author[0000-0003-0394-8377]{Karoline M. Gilbert}
\affiliation{Space Telescope Science Institute, 3700 San Martin Dr., Baltimore, MD 21218, USA}
\affiliation{The William H. Miller III Department of Physics \& Astronomy, Bloomberg Center for Physics and Astronomy, Johns Hopkins University, 3400 N. Charles Street, Baltimore, MD 21218, USA}

\author[0000-0002-7502-0597]{Benjamin F. Williams}
\affiliation{Department of Astronomy, University of Washington, Box 351580, Seattle, WA 98195, USA}

\author[0000-0002-6442-6030]{Daniel R. Weisz}
\affiliation{Department of Astronomy, University of California, Berkeley, Berkeley, CA, 94720, USA}

\author[0000-0003-0605-8732]{Evan D.\ Skillman}
\affiliation{Minnesota Institute for Astrophysics, University of Minnesota, 116 Church St.\ SE, Minneapolis, MN 55455, USA}

\author[0000-0002-1264-2006]{Julianne J.\ Dalcanton}
\affiliation{Center for Computational Astrophysics, Flatiron Institute, 162 Fifth Avenue, New York, NY 10010, USA}
\affiliation{Department of Astronomy, University of Washington, Box 351580, Seattle, WA 98195, USA}

\author[0000-0001-5538-2614]{Kristen B.~W.\ McQuinn}
\affiliation{Space Telescope Science Institute, 3700 San Martin Dr., Baltimore, MD 21218, USA}
\affiliation{Department of Physics and Astronomy, Rutgers, The State University of New Jersey, 136 Frelinghuysen Rd, Piscataway, NJ 08854, USA}

\author{Andrew E. Dolphin}
\affiliation{Raytheon, 1151 E. Hermans Road, Tucson, AZ 85756, USA} 
\affiliation{University of Arizona, Steward Observatory, 933 North Cherry Avenue, Tucson, AZ 85721, USA}

\author[0000-0002-9300-7409]{Giada Pastorelli}
\affiliation{Department of Physics and Astronomy G. Galilei, University of Padova, Vicolo dell’Osservatorio 3, I-35122, Padova, Italy}
\affiliation{Osservatorio Astronomico di Padova -- INAF, Vicolo dell'Osservatorio 5, I-35122 Padova, Italy}

\author[0000-0003-4850-9589]{Martha L. Boyer}
\affiliation{Space Telescope Science Institute, 3700 San Martin Dr., Baltimore, MD 21218, USA}

\author[0000-0001-5340-6774]{Karl D.\ Gordon}
\affiliation{Space Telescope Science Institute, 3700 San Martin Dr., Baltimore, MD 21218, USA}

\author[0000-0002-6301-3269]{L\'eo Girardi}
\affiliation{Osservatorio Astronomico di Padova -- INAF, Vicolo dell'Osservatorio 5, I-35122 Padova, Italy}

\author[0000-0003-3747-1394]{Maude Gull}
\affiliation{Department of Astronomy, University of California, Berkeley, Berkeley, CA, 94720, USA}

\begin{abstract}
We present the star formation histories (SFHs) of ten metal-poor ($\lesssim$12\% \Zsun), star-forming dwarf galaxies from the Local Ultraviolet to Infrared Treasury (LUVIT) survey. The derived SFHs exhibit significant variability, consistent with the irregular star formation expected for dwarf galaxies. Using synthetic near ultraviolet (UV) and optical CMDs with various yet targeted configurations for dust and input SFHs, we quantitatively demonstrate that simultaneous modeling of the UV and optical CMDs (``UVopt'' case) improves the precision of SFH measurements in recent time bins up to $\sim$1~Gyr, compared to the classical single optical CMD modeling (``Opt-only'' case). The UVopt case reduces uncertainties relative to the Opt-only case by $\sim$4–8\% over the past 10~Myr, $\sim$8–20\% over 100~Myr, and $\sim$8–14\% over 1~Gyr, across various dust configurations and input SFHs. Additionally, we demonstrate discrepancies in stellar models for blue core helium-burning (BHeB) stars at the low metallicity regime. This discrepancy can artificially inflate star formation rate (SFR) estimates in younger age bins by misinterpreting the evolved BHeB stars as reddened upper main-sequence (MS) stars. Incorporating UV data improves BHeB-MS separation and mitigates the limitations of current low metallicity stellar models. Comparisons of the UVopt SFHs with H$\alpha$ and FUV-based SFRs reconfirm that H$\alpha$ is an unreliable tracer over its nominal 10~Myr timescale for low-SFR galaxies, while FUV provides a more reliable tracer but yields SFR$_{\rm FUV}$ values up to twice those of CMD-based \sfravg{100}{Myr}. Our findings underscore the importance of UV data in refining recent SFHs in low-metallicity environments.
\end{abstract}


\section{Introduction}
Our understanding of star formation (SF), one the most fundamental astrophysical processes driving galaxy evolution, relies heavily on detailed studies of resolved stars within the Local Volume \citep[references therein]{Tolstoy2009}. Nearby galaxies in the Local Volume offer the unique opportunity to make accurate star formation history (SFH) measurements as well as quantitative tests of stellar evolution models. This is because individual stars in these galaxies can be simultaneously resolved across a broad range of wavelengths, from the ultraviolet (UV) to the infrared (IR). However, such panchromatic imaging has so far been largely focused on galactic environments with metallicities no lower than that of the Small Magellanic Cloud \citep[$\sim$0.2\Zsun;][]{Karakas2018}, limiting our insight into SF processes at more extreme low-metallicity regimes prevalent in early-universe galaxies.

To improve our knowledge of SF in low-metallicity environments, extending panchromatic, UV through IR resolved stellar population studies to even lower metallicity regimes ($\lesssim$0.1\Zsun) is crucial. Extremely metal-poor, low-mass, star-forming dwarf galaxies in the Local Volume serve as ideal laboratories for unraveling the evolutionary processes that have shaped cosmic history, from the formation of the first galaxies to the present day. The Local Ultraviolet to Infrared Treasury (LUVIT; PIs K.~M.~Gilbert and M.~L.Boyer, GO-15275 and GO-16162, respectively) survey is ideally positioned to fill a critical gap in the study of resolved stellar populations in this context. LUVIT, a Hubble Space Telescope (HST) program, provides panchromatic imaging from the near-UV (NUV) to the near-IR (NIR) of resolved stars in 22 star-forming dwarf galaxies within $\sim$3.8~Mpc \citep[][hereafter Paper I]{Gilbert2025}. These galaxies occupy the lower-mass ($M_{\star} < 10^{7.5}~\Msun$) and lower-metallicity ($Z < 12\%~\Zsun$) regime, a parameter space unexplored by previous large-scale NUV through NIR surveys of resolved stellar populations. Examples include the Panchromatic Hubble Andromeda Treasury in M31 \citep[PHAT;][]{Dalcanton2012}, the Panchromatic Hubble Andromeda Treasury: Triangulum Extended Region in M33 \citep[PHATTER;][]{Williams2021}, the Legacy Extragalactic Ultraviolet Survey \citep[LEGUS;][]{Calzetti2015}, and the Scylla survey of the LMC and SMC \citep{Murray2024}. 

Thus, the LUVIT data provide a unique opportunity for comprehensive, multi-wavelength studies of stellar populations and SFHs in low-mass, low-metallicity environments. By leveraging the detailed insights enabled by resolving individual stars, these data allow for precise tests of stellar evolution models, direct measurement of feedback processes, and reconstruction of SFHs with unparalleled accuracy \citepalias{Gilbert2025}.

A galaxy's SFH is imprinted in its color-magnitude diagrams (CMDs). Sophisticated CMD-fitting algorithms have successfully measured the SFHs for nearby galaxies by comparing their observed CMDs with synthetic ones generated from stellar evolution models \citep[e.g.,][]{Aparicio1996, Tolstoy1996, Holtzman1999, Olsen1999, Harris2001, Dolphin02, Skillman2003, Gallart2005, Williams2009, McQuinn2010a, Weisz2011, Choi2015, Lewis2015, Cignoni2019, Lazzarini2022, Massana2022, Savino2023, Cohen2024}. CMD-based SFHs have traditionally focused on modeling individual CMDs in optical or NIR wavelengths. Only a few studies have attempted simultaneous modeling of two CMDs, such as in near-IR (NIR) YJK$_{s}$ filters \citep{Rubele2012, Rubele2015}. These efforts, however, were hindered by photometric calibration issues in their Y filter, leading to the decoupling of the CMDs in a subsequent study by the same team \citep{Rubelel2018}. Other studies have independently modeled two CMDs \citep[e.g.,][]{Mazzi2021} or a single CMD spanning different wavelength regimes \citep[e.g., U--V in ][]{Cignoni2019}. Yet, to our knowledge, SFHs based on the simultaneous modeling of UV and optical CMDs have not been achieved. The addition of UV CMDs can offer distinct advantages. While optical and NIR CMDs are primarily sensitive to older stellar populations and provide critical distance information, UV CMDs are more effective for constraining younger populations. Therefore, simultaneously modeling both UV and optical CMDs should, in principle, enable a more comprehensive understanding of SFHs by leveraging the complementary information provided by these wavelength regimes. 

LUVIT's UV stellar photometry can play a pivotal role in analyzing young stellar populations and probing dust properties. While optical and longer-wavelength imaging broadly distinguish old and young stars, they fail to capture the bolometric flux peak of young, massive main-sequence (MS) stars (T\textsubscript{eff} $\gtrsim$ 10,000~K). This limitation is exacerbated by dust, which can obscure or alter stellar light. For instance, relying solely on optical photometry makes disentangling the effects of reddening from stellar temperature exceedingly difficult \citep[e.g.,][]{Dalcanton2012, Gordon2016, Choi2020, Lindberg2024}. These challenges are most pronounced in star-forming regions, where young stars dominate and dust is abundant. Thus, incorporating UV data alongside optical imaging in CMD analysis is essential for accurately and precisely characterizing stellar populations in star-forming galaxies. This approach will enhance our understanding of SF processes and improve empirical star formation rate (SFR) calibrations in extremely low-metallicity regimes.

This paper is organized as follows. Section~\ref{sec:data} describes the LUVIT survey data, photometric quality cuts, and artificial star tests (ASTs). Section~\ref{sec:sfh} outlines the SFH measurement process, including general CMD fitting, handling limitations of BHeB models, details of UVopt CMD modeling, and  assessing the impact of UV CMDs inclusion on SFH measurements using simulated CMDs. Section~\ref{sec:results} presents the resulting UVopt SFHs for our ten LUVIT galaxies and compares them with H$\alpha$ and FUV-based SFRs from the literature. Section~\ref{sec:UVimpact} compares the UVopt and the canonical Opt-only SFHs of the ten galaxies and explores the causes of the discrepancies between the two SFHs. Section~\ref{sec:discussion} discusses ways to improve SFH measurements in low-metallicity dwarf galaxies, and Section~\ref{sec:summary} summarizes our findings.

\begin{figure}
 \centering
      \includegraphics[width=\linewidth]{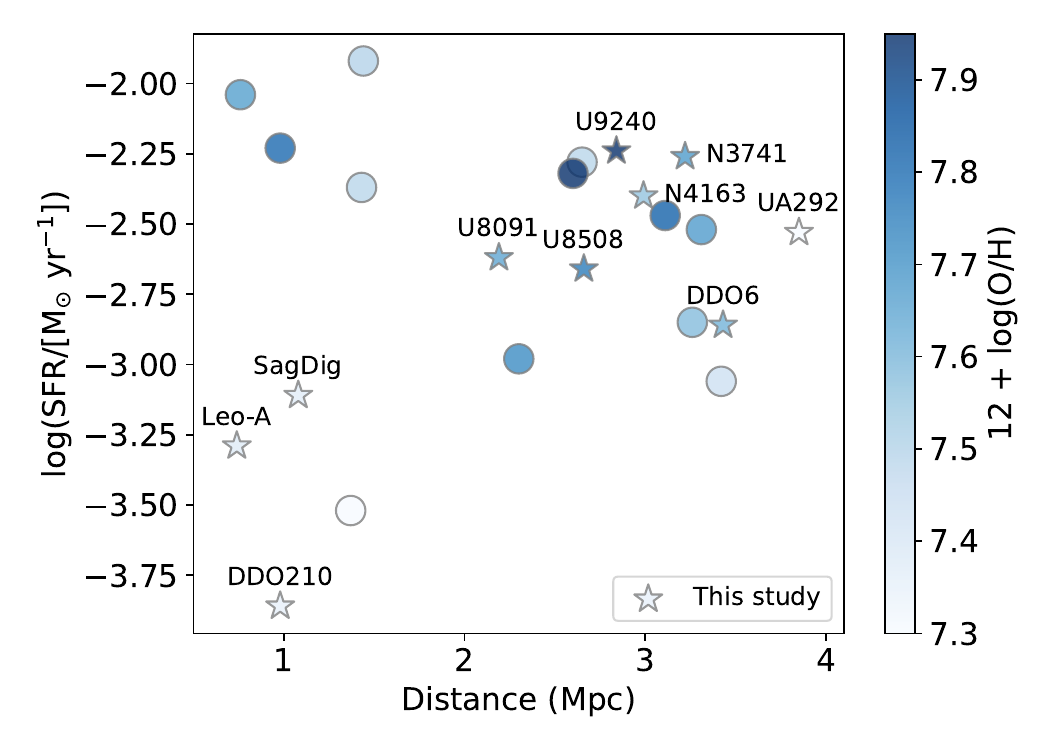}
       \caption{SFRs and distances for the full LUVIT sample, color-coded by their oxygen abundances. The quoted SFRs are calculated from extinction-corrected FUV measurements \citep{Karachentsev2013}, except for UGC 8508, which lacks GALEX FUV data and instead uses an H$\alpha$-based SFR as calculated by \citet{Karachentsev2013}. This study focuses on the ten galaxies observed with WFC/ACS (F475W, F814W) and WFC3/UVIS (F275W, F336W). These ten galaxies are depicted as stars and labeled with their corresponding names. For more details on the LUVIT sample, including basic properties and references, see Table 1 in \citetalias{Gilbert2025}. 
      \label{fig:ten_sample}}
\end{figure}

\begin{deluxetable*}{lccccccccccc}
\tablecaption{Galaxy properties, 50\% completeness in UV and optical filters, and the best-fit dust parameters for our sample. The methodology for measuring the distance modulus and dust parameters in \texttt{MATCH} is outlined in Section~\ref{sec:sfh}.
\label{tab:param_summary}}
\tablewidth{0pt}
\tablehead{
\colhead{Galaxy} &  \colhead{$m$ - $M$\tablenotemark{a}} &  \colhead{$m$ - $M$\tablenotemark{b}} &  \colhead{12 + log(O/H)\tablenotemark{c}} & \colhead{F275W} & \colhead{F336W} & \colhead{F475W} & \colhead{F814W} & \colhead{$A_{V, \rm MW}$\tablenotemark{d}} & \colhead{$A_{V, \rm MW}$\tablenotemark{e}} & \colhead{d$A_{V}$} & \colhead{d$A_{V_{\rm y}}$} \\
& (mag) & (mag) & (dex) & (mag) & (mag) & (mag) & (mag) & (mag) & (mag) & (mag) & (mag)}
\startdata
Leo A & 24.48 & 24.35 & 7.38$\pm$0.10\tablenotemark{1} & 24.97 & 25.89 & 29.04 & 27.94 & 0.025 & 0.057 & 0.0 & 0.6 \\
DDO210 & 24.95 & 24.96 & 7.35$\pm$0.14\tablenotemark{2} & 24.70 & 25.53 & 28.94 & 28.14 & 0.072 & 0.138 & 0.1 & 0.4 \\
SagDIG & 24.91 & 25.17 & 7.37$\pm$0.12\tablenotemark{3} & 24.93 & 25.76 & 28.02 & 27.74 & 0.385 & 0.338 & 0.0 & 0.6 \\
UGC 8091 & 26.68 & 26.70 & 7.65$\pm$0.06\tablenotemark{1} & 24.86 & 25.67 & 28.67 & 27.61 & 0.062 & 0.071 & 0.1 & 0.2  \\
UGC 8508 & 27.09 & 27.12 & 7.76$\pm$0.07\tablenotemark{2} & 25.28 & 26.07 & 27.86 & 27.08 & 0.033 & 0.042 & 0.1 & 0.2  \\
UGC 9240 & 27.24 & 27.27 & 7.95$\pm$0.03\tablenotemark{4} & 25.28 & 26.01 & 27.94 & 26.99 & 0.026 & 0.034 & 0.1 & 0.2 \\
NGC 4163 & 27.32 & 27.38 & 7.56$\pm$0.14\tablenotemark{2} & 25.29 & 26.02 & 27.71 & 26.60 & 0.053 & 0.055 & 0.1 & 0.2  \\
NGC 3741 & 27.62 & 27.54 & 7.68$\pm$0.05\tablenotemark{2} & 25.13 & 26.00 & 27.74 & 26.75 & 0.041 & 0.067 & 0.2 & 0.4 \\
DDO6 & 27.60 & 27.68 & 7.61$\pm$0.14\tablenotemark{5} & 25.14 & 26.02 & 28.22 & 27.10 & 0.018 & 0.047 & 0.2 & 0.7  \\
UGCA 292 & 27.71 & 27.93 & 7.32$\pm$0.06\tablenotemark{4} & 25.08 & 25.97 & 28.25 & 27.24 & 0.042 & 0.043 & 0.1 & 0.0  \\
\enddata
\tablenotetext{a}{Distance modulus measured based on our CMD modeling.}
\tablenotetext{b}{Distance modulus adapted from the literature based on the TRGB method (see Table 1 in \citetalias{Gilbert2025}).}
\tablenotetext{c}{References for the Oxygen abundances are: (1) \citet{vanZee2006b}; (2) \citet{Berg2012}; (3) \citet{Saviane2002}; (4) \citet{vanZee2006}; and (5) calculated based on the galaxy’s extinction-corrected, absolute magnitude in the Spitzer 4.5$\mu$m band, following the relationships in \citet{Berg2012} due to the absence of a direct method measurement.}
\tablenotetext{d}{MW foreground dust extinction derived from our CMD modeling.}
\tablenotetext{e}{MW foreground dust extinction from \citet{Schlegel1998, Schlafly2011}. Note that these values become less reliable toward the direction of external star-forming galaxies with infrared emission emerging from their own internal dust.}
\end{deluxetable*}

\begin{figure*}[th!]
 \centering
      \includegraphics[width=0.85\linewidth]{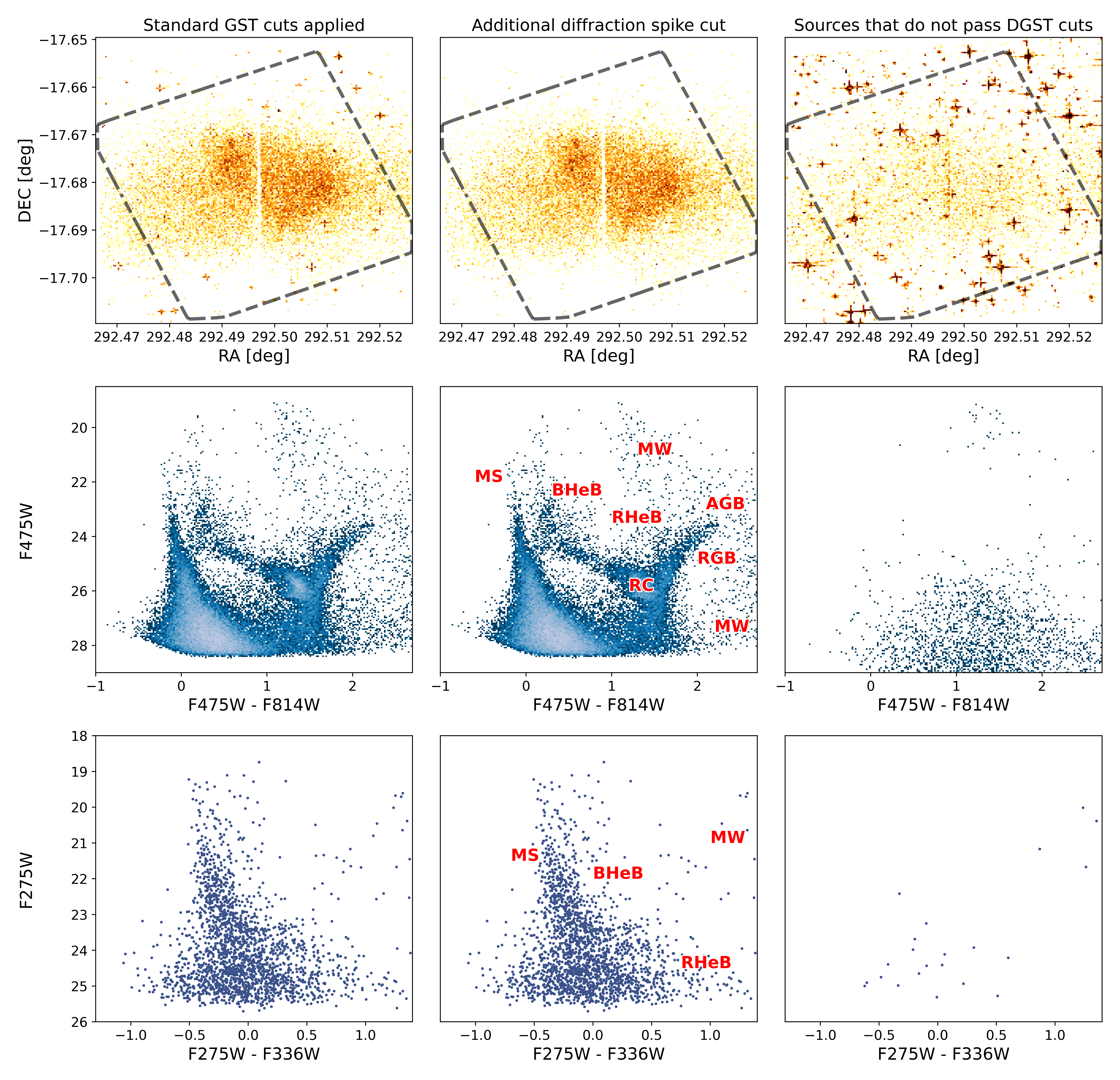}
       \caption{SagDIG as an example to illustrate our photometric quality cuts. The left column panels, from top to bottom, display the spatial distribution and optical CMD of sources that pass the default GST cuts in optical filters, and the UV CMD for sources that pass the GST cuts in UV filters. The dashed lines on the spatial maps indicate the overlapping area between optical and UV images, where SFHs are derived. The middle column shows the same data with an additional cut to remove diffraction spike artifacts (DGST cuts), which are position-based and applied across all filters. The right column shows the spatial distribution of sources excluded by DGST cuts, highlighting the location of diffraction spike artifacts in the optical and UV CMDs, respectively. Key evolutionary phases are annotated in the DGST cut-applied optical and UV CMDs. MW contaminants are noticeable in SagDIG due to its proximity to the Galactic Plane (b = -16.2879$^\circ$), whereas the other nine galaxies, located at higher Galactic latitudes, exhibit little to no MW contamination in their CMDs.
      \label{fig:dgst}}
\end{figure*}

\section{Data}\label{sec:data}
This section provides a concise overview of the LUVIT survey, the subset of galaxies analyzed in this study, and our photometry pipeline. We also detail the photometric quality cuts applied to our stellar catalogs and the ASTs. 

\subsection{Local Ultraviolet to Infrared Treasury Survey}
The LUVIT Survey combines archival HST data with two HST GO programs: GO-15275 (PI: K.~Gilbert) for NUV observations in F275W and F336W and GO-16162 (PI: M.~Boyer) for NIR observations in F110W, F127M, F139M, F153M, and F160W, as well as supplementary observations in F475W for galaxies without archival data in F475W. We simultaneously reduce the new NUV and NIR observations along with archival HST imaging data in broadband optical and NIR filters. In this study, we specifically use the NUV and optical photometry. Details of the survey strategy, observations, data reduction, and ASTs are presented in \citetalias{Gilbert2025}.

In brief, the LUVIT program targeted 22 low-metallicity, star-forming dwarf galaxies with distances $\lesssim$\,3.5~Mpc (as measured at the time of sample selection) and that had existing broadband optical images with HST as well as rich multi-wavelength ancillary datasets. By maximizing coverage of star-forming regions and overlap with the existing HST optical observations, we obtained additional HST NUV (NIR) images of these 22 (19) nearby dwarf galaxies to understand resolved stellar populations and SFHs in the low mass (\Mstar $\lesssim 10^{7.5}$~\Msun) and low metallicity ($\lesssim$ 0.12~\Zsun) regime.  

Figure~\ref{fig:ten_sample} shows the GALEX FUV-based\footnote{The H$\alpha$-based SFR is quoted for UGC 8508 due to the absence of GALEX/FUV observation.} SFR, calculated by \citet{Karachentsev2013}, versus the tip of the red giant branch (TRGB)-based distance measured in F814W for the entire sample, color-coded by their oxygen abundance. All of these values are adopted from Table 1 in \citetalias{Gilbert2025}. In this study, we focus on analyzing the global (i.e., spatially unresolved) SFHs of the ten galaxies that were observed with WFC/ACS in F475W and F814W, and WFC3/UVIS in F275W and F336W, to minimize any systematic uncertainties arising from varying camera and filter combinations. These ten galaxies are representative of the entire LUVIT sample in terms of data depth, SFR, distance, and metallicity.

The LUVIT data were reduced using the photometric reduction pipeline developed for the PHAT survey \citep{Williams2014} and refined for the PHATTER survey \citep{Williams2021}. This pipeline uses the DOLPHOT software package\footnote{http://americano.dolphinsim.com/dolphot/} \citep{Dolphin2016} and produces simultaneous multi-band point spread function fitting (PSF) photometric measurements. We refer the reader to \citetalias{Gilbert2025} for survey design, observing strategy, archival data, astrometric alignment, image processing, and PSF photometry and its quality for the full sample in detail.

\subsection{Photometric Quality Cuts}
The standard DOLPHOT output catalogs return all of the photometric measurements on every PSF fit along with several photometry quality metrics, including \texttt{sharpness} and \texttt{crowding}. Following \citet{Williams2014}, we first trim the DOLPHOT output catalogs, retaining only sources with a signal-to-noise ratio (SNR) $\geq$ 4 in at least one filter, and save them as the \texttt{ST} (``star'') catalog. We then apply the \texttt{GST} (``good star'') quality criteria to remove non-stellar objects such as cosmic rays, background galaxies, artifacts, and noise spikes, maximizing the quality of the resulting CMDs. For this study, the \texttt{GST} criteria require SNR $\geq$ 4, \texttt{sharpness}$^2$ $<$ 0.2 (0.15), and \texttt{crowding} $<$ 2.25 (1.3) for ACS/WFC (WFC3/UVIS) in each filter independently. For any measurements that do not pass these cuts, we replace their reported magnitudes with 99.999. A complete description of the \texttt{GST} quality criteria for all cameras used in LUVIT can be found in Table~6 in Paper I. 

In Figure~\ref{fig:dgst}, we use SagDIG as an example to illustrate our photometric quality cuts.
In the left column of Figure~\ref{fig:dgst}, we show the spatial distribution (top) and the optical CMD (middle) of the objects that pass the \texttt{GST} cuts in the two optical filters (F475W and F814W), and the UV CMD (bottom) of the objects that pass the \texttt{GST} cuts in the two UV filters (F275W and F336W). Although these optical and UV CMDs have well-defined features, the spatial map still shows remaining features of diffraction spikes due to bright foreground stars, indicating the \texttt{GST} cuts are not sufficient to remove all artifacts in our galaxies. 

We therefore introduce an extra quality flag to help remove the diffraction spike artifacts. Since diffraction spikes typically manifest as artificially induced local overdensities, and this flag operates independently of the \texttt{GST} cuts, we identify spurious density peaks using a star count map generated from the \texttt{ST} catalogs. For each galaxy, we divide the star count map into 200$\times$200 sub-regions, corresponding to physical scales of $\sim$5--30~pc on one axis depending on the galaxy's distance. We then apply a median filter to the map using \textsc{scipy.signal.medfilt} with a kernel size of 15. No additional adjustments are applied to the constructed star count map. We identify individual pixels where the difference between the star count map and its median filter map exceeds $N$ times the expected statistical fluctuation in the star counts (i.e., the Poisson noise), effectively applying an $N\sigma$ noise filter. The threshold $N$ varies from 3 to 5 from galaxy to galaxy to make sure not to exclude likely real stars in crowded star-forming regions while effectively removing diffraction spike artifacts.

The middle-column panels in Figure~\ref{fig:dgst} present the stellar density map, optical and UV CMDs after applying the diffraction spike cut in addition to the default \texttt{GST} cuts. We refer to the combined cuts as the \texttt{DGST} (``diffraction spike-flagged good star'') cuts. The top right panel shows what objects are removed spatially by applying the \texttt{DGST} cuts, and positions of the identified diffraction spike artifacts in the optical and UV CMDs are shown in the middle and bottom panels in the right column, respectively. The culled measurements are mostly faint and spurious, corroborating the effectiveness of the \texttt{DGST} cuts. Given the faintness of the removed detections, the cuts have larger effects on the deeper optical CMDs than the shallower UV CMDs. 

The dashed gray line on each spatial map in Figure~\ref{fig:dgst} denotes the area where the UV and optical observations overlap. To make an apple-to-apple comparison between SFHs derived from a single optical CMD and those from the combined UV and optical two-CMD approach discussed in Section~\ref{sec:UVimpact}, our analysis is confined to this overlapping region.

\subsection{Artificial Star Tests}\label{sec:ASTs}
The correct interpretation of crowded-field stellar photometry requires comprehensive understanding of the complex biases and uncertainties that are inherent when analyzing stellar photometry extracted from highly blended images. To characterize the quality of our photometry, we conduct extensive ASTs. We inject a series of artificial stars, each with known properties and sky coordinates, into the appropriate location on each overlapping input image, with the appropriate PSF and flux. The stack of images are then processed through the identical photometry pipeline as used for the unaltered data. An artificial star is considered as ``recovered'' if $|m_{\text{input}} - m_{\text{ouput}}| \leq 0.75$~mag, the star is found within 2 reference pixels of its original position, and its photometric measurement meets all \texttt{DGST} cuts. By comparing the input and recovered properties of the artificial stars, we quantify our photometric accuracy, precision, and completeness per each filter as well as per CMD. 

The input stars for the ASTs are generated simultaneously in $N$ bands as follows using the capability in the Bayesian Extinction and Stellar Tool \citep[BEAST;][]{Gordon2016}. The BEAST maps a stellar evolution library onto a stellar atmospheric model based on the stellar initial mass, age, metallicity, effective temperature, and surface gravity. When generating input ASTs, we populate a model grid using PARSEC version 1.2S isochrones \citep{Bressan2012,Tang2014,Chen2015}. For the stellar atmospheric models, we adopt the grid by \citet{Castelli2003} for local thermodynamic equilibrium (LTE) models and the TLUSTY OSTAR and BSTAR grids by \citet{Lanz2003,Lanz2007} for non-LTE models. The BEAST provides a complete range of dust properties in terms of dust extinction ($A_{V}$), average dust grain size ($R_{V}$), and the 2175\AA\, bump strength. Combinations of requested dust properties are then applied to an intrinsic stellar spectral energy distribution (SED) to compute extinguished SEDs. 

Equipped with these capabilities from the BEAST, for each galaxy, we first create an extinguished stellar model grid that appropriately covers the galaxy's stellar and dust parameter space. We randomly select at least 70 model SEDs from the extinction added model grid per flux bin in all bands. The number of flux bins ranges 40-70 depending on the depth of data. We also create a stellar number density map with a 5$''$$\times$5$''$ resolution, using stars bright enough in F475W that they are likely to be recovered with close to 100\% completeness. Depending on each galaxy's level of crowding, we divide the spatial distribution of ASTs into 5--10 stellar number density bins. Repeating an identical set of randomly selected artificial stars over different stellar number density bins guarantees more rigorous and numerous ASTs in more crowded regions, while broadly following the observed stellar spatial distribution of a galaxy. 

This initial selection of input artificial stars is designed to cover the entire physics model grid, which is required for the BEAST to perform individual stellar SED fitting (Y. Choi et al., in prep). To make ASTs appropriate for the CMD modeling as well, we supplement at least 4,000 (up to 20,000) model SEDs per stellar density bin, focusing on the observed color and magnitude ranges. This ensures that there are sufficient artificial stars to compute completeness as a function of color and magnitude, which is essential for the CMD modeling. The number of flux bins, number of model SEDs per flux bin, number of stellar number density bins, and number of supplementing model SEDs per stellar number density bin vary with individual galaxies' properties (e.g., compactness, distance) and observations (e.g., depth), in all cases targeting $\sim$1.2--2$\times$ more artificial stars than the real stars. 

\begin{figure}[t!]
 \centering
      \includegraphics[width=\linewidth]{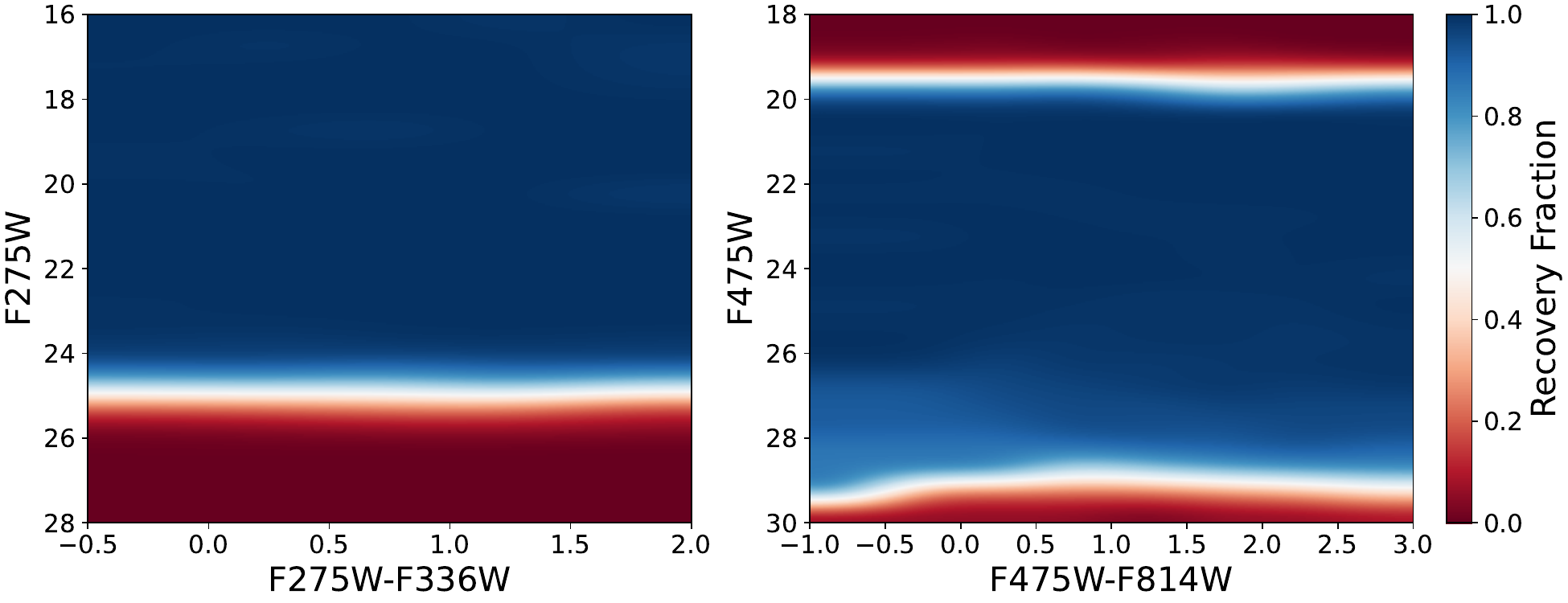}
      \includegraphics[width=\linewidth]{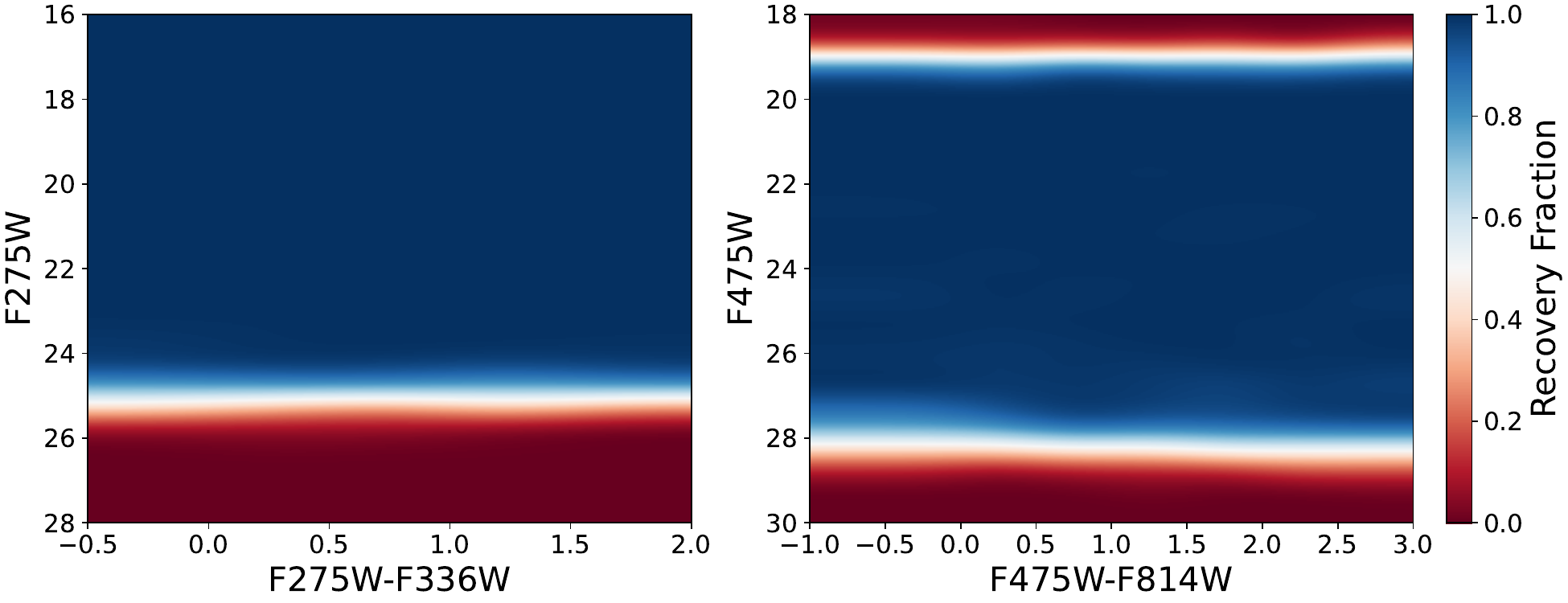}
       \caption{Recovery fraction of our photometry as a function of color and magnitude in the UV (left) and optical (right) for the closest galaxy Leo-A (top) and the farthest galaxy UGCA 292 (bottom) in our sample, respectively. Low completeness at the bright end in the optical CMDs indicates saturation due to the absence of short exposures.  
      \label{fig:compl2d}}
\end{figure}

Figure~\ref{fig:compl2d} presents example completeness functions on the UV and optical CMDs, showing the recovery fraction of input artificial stars as a function of color and magnitude, for the nearest galaxy (Leo-A) and the most distant galaxy (UGCA 292) in our sample. The resulting ASTs aid CMD modeling by allowing us to determine the 50\% completeness both at the bright and faint magnitude ends per filter and to apply photometric noise to synthetic simple stellar population models. No major dependence in completeness on color is detected both in the UV and optical CMDs -- only minor dependence towards bluer color in the optical CMD. The absence of short exposures in the archival optical observations leads to saturation, and is imprinted as low completeness at the bright magnitude end. The AST results indicate that our UV CMDs probe lookback times of at least $\sim$150~Myr for the most distant galaxy, UGCA 292, and up to $\sim$1~Gyr for the nearest galaxy, Leo-A. Notably, our UV data reach significantly further back in time compared to the conservative 60~Myr limit for the LEGUS UV data, primarily due to the greater distances of their sample galaxies, ranging from 3.3--12.7~Mpc \citep{Cignoni2019}. Throughout the paper, we will present our SFH results up to the maximum lookback time of 1~Gyr, which represents the oldest age that our UV data can reliably probe for Leo A for consistency. However, we note that galaxies at greater distances have reduced UV sensitivity. 

\citetalias{Gilbert2025} provides an in-depth discussion of the results from the ASTs for the LUVIT data, including completeness variations with spatial dependence. Accounting for spatially varying completeness as a function of stellar density within a galaxy could enhance SFH interpretation, especially for more crowded galaxies. However, deriving spatially resolved SFHs is beyond the scope of this paper. Among the 10 targets in this study, NGC 4163 exhibits the largest difference in the 50\% completeness limits between the lowest and highest spatial density bins in the optical filters used (see Sec. 5.1 in \citetalias{Gilbert2025}), with the difference in the 50\% completeness limits exceeding the photometric uncertainties at these magnitudes. This is followed by UGC 9240, NGC 3741, and UGC 8508. These four galaxies are the most likely to have global SFHs influenced by crowding effects, which will be briefly discussed in Section~\ref{sec:UVimpact}. Future work will focus on spatially resolved SFHs to refine these interpretations across the entire LUVIT galaxy sample  (Y. Choi et al., in prep).

\section{Deriving Star Formation Histories}\label{sec:sfh}
The distribution of resolved stellar populations on a CMD encodes the formation and evolution of stars in a galaxy, namely how many stars form at a specific time with a specific metallicity, following an initial stellar mass distribution. It also provides information about how far the galaxy is and how much dust the member stars experience before being observed, i.e., the combination of Milky Way foreground dust and internal dust within the galaxy. Thus, the galaxy's SFH can be determined by constructing a synthetic CMD that most closely reproduces its observed CMD. The CMD modeling techniques to extract this information have been established and extensively used for more than 30 years \citep[e.g.,][]{Tosi1991, Aparicio1996, Harris2001, Gallart2005}. The SFHs presented in this paper are calculated using the CMD fitting code called \texttt{MATCH} \citep{Dolphin02, Dolphin2012, Dolphin2016}.  

Below, we list the key assumptions for our SFH measurements using \texttt{MATCH}.
\begin{itemize}
    \item Utilize PARSEC version 1.2S stellar isochrones \citep{Bressan2012, Tang2014, Chen2015}.
    \item Assume a Kroupa initial mass function \citep{Kroupa2001} from 0.1 to 350\Msun\,with a binary fraction of 0.35. The secondary masses follow a flat distribution, constrained to ensure that the secondary star is always less massive than its primary companion. 
    \item Let metallicity monotonically increase over cosmic time from [M/H] = -2.2 up to 0.1 dex at a resolution of 0.1 dex, by setting \texttt{-zinc} flag. Although we allow the maximum [M/H] = 0.1, the highest metallicities inferred for the youngest age bin in the SFH solutions for most galaxies do not exceed $\simeq$ -1, which is consistent with their H\textsc{ii} region oxygen abundances (see Table~\ref{tab:param_summary}).   
    \item Measure SFRs in 64 log-spaced age (t) bins between log$_{10}$(t/[yr]) of 6.0 and 10.15 (0.001--14.125~Gyr) with a resolution of 0.3~dex between log$_{10}$(t/[yr]) of 6 and 6.6, 0.1~dex between log$_{10}$(t/[yr]) of 6.6 and 7.5, and 0.05~dex for the rest of the age range. With the 1~Myr isochrone set as the youngest in this study, all SF occurring within the 0--1~Myr interval is effectively consolidated into the first 1~Myr-wide age bin, leading to an apparent doubling of the SFR for that bin. This adjustment has been consistently accounted for in all the results presented in this paper.
    \item Exclude the region redward of the TRGB's red edge in the optical CMD, primarily populated by Asymptotic Giant Branch (AGB) stars, due to the limited understanding of their evolution, particularly at low metallicities.   
    \item Apply dust extinction assuming the \citet{Cardelli1989} relationship with $R_{V}$ = 3.1.
\end{itemize}

\texttt{MATCH} offers three key options for dust treatment based on stellar ages: Milky Way foreground extinction applied to all stellar populations (\Avmw), differential extinction applied to all stellar populations (\dAv), and additional differential extinction applied exclusively to young stars (\dAvy). Specifically, \Avmw\,applies a uniform dust screen to all stars, \dAv\,sets a maximum differential extinction value for a flat distribution across all populations, and \dAvy\,sets the maximum additional extinction applied only to stars younger than 100~Myr, with the maximum extra extinction linearly decreasing to zero between 40~Myr and 100~Myr. 

\begin{figure*}[ht]
 \centering
      \includegraphics[width=\linewidth]{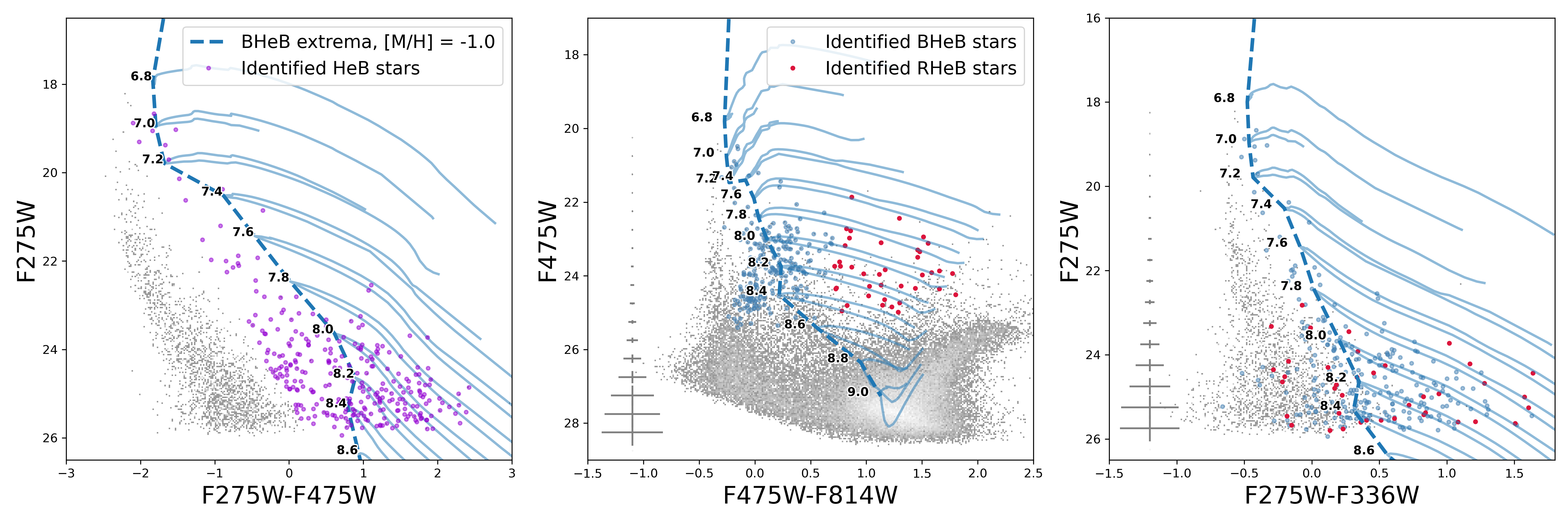}
      \includegraphics[width=\linewidth]{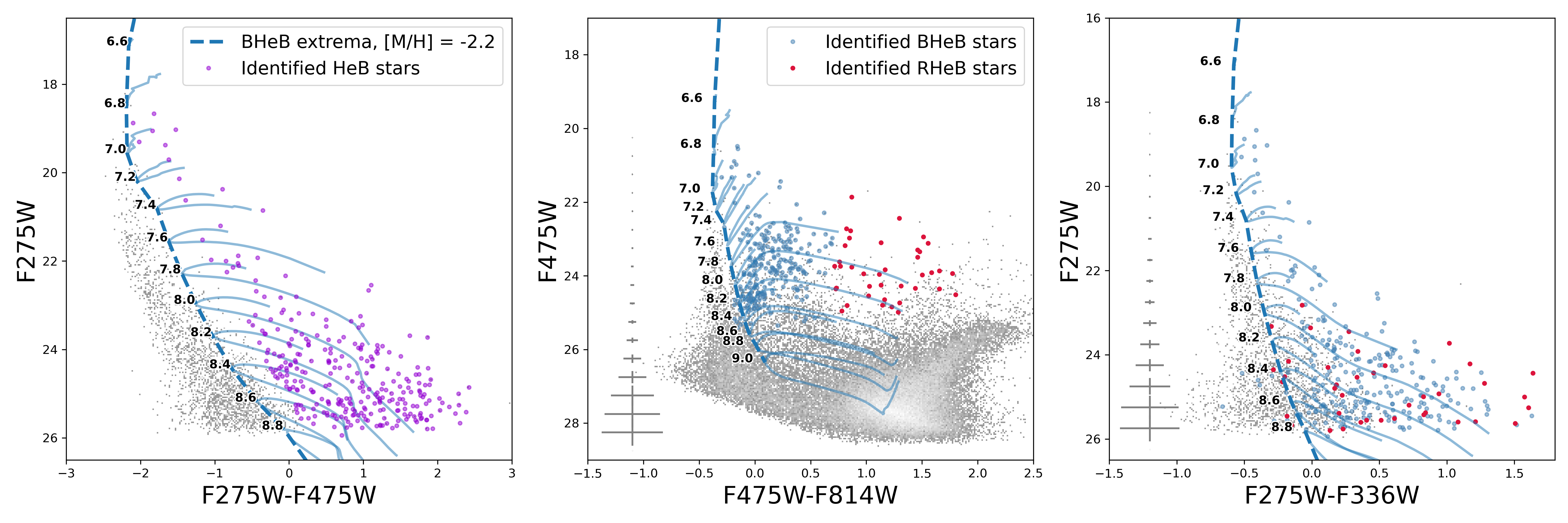}
       \caption{\textit{Top:} Comparison between the observed CMDs, constructed from the stars passing the \texttt{DGST} cuts, and the PARSEC models for HeB stars in UGC 8508 as an example. The HeB star selection is performed in a (F275W, F475W) CMD space where upper main-sequence stars are more clearly separated from HeB stars (left column). The PARSEC isochrones span ages from log$_{10}$(t/[yr]) = 6.0 to 9.0 in steps of 0.2, assuming [M/H] $= -1$ with \Avmw\ = 0.04~mag. The 10\% solar metallicity is consistent with the reported oxygen abundance of the star-forming region in this galaxy (see Table~\ref{tab:param_summary}). We plot only the HeB phase of the isochrones. The blue dashed line denotes the blue extrema of the HeB tracks, which is much redder than the observed BHeB sequence both in the optical (middle column) and UV (right column) CMDs. The error bars on the left side of each optical and UV CMD represent the average photometric uncertainties in each 0.5 magnitude-wide bin, as derived from ASTs. \textit{Bottom:} Same as the upper panels, but overlaid with the most metal-poor PARSEC isochrones with [M/H] $= -2.2$, which, in fact, shows a great match to the observed blue extrema. However, this very low metallicity does not physically make sense for these young stellar populations. 
      \label{fig:bhebmodels}}
\end{figure*}

\subsection{Blue Helium Burning Stars at Low Metallicity} \label{sec:bheb}
Stars born with intermediate initial masses ($\simeq$2--15~\Msun) go through a phase of core helium-burning (HeB). In this phase, HeB stars are bright and young, and have just evolved off the main sequence with a hydrogen-burning shell, resulting in an expanding envelope with cooler surface temperature. These stars populate the red HeB (RHeB) branch in the optical CMD. As the burning of helium in their core progresses, they move blueward by traversing the CMD along the ``blue loop'' until they reach the maximum temperature and populate the blue HeB (BHeB) branch, which is more prominent in low-metallicity galaxies, due to the effects of metallicity on the extent and duration of the blue loop phase \citep[e.g.,][]{Tang2014, Walmswell2015, Tang2016}. Due to the rapid evolution of stars from the RHeB branch to the BHeB branch, most HeB stars are found at their blue and red color extrema in the CMD. They will either evolve into thermally-pulsating AGB stars or explode as supernovae, depending on their initial mass, with a critical threshold around $\sim$8~\Msun\ \citep[e.g.,][]{Smartt2009}.

The luminosity of an HeB star is mainly set by its initial mass and therefore its age, making HeB stars a powerful age probe. Together with MS stars, which are abundant and therefore provide robust statistics, it should be possible to measure (especially recent) SFH with high age resolution in principle \citep{Dohm-Palmer1997}. 

Unfortunately, despite the importance and usefulness of HeB stars, their theoretical models are poorly understood, especially at low metallicities. Significant color offsets in the HeB sequence between the observations and theoretical predictions have been reported for low metallicity systems, but have been limited to the optical regime \citep[e.g.,][]{McQuinn2011, Tang2014, Tang2016, Cignoni2018, Lescinskaite2022}. In four local-volume dwarf galaxies, including SagDIG, \citet{Tang2014, Tang2016} demonstrate that reproducing the observed extension of blue loops at low metallicity requires overshooting at the base of the convective envelope to be a couple times stronger than that used in standard PARSEC models. The combined UV (from WFC3/UVIS F336W) and optical properties of HeB stars have been studied in \citet{Cignoni2019}, but only in dwarf galaxies with higher metallicities than our sample, where no discrepancy is reported.

Figure~\ref{fig:bhebmodels} compares the observed UV and optical CMDs of UGC 8508 with the core HeB phase from PARSEC isochrones at two different metallicities, as an example. HeB stars are identified in the (F275W-F475W, F275W) CMD, where they are distinctly separated from upper MS stars \citep{Gull2022}. While the majority of the selected HeB stars are BHeB stars, a small number of bright RHeB stars are also included, although most RHeB stars are too faint to be detected in the LUVIT F275W observations. We note that [M/H] $= -1$ aligns with the galaxy's reported oxygen abundance, while [M/H] $= -2.2$ represents the lowest metallicity available in the PARSEC models. No HeB stars are predicted by the isochrones for ages younger than $\sim$6~Myr (log$_{10}$(t/[yr]) $\simeq$ 6.8).

It is clear, from the top panels in Figure~\ref{fig:bhebmodels}, that the observed BHeB stars can be significantly bluer than the theoretical blue extrema set by the [M/H] $= -1.0$ isochrones, and the color discrepancy increases with magnitude both in the UV and optical CMDs. Interestingly, as shown in the bottom panels, the observed BHeB stars show good agreement with much lower metallicity isochrones with [M/H] $= -2.2$, in both UV and optical CMDs across the entire magnitude ranges. However, such a low metallicity is far too metal poor to explain young stellar populations even in these low-mass galaxies. The same behavior is also demonstrated in \citet{Tang2014}. 

Moreover, these extremely metal-poor isochrones disagree with, and are significantly fainter and bluer than the observed RHeB stars in the optical CMD, where [M/H] $= -1.0$ models show much better agreement. We observe the same pattern with the MIST isochrones with v/vcrit =0.4 \citep{Choi2016, Dotter2016}, and the conclusion holds for the rest of galaxies in our sample. This corroborates that current state-of-art stellar evolutionary models have deficiencies that make them inadequate for modeling HeB stars in dwarf galaxies \citep{McQuinn2011, Cignoni2018, Lescinskaite2022}, and further confirms that it remains true for dwarf galaxies including those in the weaker SF and lower metallicity regime. Model calibrations for HeB stars will be addressed in future work using the LUVIT galaxy sample (G. Pastorelli et al., in prep.).

The model offset, particularly in the BHeB stars, causes an issue for our SFH measurements. Given much redder BHeB models at relevant metallicities for young stellar populations, we find that \texttt{MATCH} tends to interpret the observed BHeB stars as highly reddened young MS stars rather than extremely metal-poor ([M/H] $\simeq$ -2.2) BHeB stars, because we require monotonically increasing metallicity in our SFH measurement process. Specifically, \texttt{MATCH} achieves improved CMD residuals by adding an unreasonable amount of extra dust to young MS stars to account for the stars near blue extrema of the BHeB branch, thus spreading young MS stars across a larger color and magnitude range. To simultaneously obtain a good match to the observed stellar counts along the blue edge of the upper MS, \texttt{MATCH} must increase the total number of young and massive MS stars. This results in artificially enhanced SFRs in the youngest age bins. We observe this behavior in almost half of our targets; either the fits fail to converge even with \dAvy $=$ 2.0~mag, or they converge at \dAvy $\simeq$ 2.0~mag, resulting in poor fits on the upper MS despite an improved global fit value. 

This unfortunate impact of the current BHeB models on deriving SFHs for our targets requires us to exclude BHeB stars and the BHeB CMD region from the UVopt case SFH measurement. Specifically, we identify HeB stars in the (F275W-F475W, F275W) CMD for each galaxy \citep{Gull2022}, remove them (mostly BHeB stars) from the stellar catalogs used in the CMD modeling, and also mask the corresponding regions in both the UV and optical CMDs to prevent \texttt{MATCH} from being misled by their absence. Masking these regions in both CMDs ensures that excluding the BHeB stars has minimal effect on the overall fitting process. This approach prevents \texttt{MATCH} from fitting the CMD with unreasonably large d$A_{V_{\rm y}}$ values. Additional details are provided in Section~\ref{sec:uvoptSFHs}.

The HeB sequence can be a valuable probe of the recent SFH, especially in spatially resolved studies where it provides complementary age constraints, capturing local SF variations when its spatial distribution differs from MS stars \citep[e.g.,][]{Dohm-Palmer1997, DohmPalmer1998, DohmPalmer2002, Lescinskaite2022}. In our case, this factor is less relevant since we are measuring global SFHs. Thus, excluding BHeB stars from CMD modeling causes no significant loss of time information in our study because their age data is redundantly imprinted in contemporaneous fainter MS stars. Although there is a lack of adequate BHeB models for low-metallicity galaxies, we underscore that they can still offer reasonable constraints on the recent SFH, particularly when they are well-populated and clearly distinguishable from upper MS stars in higher metallicity systems, more similar to the SMC, thereby reducing ambiguity in color-magnitude diagram modeling \citep[e.g.,][]{McQuinn2011}.

\begin{figure*}[ht!]
 \centering
      \includegraphics[width=\linewidth]{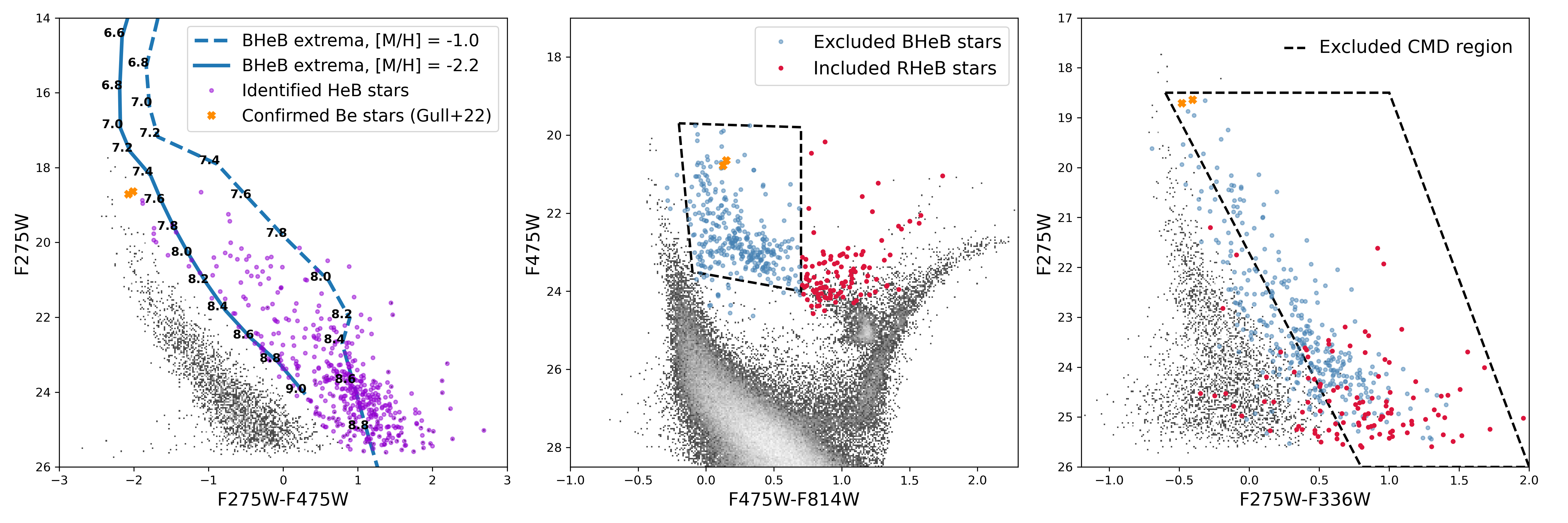}
       \caption{Demonstration of the exclusion of BHeB stars from CMD modeling, using Leo A as an example. \textit{Left:} Identification of HeB stars (purple circles) in the (F275W-F475W, F475W) CMD based on isochrones adapted for Leo A, as shown in Fig.~\ref{fig:bhebmodels}. Crosses represent spectroscopically confirmed Be stars from \citet{Gull2022}, which are included in our final CMDs. However, they fall within the excluded CMD regions, ensuring no impact on our SFH measurements. Notably, these confirmed Be stars are indistinguishable from BHeB stars based solely on their optical photometry. \textit{Middle and Right:} The identified HeB stars are roughly divided into blue HeB (BHeB) and red HeB (RHeB) stars based on their optical colors. BHeB stars, which are excluded from the stellar catalogs for the SFH fits, are marked with blue circles, while retained RHeB stars are marked with red circles in both the optical and UV CMDs used for the UVopt case. Black dashed lines indicate the regions of the CMDs which are excluded from the SFH fitting calculations, ensuring that MATCH does not misinterpret the absence of BHeB stars during the CMD modeling process.
      \label{fig:bheb_exclusion}}
\end{figure*}

\subsection{Deriving SFHs from Simultaneous UV-Optical Two-CMD Modeling}\label{sec:uvoptSFHs}
In simultaneous UV-optical two-CMD modeling, \texttt{MATCH} generates synthetic model CMDs in the (F275W, F336W) and (F475W, F814W) filters. During the fitting process, both CMDs are constructed for a given SFH, and the overall fit statistic is calculated as the sum of the individual fit statistics for the two CMDs. It is important to note that this method treats the two CMDs as statistically independent, meaning the algorithm does not account for the fact that a particular star in the UV CMD corresponds to the same star in the optical CMD. This approach prioritizes computational efficiency, as simultaneously modeling two separate CMDs approximately doubles the computational cost (in terms of time and memory), whereas directly fitting the four-band photometry would increase computational demands by nearly three orders of magnitude to accommodate the extra dimensions. While it sacrifices star-by-star correlations, it achieves significant gains in the SFH measurements with minimal computational overhead relative to direct four-band fitting by leveraging the complementary strengths of UV and optical CMDs. The SFHs derived from simultaneous UV-optical two-CMD modeling (i.e., UVopt case) serve as our fiducial results. 

As the first step, we identify HeB stars in the (F275W-F475W, F275W) CMD for each galaxy and exclude the BHeB stars from the CMD modeling, as discussed in Section~\ref{sec:bheb}. However, we retain RHeB stars as much as possible, as they align well with models at the relevant metallicity in the optical CMD (see Figure~\ref{fig:bhebmodels}). To prevent \texttt{MATCH} from misinterpreting the absence of BHeB stars, we also mask and exclude the corresponding regions in the UV and optical CMDs as closely as possible. In \texttt{MATCH}, these exclusion regions are quadrilaterals defined by four color and magnitude points, and serve to omit areas with unreliable model predictions from the CMD fitting process. While setting a more complex exclusion region is possible with multiple quadrilaterals, we have chosen to keep the exclusion regions as simple as possible.  

Figure~\ref{fig:bheb_exclusion} illustrates this process using Leo-A as an example. Purple circles in the (F275W-F475W, F275W) CMD represent HeB stars, identified by their separation from upper MS stars based on isochrones in the UV-opt, UV, and optical CMDs. HeB stars identified in this CMD are further classified by their optical color into BHeB (blue circles) and RHeB (red circles) in the optical CMD. This division is flexible and varies across galaxies, depending on how well the BHeB and RHeB stars are populated (although identified HeB stars bluer than F475W-F814W = 0.5 are always classified as BHeB stars). Since HeB stars provide redundant age information to their contemporaneous MS stars, and RHeB stars are relatively rare in our galaxies, the precise placement of the dividing line has no significant impact on the derived SFH results. Furthermore, not all RHeB stars are visible in the (F275W-F475W, F275W) CMD, as some are too faint to be detected in F275W. We use the positions of the BHeB stars in both the UV and optical CMDs to define exclusion regions that mask the BHeB area without affecting the upper MS region as much as possible (dashed quadrilaterals in the middle and right panels). 

The orange crosses in Figure~\ref{fig:bheb_exclusion} represent two spectroscopically confirmed Be stars by \citet{Gull2022}. They fall within our excluded CMD regions both in the UV and optical CMDs, and thus do not affect the SFH measurements.  This level of detailed classification within the LUVIT footprint is only available for Leo A. However, we confirm that the resulting SFHs remain consistent regardless of the inclusion or exclusion of these two objects. This indicates that any similarly minor misclassification of BHeB stars in other galaxies is likely to have a minimal impact. 

For the purposes of CMD modeling, we determine the optimal distance for each galaxy, rather than relying on fixed values from the literature. As a baseline, we first derive the best-fit SFH using only the optical CMD for each galaxy and adopt its best-fit distance modulus for the simultaneous two-CMD modeling to enhance fitting efficiency, given the lack of distance-sensitive features in the UV CMD. Specifically, we begin with a broader grid search around each galaxy’s TRGB-based $\mu$ value \citep{Jacobs2009}, noted in Table~\ref{tab:param_summary}. We then iteratively refine the range and step size for $\mu$, using a final search interval of 0.01~mag. Our best-fit $\mu$ values are not significantly different from those based on the TRGB, with a median difference of 0.07~mag. This difference is substantially smaller than the systematic offsets of $\sim$0.15~mag reported in isochrone-based calibrations of the TRGB absolute magnitude \citep[e.g.,][]{Durbin2020}.

For the UVopt case, we also solve for all three dust parameters ($A_{V, \rm MW}$, d$A_{V}$, d$A_{V_{\rm y}}$), fully leveraging the diagnostic potential of the UV data. While still significantly simplified compared to realistic star/dust geometry, our age-dependent dust modeling provides an incremental step toward more accurately constraining recent SFHs through CMD modeling.   

It is known that the galactic dust map based on far-infrared emission becomes unreliable for external star-forming galaxies due to additional dust emission from the galaxies themselves, which can distort MW foreground estimates \citep{Schlegel1998}. To address this, we perform an initial grid search with broader intervals around the reported MW foreground extinction values for each galaxy \citep{Schlafly2011}, as listed in Table~\ref{tab:param_summary}. We progressively refine the search ranges and intervals for dust parameters, using a 0.001~mag interval for $A_{V, \rm MW}$ and a 0.1~mag interval for d$A_{V}$ and d$A_{V_{\rm y}}$ in the final searches. This iterative grid search provides \texttt{MATCH} some flexibility, which can help mitigate potential discrepancies between the observations of low-metallicity dwarf galaxies and the model isochrones. 

\begin{figure}[ht]
 \centering
      \includegraphics[width=\linewidth]{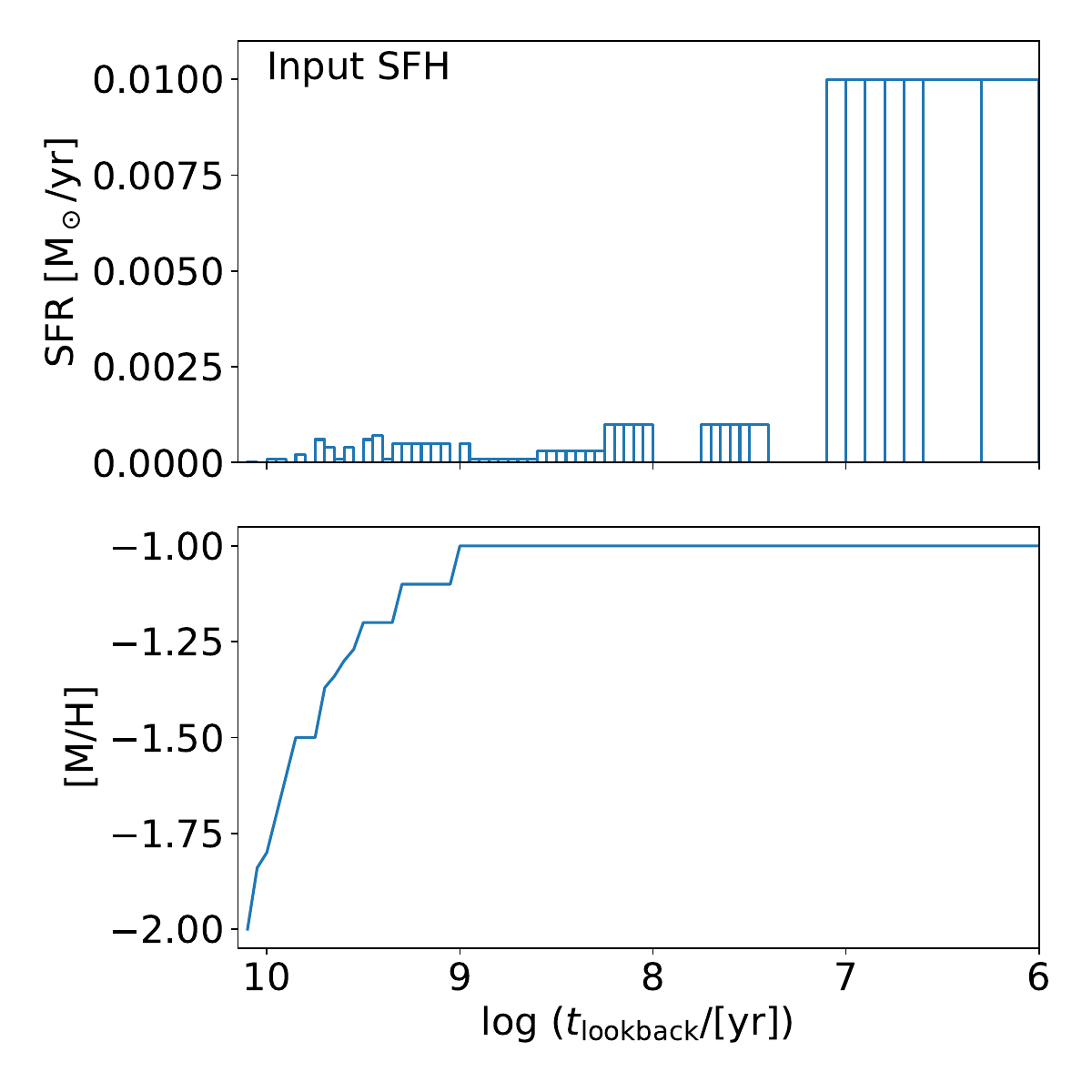}
       \caption{Input SFH and chemical enrichment history used to simulate UV and optical CMDs. The input SFH is based on a modified version of Leo A's best-fit SFH and chemical enrichment history from the UVopt solution (Section~\ref{sec:results}), with a constant SFR of 0.01~\Msun~yr$^{-1}$ over the past 10~Myr to enhance the upper MS, facilitating a clearer assessment of the impact of UV inclusion.
      \label{fig:inputsfh}}
\end{figure}

\begin{figure*}[ht]
 \centering
      \includegraphics[width=\linewidth]{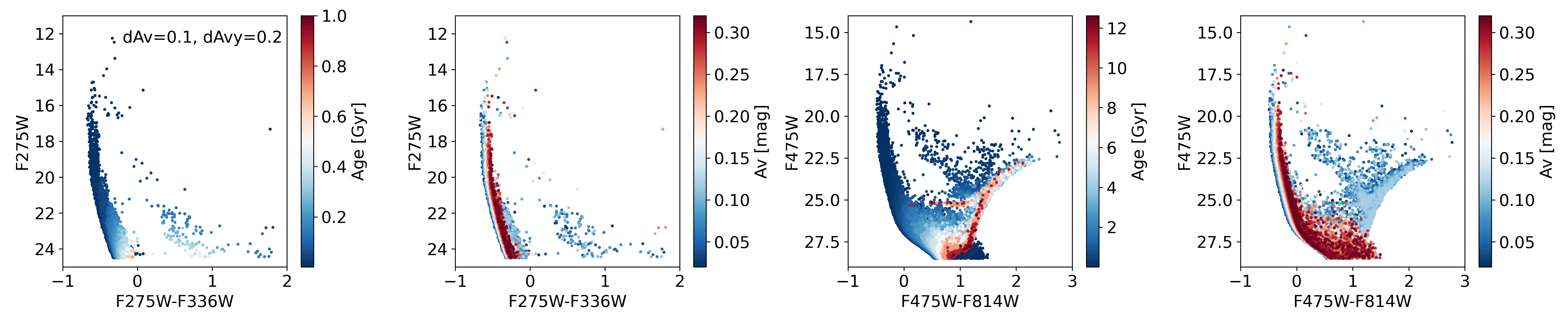}
      \includegraphics[width=\linewidth]{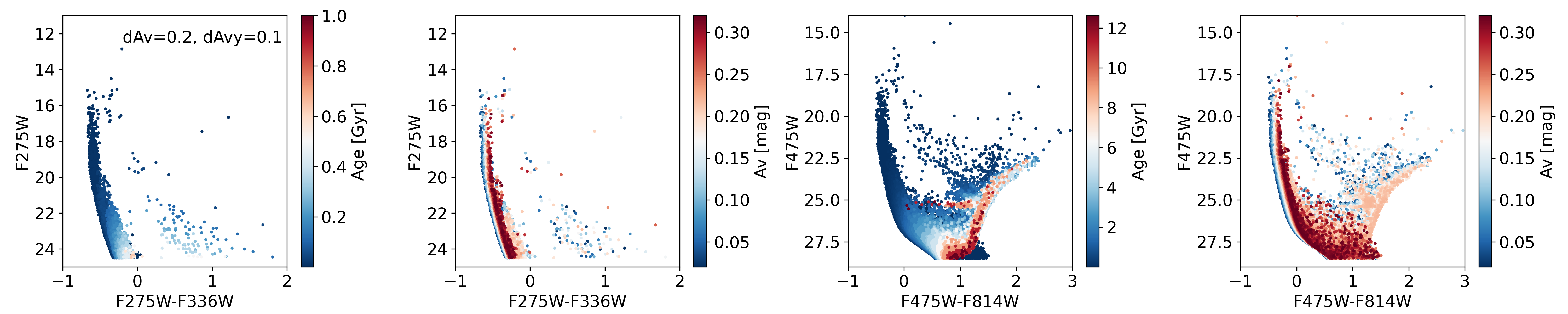}
      \includegraphics[width=\linewidth]{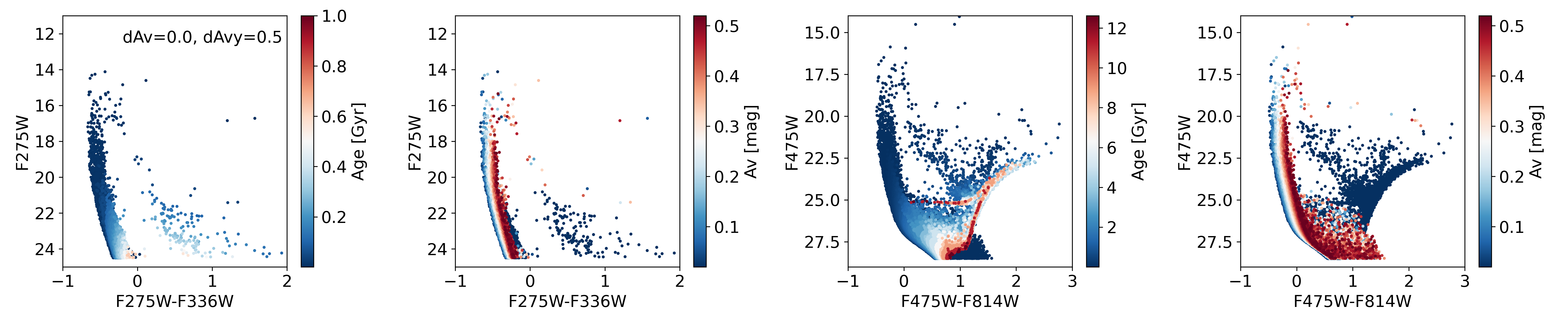}
      \includegraphics[width=\linewidth]{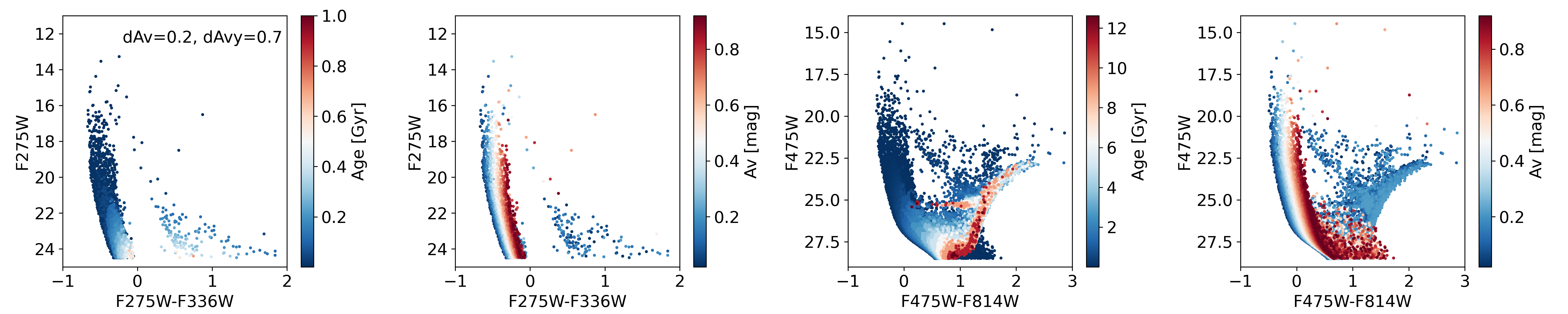}
       \caption{Each row displays an example realization of synthetic UV and optical CMDs for a Leo A-like galaxy, with exaggerated recent SF activity, modeled under different dust extinction scenarios: (\dAv, \dAvy) $=$ (0.1, 0.2), (0.2, 0.1), (0.0, 0.5), and (0.2, 0.7) from top to bottom. The left two panels display UV CMDs color-coded by stellar age and total dust, and the right two panels show optical CMDs with the same color coding. This layout highlights the impact of dust on the appearance of UV and optical CMDs, as well as the effect of stochastic sampling in the high end of the IMF. Note that these CMDs are shown prior to applying photometric uncertainties as well as completeness limits to maintain clarity in presenting the impacts of stellar age and total dust on the locations of stars in the CMDs.
      \label{fig:synCMDs}}
\end{figure*}

\begin{figure*}[ht]
 \centering
      \includegraphics[width=\linewidth]{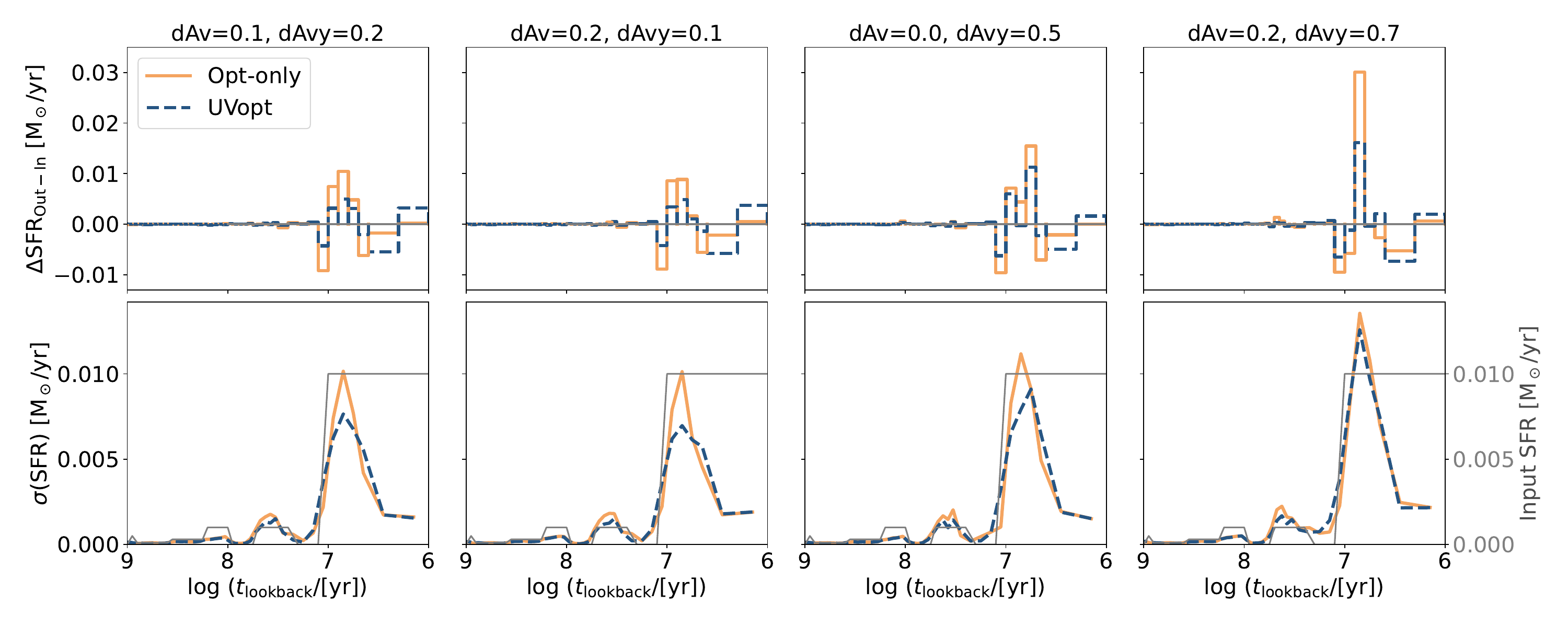}
       \caption{Test results from the simulated CMD fitting. The top row presents the differences between the recovered and input SFHs for four distinct dust configurations. The recovered SFHs closely match the input SFH for ages older than 1~Gyr, so the plots are limited to a 1~Gyr lookback time. The bottom row displays the standard deviation of SFRs in each time bin, calculated from the 100 best-fit SFHs for each dust configuration. The gray line denotes the input SFH. Across all cases, the UVopt scenario consistently demonstrates reduced deviation from the input SFH on average and smaller uncertainties in the resulting SFHs compared to the Opt-only case, underscoring the enhanced precision gained by including UV data. As total dust content increases, both deviations and uncertainties also increase, though the UVopt case continues to recover the input SFH more accurately than the Opt-only case.
      \label{fig:res_synCMDs}}
\end{figure*}

\begin{figure}[ht]
 \centering
      \includegraphics[width=\linewidth]{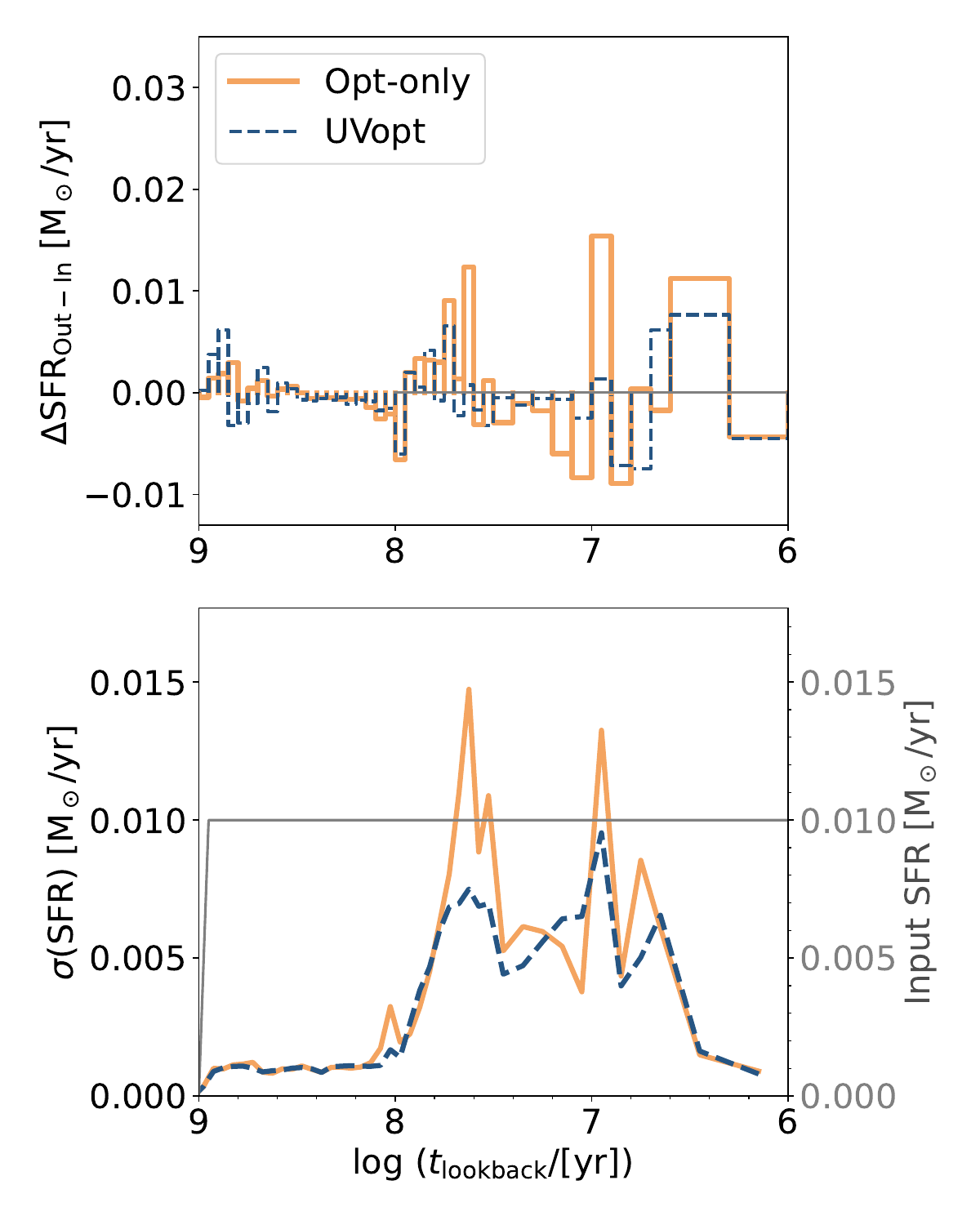}
       \caption{Same as Figure~\ref{fig:res_synCMDs}, but for the constant SFH with (\dAv, \dAvy) $=$ (0.0, 0.5) as an input, where the SFR is set to 0.01~~\Msun~yr$^{-1}$ over the past 1~Gyr and 0 at earlier times. 
      \label{fig:res_constSFH}}
\end{figure}

\begin{figure*}[ht]
 \centering
    \includegraphics[width=\linewidth, trim={0cm 0cm 0cm 2cm}, clip]{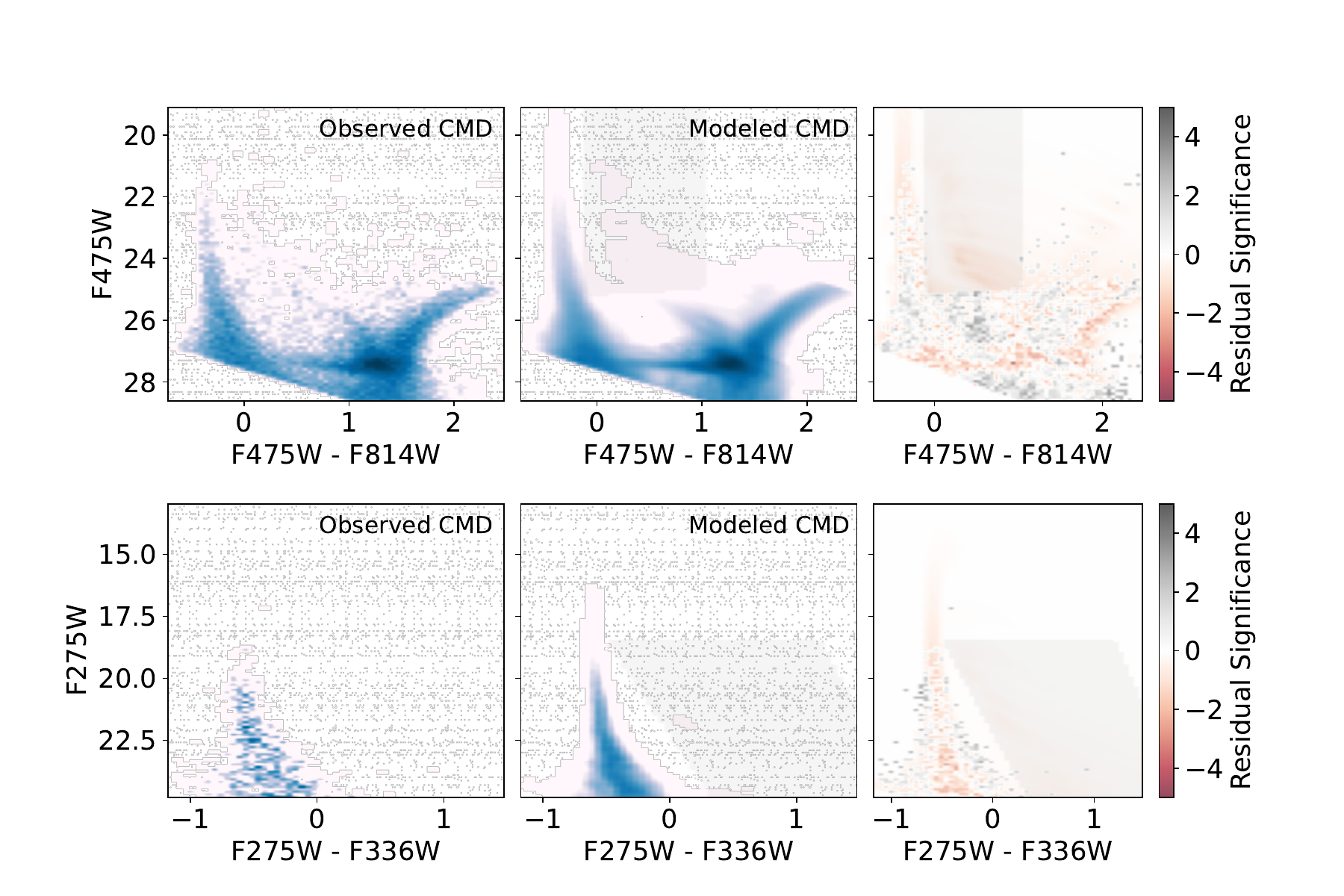}
    \caption{As an example, for UGC 8091, we present Hess diagrams for the optical (top row) and UV (middle row) CMDs, showing the observed data, best-fit model, and residual significance, with the exclusion region mask applied. The residual significance diagram quantifies deviations between observed and modeled distributions, where positive values indicate an excess of observed stars in a given color-magnitude bin, and negative values denote model overpredictions. Fewer than 1\% of color-magnitude bins exhibit residual significance exceeding $\pm$3 in both the optical and UV CMDs across all 10 galaxies. The complete figure set (10 images) is available in the online journal. \label{fig:HessDiagrams}}
\end{figure*}

\begin{figure*}[ht]
 \centering
    \includegraphics[width=\linewidth]{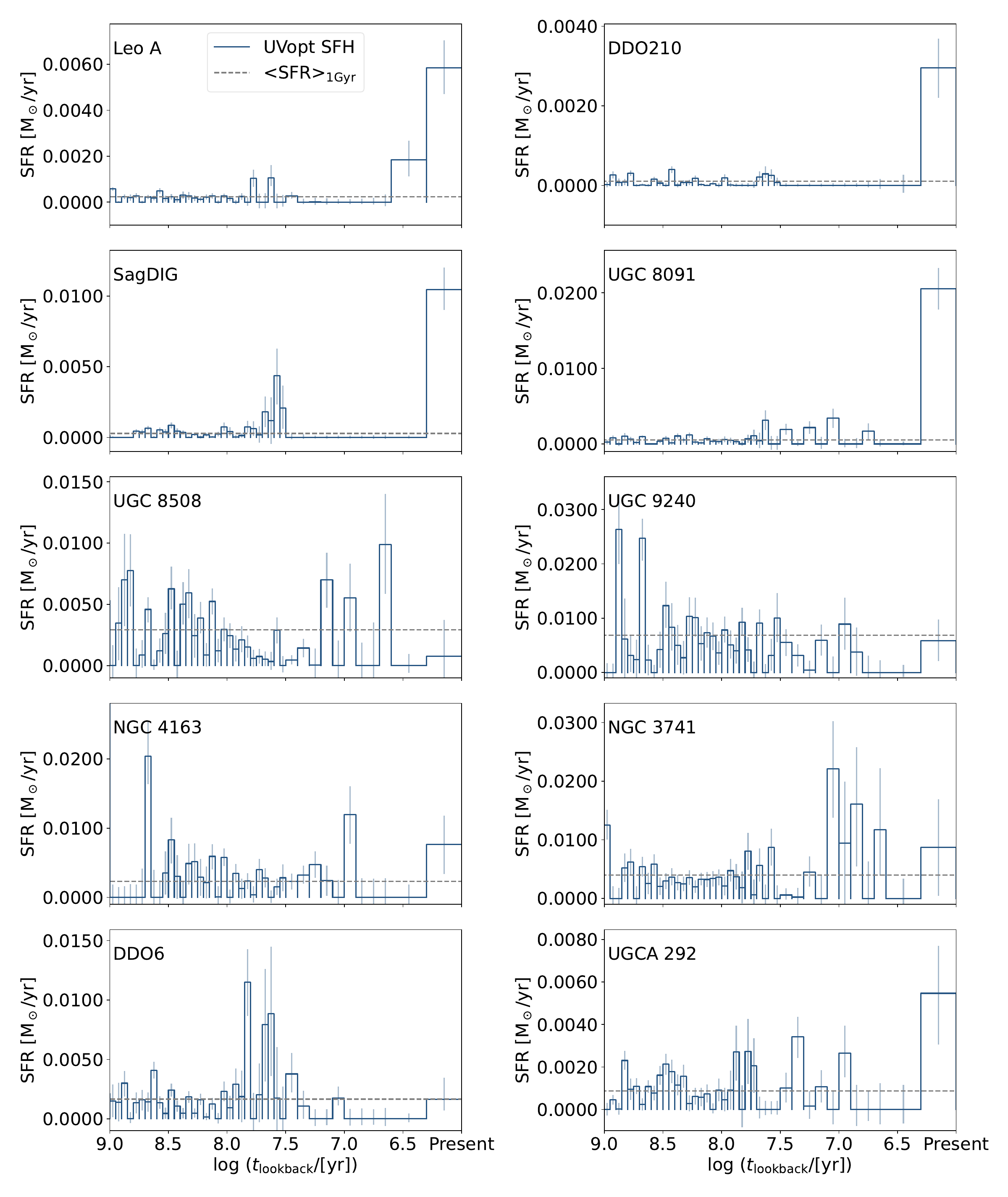}
    \caption{Differential SFHs for the 10 galaxies over the past 1~Gyr, with error bars indicating 1-$\sigma$ random uncertainties. For each galaxy, the gray dashed lines represent the average SFR over the past 1~Gyr. The SFHs exhibit stochastic fluctuations, with all 10 galaxies showing multiple age bins where the SFRs exceed their \sfravg{1}{Gyr} by a factor of a few to several. This underscores the episodic nature of SF in these metal-poor dwarf galaxies.
    \label{fig:uvoptSFH}}
\end{figure*}

\begin{figure*}[ht!]
 \centering
    \includegraphics[width=\linewidth]{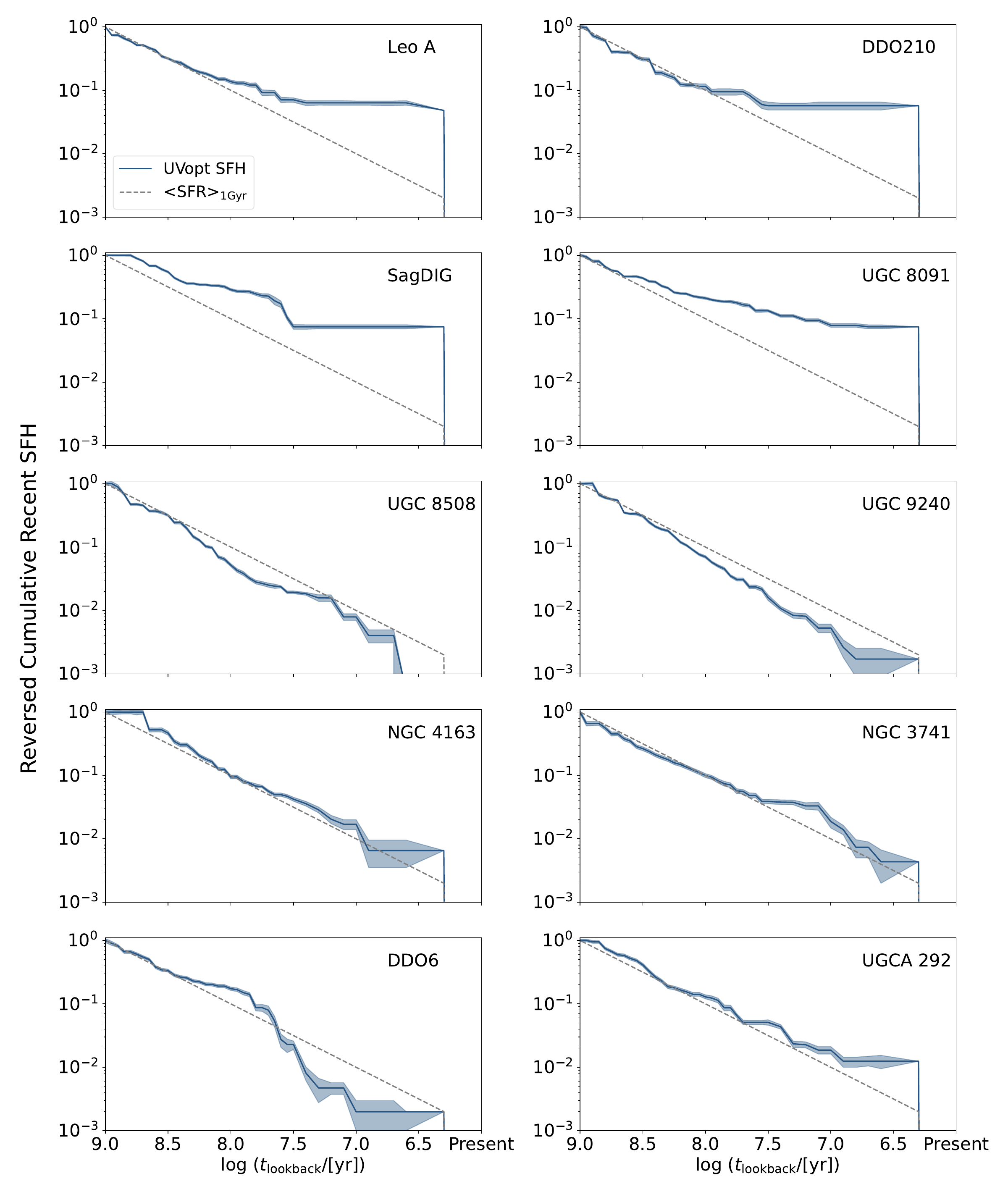}
    \caption{Reversed cumulative SFHs of the past 1 Gyr for the 10 galaxies, with a shaded envelope indicating 1-$\sigma$ random uncertainties. To better visualize the fraction of stellar mass formed in recent time bins, we reverse the cumulative calculation backward in time, setting the cumulative stellar mass fraction to 0 at the present and 1 at 1~Gyr ago, and display the Y-axis on a logarithmic scale. For each galaxy, the gray dashed lines represent a constant SFH scenario, illustrating the average SFR over the past 1~Gyr and appearing as a straight line. The abrupt drop below log$_{10}$(t/[yr]) = 6.3 arises from the challenge of representing a value of 0 on a logarithmic scale. 
    \label{fig:uvoptcumSFH}}
\end{figure*}

\subsection{Refining recent SFHs with UV CMD Integration: Insights from Simulated CMDs}\label{sec:simulations}
It has been proposed that adding UV observations improves the time resolution in measuring the recent SFH \citep[e.g.,][]{Calzetti2015, Cignoni2019}. However, to the best of our knowledge, no rigorous studies have quantitatively assessed the extent to which incorporating UV CMDs into the CMD modeling process can reduce SFH uncertainties or enhance time resolution.

To evaluate the impact of incorporating a UV CMD on SFH measurements, we recover SFHs from synthetic photometry using two different approaches. The first approach models only an optical (F475W-F814W, F475W) CMD, referred to as the ``Opt-only'' case. The second approach simultaneously fits both a corresponding UV (F275W-F336W, F275W) CMD and the optical CMD, referred to as, again, the ``UVopt'' case. Since this test is based on synthetic CMDs, there are no discrepancies in the HeB phases between the synthetic input CMDs and the modeled CMDs. Therefore, the BHeB exclusion step is omitted in this test. The only difference between the Opt-only and UVopt cases in this test is the inclusion of a UV CMD, which is added exclusively for the UVopt case, while all other fitting parameters remain identical for both cases.

Figure~\ref{fig:inputsfh} presents our input SFH and chemical enrichment history used to simulate UV and optical CMDs. These inputs are based on a benchmarked and simplified version of Leo A's best-fit SFH and chemical enrichment history, as determined from the UVopt solution (Section~\ref{sec:results}). In particular, we set a constant SFR of 0.01~\Msun~yr$^{-1}$ over the past 10~Myr to enhance the upper MS. This approach reduces stochastic effects in the upper MS and allows for a clearer assessment of the impact of UV inclusion. 

Using these inputs, we generate 100 pairs of UV and optical CMDs simultaneously using the \texttt{fake} routine from \texttt{MATCH} in the same four filters where our galaxies are observed: WFC3/UVIS F275W and F336W, and WFC/ACS F475W and F814W. For these simulations, we assume a Leo A-like galaxy and adopt the same Kroupa IMF, distance, MW foreground extinction, binary fraction, and photometric depth and quality (see Table~\ref{tab:param_summary}). After generating synthetic photometry in the four filters, we apply the real AST results to individual synthetic stars to assign realistic photometric uncertainties. We repeat the generation of 100 pairs of CMDs for various combinations of \dAv\,and \dAvy. Here we present the results that cover the relevant range of the dust extinction to our sample: (\dAv, \dAvy) $=$ (0.1, 0.2), (0.2, 0.1), (0.0, 0.5), and (0.2, 0.7). 

In Figure~\ref{fig:synCMDs}, each row presents, from left to right, an example realization of synthetic UV CMDs color-coded by stellar age and total applied dust extinction, followed by synthetic optical CMDs with the same color scheme. These synthetic CMDs showcase one of 100 realizations for each dust combination. It is clear that dust alone can cause a significant increase in the width of a given CMD. Additionally, the variation in the distribution of bright MS stars across the example CMDs, which cannot be attributed to differences in the amount of applied dust, highlights the stochastic sampling of the high-mass end of the IMF, despite all being generated from the same input SFH. The relative standard deviation (RSD) map, defined as the standard deviation-to-mean ratio per color and magnitude bin across 100 synthetic CMDs, indeed reveals regions where stochastic IMF sampling effects are significant--primarily the upper MS and blue loop, which are dominated by younger stars. High RSD values also appear on the red side of the RGB, largely due to the stochastic application of total dust extinction. These trends in the RSD maps remain consistent across different dust extinction scenarios.

We use the AST-applied synthetic CMDs as input for \texttt{MATCH}, treating them as if they were observed CMDs, to measure SFHs for both the Opt-only and UVopt cases across 100 realizations for each \dAv\,and \dAvy\,combination. We fit the three dust parameters (\Avmw, \dAv, \dAvy) for both cases, by searching a grid space with a step size of (0.01 dex, 0.1 dex, 0.1 dex) for each parameter, while keeping the distance modulus fixed. This approach leverages the UV CMD’s strength in constraining young stars and dust, acknowledging its limited effectiveness in refining distance measurements. The primary goal of this experiment is to assess how incorporating a UV CMD reduces SFH uncertainties in real galaxies while minimizing the impact of other variables. Note that, while the SFH solutions from \texttt{MATCH} use the same settings as those applied to real galaxies, including metallicity grid ranges and the \texttt{-zinc} flag, the metallicity distribution function used to create the simulated CMDs is different from that used for the solutions. Specifically, the \texttt{fake} routine populates stars uniformly within a specified [M/H] range for each age bin. In contrast, the \texttt{calcsfh} routine with the \texttt{-zinc} flag models the CMDs using a Gaussian metallicity distribution with a fixed width of 0.15~dex per age bin. Other potential sources of systematic error, such as uncertainties in the isochrones, are not accounted for in these test results.

For each dust combination, we measure a best-fit SFH for each of the 100 realizations for both the Opt-only and UVopt cases. We then calculate the average Opt-only and UVopt SFHs and their standard deviation from the 100 best-fit SFHs for each case. In both the Opt-only and UVopt cases, the recovered SFHs and \dAv\, and \dAvy\,values align well with the input values. While the UVopt case more accurately recovers the \Avmw\,parameter, the difference in performance is nearly equivalent to our grid step size. This overall consistency reaffirms the robustness and reliability of the CMD modeling technique, demonstrating its effectiveness in accurately reconstructing SFHs and dust properties both with and without a UV CMD, particularly when there is no mismatch between observed BHeB stars and stellar evolutionary models. Unfortunately, we face this challenge in real galaxies at low-metallicity environments, as discussed in the previous two sections. 

Figure~\ref{fig:res_synCMDs} illustrates the difference between the mean recovered SFH and input SFH over the past 1~Gyr (top panels), along with the standard deviation of SFRs in each time bin (bottom panels) for each dust combination. While SFH recovery seems robust for both cases, the UVopt case consistently achieves smaller absolute deviations from the input SFH across all dust combinations. This improvement stems from the inclusion of a UV CMD, which strengthens constraints on young, hot stars and helps break the age–metallicity degeneracy. As a result, the ambiguity between older, metal-poor and younger, metal-rich solutions is reduced, leading to more accurate SFH reconstructions. Notably, these deviations increase with the total applied dust and become more pronounced at ages below 100~Myr. As dust content increases in the model, the enhanced extinction within the stellar populations exacerbates the challenge of disentangling age and metallicity. This, in turn, introduces greater uncertainties and larger deviations in the SFH recovery, particularly for young populations, where dust extinction more significantly alters the observed CMD. 

To quantitatively assess the impact of UV addition on constraining recent SFH, we adapt the ``Wasserstein distance'' as a metric to evaluate how much a recovered SFH is different from an input SFH. If two SFHs are identical, the Wasserstein distance should be zero. A larger Wasserstein distance indicates that the recovered SFH is more different from the input SFH. We compute the Wasserstein distance for the Opt-only and UVopt cases over the 10~Myr, 100~Myr, and 1~Gyr timescales for each dust combination, then take the ratio of Opt-only to UVopt. This approach is used because the Wasserstein distance is consistently close to zero (on the order of 10$^{-5}$) for all cases prior to 1~Gyr ago, indicating excellent SFH recovery in older time bins. This result is expected, given the great depth of the simulated CMDs (see Figure~\ref{fig:synCMDs}). The computed ratios for dust combinations (\dAv, \dAvy) = (0.1, 0.2), (0.2, 0.1), (0.0, 0.5), and (0.2, 0.7) are as follows: 1.404, 1.512 and 1.404, 1.353, 1.473 and 1.272, 1.301, 1.286 and 1.318, and 1.569, 1.431 and 1.431, for the 10~Myr, 100~Myr, and 1~Gyr timescales, respectively. In summary, the Opt-only SFHs deviate more from the input SFH than the UVopt SFHs across all dust cases and timescales. These results confirm that incorporating UV data into CMD modeling improves the recovery of the true recent SFHs. 

As shown in the bottom row in Figure~\ref{fig:res_synCMDs}, our results from simulated CMDs indicate that the UVopt case achieves reduced uncertainties in recovered SFHs compared to the Opt-only case, particularly in recent time bins ($<$100~Myr), underscoring the improved precision gained by incorporating UV data into SFH measurements. More specifically, at the peak uncertainty (found between 5--10~Myr), the Opt-only cases' uncertainties are approximately 33\%, 45\%, 23\%, and 8\% higher than those of the UVopt cases for dust combinations (\dAv, \dAvy) = (0.1, 0.2), (0.2, 0.1), (0.0, 0.5), and (0.2, 0.7), respectively. However, as total dust content increases, uncertainties also rise. 

When averaged over the past 10~Myr, 100~Myr and 1~Gyr, incorporating a UV CMD reduces overall SFH uncertainties by approximately 4--8 \%, 8–20\% and 8–14\%, respectively, depending on the dust content. The corresponding uncertainty ratios of the Opt-only to UVopt cases on the 10~Myr, 100~Myr and 1~Gyr timescales are (1.074, 1.248, 1.163), (1.086, 1.230, 1.167), (1.076, 1.149, 1.125), and (1.041, 1.090, 1.088) for each dust combination. The improvements are more pronounced in cases with lower total dust content. In the highest dust scenario (\dAv, \dAvy) = (0.2, 0.7), the inclusion of UV data yields only a modest overall precision improvement of $\sim$4\% at $\tau$ =  10~Myr and $\sim$8\% at $\tau$ = 100~Myr and 1~Gyr. However, the UVopt case still demonstrates higher accuracy in recovering the input SFH compared to the Opt-only case. Notably, the most significant improvements are observed on 100~Myr timescales, rather than at 10~Myr or 1~Gyr in all dust cases. 

Finally, an intriguing pattern emerges in the past 100~Myr, where uncertainties are particularly larger in the 4--10~Myr and 30--60~Myr age ranges. The smaller uncertainty inflation (but large fractional uncertainty) observed in the 30--60~Myr range is likely attributed to the Leo A-like input SFH, which has a lower SFR in that age range. To assess whether this behavior is specific to our input SFH, we conduct an additional test with a constant input SFH with (\dAv, \dAvy) $=$ (0.0, 0.5). As shown in Figure~\ref{fig:res_constSFH}, the input SFR for this test is set to 0.01~~\Msun~yr$^{-1}$ over the past 1~Gyr and 0 at earlier times. The results confirm the robustness of our findings: elevated uncertainties in the 4--60~Myr age range, encompassing the two age bins that stood out in tests with the more realistic input SFH, persist in the constant SFH tests. This consistent pattern of increased uncertainties within the same age range, regardless of the input SFH, suggests that these uncertainties are primarily driven by degeneracies in the SEDs of stars corresponding to these ages, rather than by the input SFH or uncertainties in the stellar evolutionary models, as the same models were used to produce and fit the CMDs. These degeneracies, particularly at younger ages, arise from the fact that similar SEDs can result from different combinations of stellar age, mass, metallicity, and dust extinction. The inclusion of UV data helps mitigate these degeneracies, leading to the smaller uncertainties observed in the UVopt case.

In conclusion, while the degree of improvement varies across individual time bins, timescales, and dust configurations--making it difficult to generalize the precision gain--our results from synthetic CMDs clearly demonstrate, for the first time, the substantial impact of incorporating a UV CMD on the precision and accuracy of CMD-based SFH measurements, especially in recent time bins.

\section{Results}\label{sec:results}
In this section, we present and discuss the UVopt SFHs of 10 galaxies, and compare the CMD-based average SFRs with the two widely-used SFR indicators based on H$\alpha$ and FUV fluxes. 

\subsection{Differential and Cumulative UVopt SFHs}
Figure~\ref{fig:HessDiagrams} illustrates the CMD modeling for UGC 8091, as an example, with Hess diagrams of the optical and UV CMDs showing observed data, best-fit model, and residual significance, with the exclusion region mask applied. The residual significance quantifies the degree of deviation between observed and modeled Hess diagrams. Positive values reflect an excess of observed stars in a given color-magnitude bin, while negative values indicate model overpredictions. Across all 10 galaxies, fewer than 1\% of color-magnitude bins (and less than 0.5\% for the majority) exhibit residual significance exceeding $\pm$3 in their optical and UV CMDs, confirming the reliability of our fits. Figures for the remaining galaxies are available in the online journal. 

Figure~\ref{fig:uvoptSFH} displays the differential SFHs for all 10 galaxies as a function of lookback time over the past 1~Gyr. The youngest age bin in our solutions, log$_{10}$(t/[yr]) = 6.0--6.3 (1--2~Myr), integrates all SF from 0 to 2~Myr, effectively doubling the reported SFR within its 1~Myr-wide bin. To correct for this, we adjust the SFR in this bin by applying a factor of 0.5, ensuring it accurately represents the true SFR over this timescale. Consequently, the adjusted SFR is displayed for the bin spanning log$_{10}$(t/[yr]) = 6.3 to the present day. This correction is consistently applied throughout the analysis presented in this paper. In each panel, the gray dashed lines mark the average SFR over the past 1~Gyr for each galaxy, denoted as \sfravg{1}{Gyr}. We find that all galaxies exhibit significant temporal fluctuations in their SFHs, which has been found to be a characteristic of low-mass star-forming galaxies \citep[e.g.,][]{McQuinn2009, McQuinn2010a, McQuinn2010b, Weisz2014}. Even in galaxies with minimal recent SF activity, like Leo A, certain age bins show SFRs substantially exceeding their \sfravg{1}{Gyr} over the past 1~Gyr. 

In these plots, we display only the 1-$\sigma$ random uncertainties (68\% confidence interval around the best-fit SFH solution), because these uncertainties are most relevant when comparing each galaxy's SFH to its corresponding Opt-only SFH (Sec.~\ref{sec:UVimpact}). \texttt{MATCH} estimates random uncertainties in the SFH measurements by generating samples proportional to the posterior probability density of the SFH parameters using the hybrid Markov Chain Monte Carlo algorithm \citep{Duane1987}. The random uncertainties are particularly important for sparsely populated CMDs, such as those of the low-mass galaxies studied here \citep{Dolphin13}. Their CMD regions corresponding to young stellar populations are particularly underpopulated due to the sparse sampling of the upper end of the stellar IMF, a consequence of both the low SFR and the stochastic nature of massive SF. 

Figure~\ref{fig:uvoptcumSFH} displays the reversed cumulative SFHs for all 10 galaxies, focusing on the past 1~Gyr. The reversed cumulative SFH plot offers a clearer view of the fraction of total stellar mass formed in recent time bins. Contrary to the classic cumulative SFH, it starts at a value of 0 for the present day and increases to 1 at 1 Gyr ago. In each panel, the gray dashed line represents the cumulative SFH under the assumption of a constant SFR over the past 1~Gyr. A steeper slope relative to this line in any time interval indicates a SFR $>$ \sfravg{1}{Gyr}, while a shallower slope denotes a SFR $<$ \sfravg{1}{Gyr} within that particular time interval. Some galaxies (e.g., SagDIG, UGCA 292) show minimal to no overlap with the constant SFR case across all time bins, while others roughly align with the constant SFH in at least some time periods. For example, Leo A exhibits SFRs that are consistent with the constant SFH case in older age bins($>$ 300~Myr). Overall, all 10 galaxies show deviations from the constant SFH over the past 1~Gyr, though the timing, extent, and duration of these deviations vary among them.

Compared to the constant SFR scenario, four galaxies (Leo A, DDO 210, SagDIG, UGC 8091) have formed a significant fraction (almost 10\%) of their stellar mass in the most recent few Myr, as indicated by steep slopes near log$_{10}$(t/[yr]) = 6.3. Another three galaxies (NGC 4163, NGC 3741, UGCA 292) exhibit a similar trend, though to a much lesser extent. In contrast, UGC 8508, UGC 9240, and DDO 6 have formed a smaller fraction of their stellar mass compared to the constant SFR case during the past 10~Myr. This trend persists up to $\sim$300~Myr ago for UGC 8508 and UGC 9240, indicating a relative dominance of older stellar populations relative to the constant SFR case in these galaxies. Meanwhile, DDO 6 has formed a substantial fraction of its stellar mass during the period between 10--100~Myr ago.

\begin{figure*}[ht!]
 \centering
      \includegraphics[width=\linewidth]{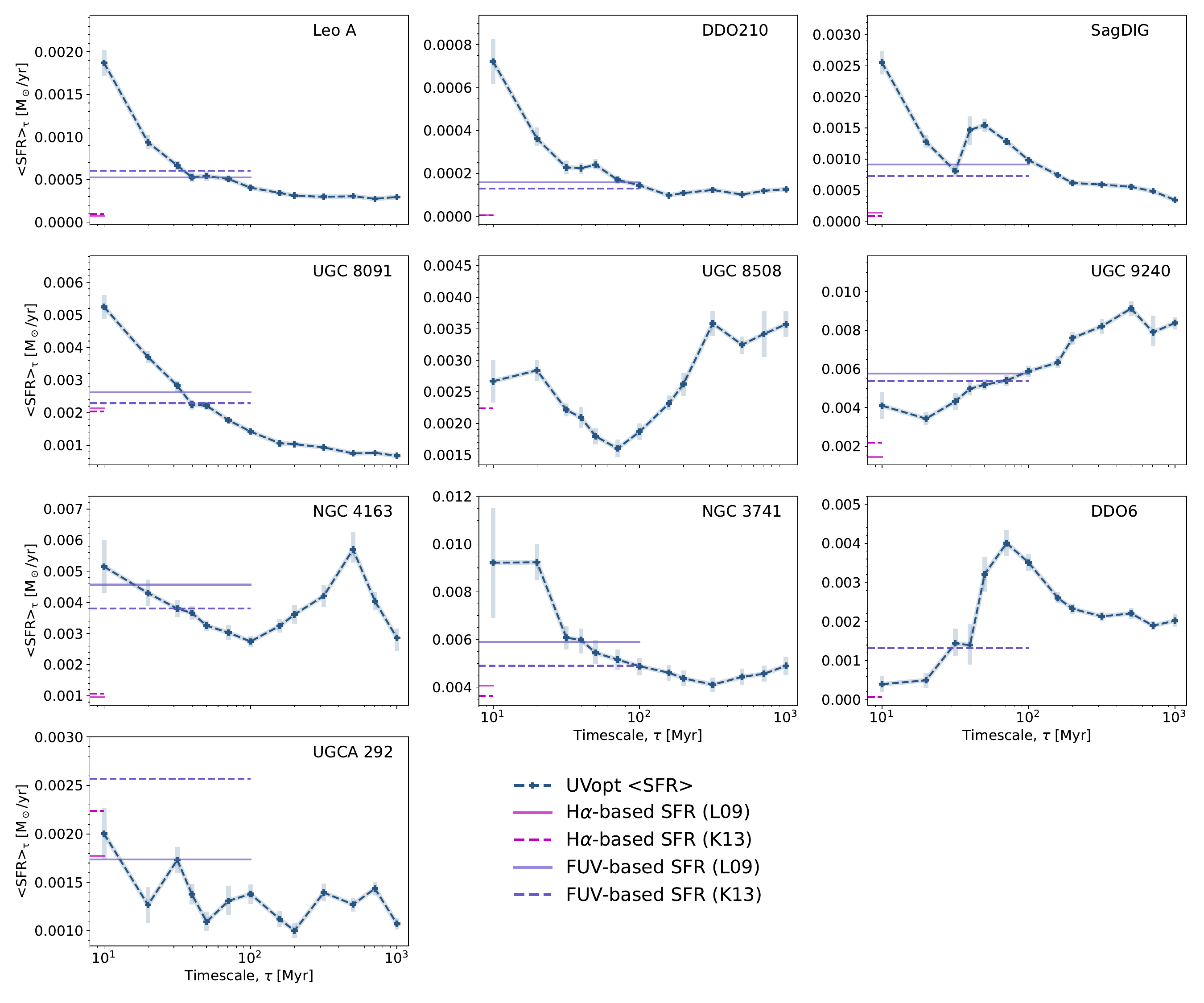}
       \caption{CMD-based average SFRs from the UVopt SFH across various timescales (10~Myr, 20~Myr, 30~Myr, 40~Myr, 50~Myr, 70~Myr, 100~Myr, 160~Myr, 200~Myr, 300~Myr, 500~Myr, 700~Myr, and 1~Gyr) are illustrated. We also plot the integrated flux-based SFRs for each galaxy, adopted from the literature. Magenta lines indicate dust-corrected H$\alpha$-based SFRs, whereas purple lines indicate dust-corrected FUV-based SFRs. The solid lines are from \citet{Lee2009} and the dashed lines are from \citet{Karachentsev2013}. These literature SFRs are listed in Table~\ref{tab:averageSFRs} along with those computed from our CMD-based SFHs. 
      \label{fig:avgSFR}}
\end{figure*}

\subsection{Average SFRs Across Different Timescales}\label{sec:avgSFR}
In this section, we present the average SFRs derived from the UVopt SFHs over various timescales ($\tau$), spanning 10~Myr to 1~Gyr. We compute the average SFRs as a function of $\tau$ by dividing the total stellar mass formed between 0--$\tau$~Myr ago by $\tau$. The upper and lower uncertainties are determined based on the respective upper and lower total stellar masses formed during the same period. We then compare these results with the literature SFRs derived from H$\alpha$ and FUV emission. To ensure the comparison is as fair as possible, we scale the UVopt SFHs by multiplying by 1.03493 to account for differences in IMF assumptions and stellar mass limits between our study (Kroupa, 0.1--350~\Msun) and those used in the literature for SFR conversion factors \citep[Salpeter, 0.1--100~\Msun;][]{Kennicutt1998}.

H$\alpha$ emission and FUV flux are two widely used SFR indicators \citep[e.g.,][]{Kennicutt1994, Kennicutt1998, Kennicutt2012}. H$\alpha$ emission comes from ionized gas surrounding young, massive stars (mostly O-type and early B-type stars). These stars emit significant amounts of ionizing photons that ionize the surrounding neutral hydrogen gas, leading to H$\alpha$ recombination emission. The lifetimes of these massive stars are short, on the order of a few million years. For an instantaneous SF event, their ionizing photon production rate declines sharply, reaching approximately 1\% of its initial value by $\sim$10~Myr \citep[e.g.,][]{Leitherer99}. Therefore, H$\alpha$ emission traces the most recent, short-term SF activity, typically within the last 10~Myr, a nominal timescale. We adopt the H$\alpha$-based SFRs from \citet[hereafter L09]{Lee2009} and \citet[hereafter K13]{Karachentsev2013}.

In contrast, FUV emission originates from a broader population of stars, including both massive O- and late B-type stars (with masses down to $\sim$3~\Msun), which contribute to non-ionizing UV radiation. Since lower mass stars have longer lifetimes than the most massive stars, FUV flux can trace SF over a longer period, up to about 300~Myr \citep[e.g.,][]{Weisz2012}, with a nominal timescale of 100~Myr. We adopt the FUV-based SFRs from \citetalias{Lee2009} and \citetalias{Karachentsev2013} as well. 

Because the SFR measurements from the literature involve their own complex process (data collection and reduction, various assumptions for choice of models, dust extinction correction, etc.), we do not delve into the detailed differences here. Readers are referred to the respective studies for a comprehensive explanation of their SFR estimation methods. In summary, the fractional differences in SFR$_{\rm H\alpha}$ and SFR$_{\rm FUV}$ between \citetalias{Lee2009} and \citetalias{Karachentsev2013} show median offsets of $\sim$19\% and $\sim$17\%, with maximum offsets of $\sim$38\% and $\sim$48\%, respectively. 

Furthermore, calibrating these SFR indicators requires several key assumptions. For instance, \citet{Kennicutt1998} adopted a Salpeter IMF with mass limits between 0.1 and 100~\Msun, assuming solar metallicity, constant SFR, and Case B recombination conditions \citep{Osterbrock1989}. Additionally, converting these observed fluxes to SFRs involves various corrections, such as accounting for internal dust attenuation and [NII] contamination in H$\alpha$ narrow-band imaging. While previous studies have made significant efforts to apply these corrections as accurately as possible, galaxy-to-galaxy uncertainties remain, which may contribute to the discrepancies discussed below.

Figure~\ref{fig:avgSFR} shows our average SFRs as a function of $\tau$, ranging from 10~Myr to 1~Gyr, as well as flux-based SFRs from H$\alpha$ and FUV. These flux-based literature SFRs are dust-corrected values, and are listed in Table~\ref{tab:averageSFRs}, alongside the CMD-based SFRs computed from this study. We plot the SFRs from the literature as horizontal lines, extending only across their respective nominal timescales to aid comparison with the CMD-based SFRs at those nominal timescales.

\subsubsection{CMD-based \sfravg{10}{Myr} vs. SFR$_{\rm H\alpha}$}\label{sec:comp_ha}
\begin{figure}[ht]
 \centering
      \hspace{-2mm}
      \includegraphics[width=\linewidth]{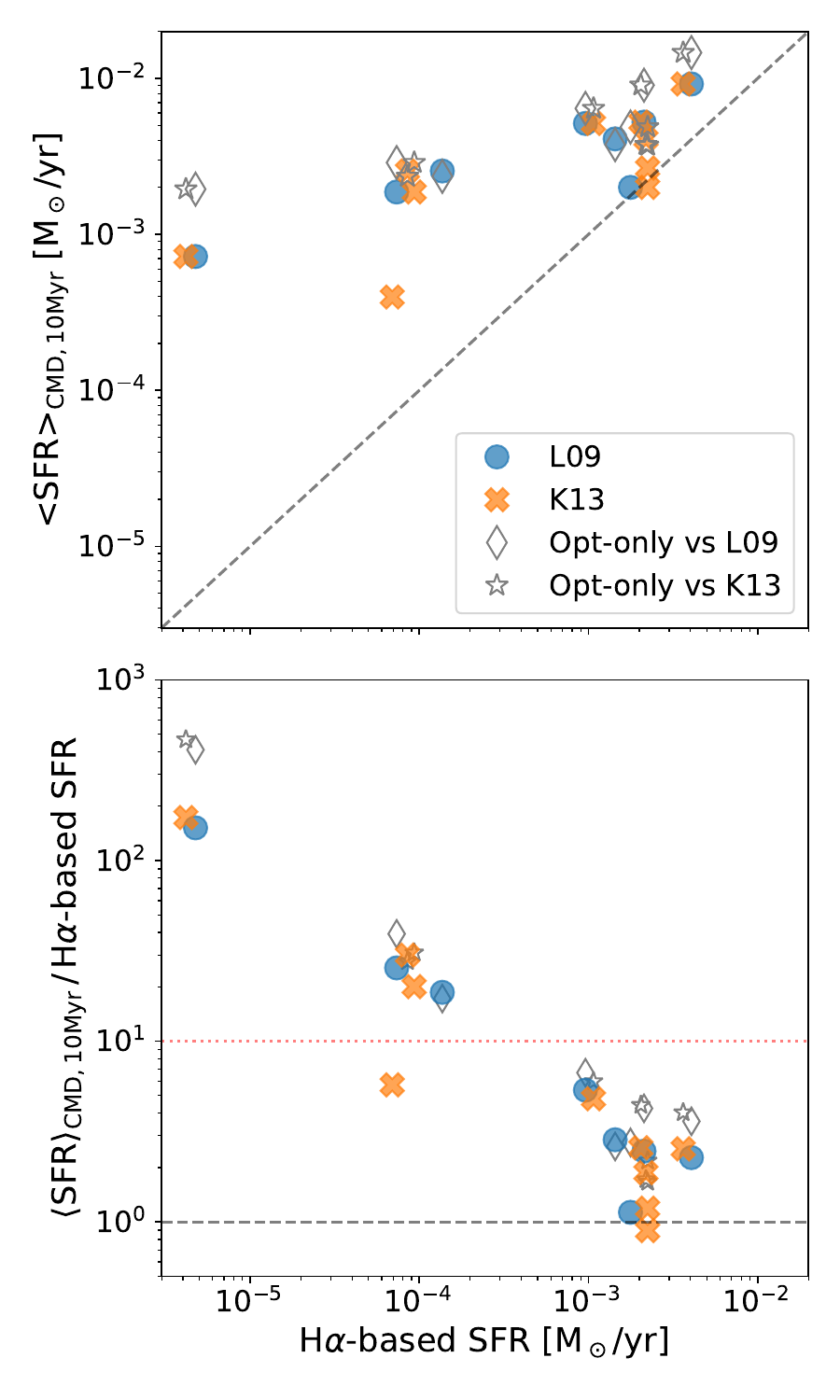}
      \hspace{-2mm}
       \caption{\textit{Top:} Comparison of the CMD-based \sfravg{10}{Myr} with the H$\alpha$-based SFRs from L09 (circles) and K13 (crosses). The dashed line represents the one-to-one relationship. We also present the comparison against \sfravg{10}{Myr} from the Opt-only SFHs (gray open symbols). \textit{Bottom:} The ratio of the CMD-based \sfravg{10}{Myr} to the H$\alpha$-based SFRs is plotted against the H$\alpha$-based SFRs, with the dashed line indicating a ratio of unity. The discrepancy between the H$\alpha$-based SFRs and the CMD-based \sfravg{10}{Myr} increases at lower SFRs. This trend is expected, as converting a global H$\alpha$ to a SFR becomes unreliable in this low-SFR regime, where the assumptions of a fully populated IMF and a constant SFR over the most recent 10~Myr break down \citep[e.g.,][]{Kennicutt2008}. The dotted line represents a ratio of 10, an estimated upper limit on the underestimation of SFR$_{\rm H\alpha}$ assuming the observed H$\alpha$ luminosity predominantly originates from a single starburst within the past 1 Myr.
      \label{fig:HaSFRs}}
\end{figure}

In Section~\ref{sec:simulations}, we show that the maximum fractional uncertainties for the UVopt case vary from $\sim$70 to $\sim$125\% in the most recent 10~Myr time bins, over which we assume a constant input SFR (see Figure~\ref{fig:res_synCMDs}). As shown in Figure~\ref{fig:HaSFRs}, even after accounting for this large uncertainty, all but two galaxies (UGC 8508 and UGCA 292) exhibit significantly higher \sfravg{10}{Myr} values compared to at least one of the two reported SFR$_{\rm H\alpha}$ measurements, with differences ranging from a factor of $\sim$2 to $\sim$170. The same behavior is observed in the comparison with the Opt-only SFHs (open symbols), which will be presented in Section~\ref{sec:UVimpact}. This significantly lower SFR$_{\rm H\alpha}$ than the CMD-based \sfravg{10}{Myr} can be partially explained by fluctuations in the SFHs. As illustrated in the differential SFHs (Figure~\ref{fig:uvoptSFH}), the majority of SF in these galaxies over the past 10~Myr occurred predominantly within the most recent few Myr. Thus, the assumption of a constant SFH used to calculate the conversion factor for SFR$_{\rm H\alpha}$ is invalid for these low-mass galaxies, which exhibit time-variable SFHs.

More specifically, the nominal timescale of 10~Myr for the H$\alpha$-based SFRs is non-universal at the low-mass regime because it holds true only under certain conditions, such as a fully populated IMF, a constant SFH with solar metallicity, no escape of ionizing photons from a star-forming system. For example, \citet{FloresVelazquez2021} investigated dwarf galaxies with highly time-variable SFHs in the FIRE simulations and found the typical timescale probed by H$\alpha$ to be $\sim$5~Myr, which is half of the nominal 10~Myr timescale. If the total observed H$\alpha$ luminosity were assumed to originate from just the past few Myr rather than 10~Myr, the discrepancy between our CMD-based \sfravg{10}{Myr} and SFR$_{\rm H\alpha}$ could be reduced by up to an order of magnitude when using the constant SF conversion factor \citep[e.g.,][]{Kennicutt1998}, bringing these estimates into closer agreement. Five of the remaining eight galaxies roughly fall into this category -- UGC 8091, UGC 9240, NGC 4163, NGC 3741, and DDO6. They fall below the dotted line in Figure~\ref{fig:HaSFRs}, which represents a ratio of 10, set as an estimated upper limit on the potential underestimation of SFR$_{\rm H\alpha}$ under the assumption that the H$\alpha$ luminosity originates from an instantaneous burst that occurred 1~Myr ago.

On the other hand, the other three galaxies above the dotted line --Leo A, DDO 210, and SagDIG-- clearly require additional factors beyond the constant SF assumption to explain the remaining discrepancy. Before delving into the potential causes, we briefly compare our SFHs for these galaxies with previous studies. Our SFHs generally align well with earlier findings, particularly the delayed SFH reported by \cite{Cole2007} and \cite{Lescinskaite2022} for Leo A, \citet{Cole2014} for DDO 210, and \citet{Weisz2014} for SagDIG. However, the significantly lower time resolution in their recent time bins and the older minimum stellar ages used to model CMDs in these studies make direct comparisons challenging. Nevertheless, their recent SFR results are consistent with ours. For example, for DDO 210, the SFR in the youngest time bin ($<$100~Myr) reported by \citet{Cole2014} aligns with our \sfravg{100}{Myr}. Similarly, \citet{Lescinskaite2022} measured the recent SFH of Leo A using individual stellar ages and reported an SFR of 0.0006\Msun\,/yr$^{-1}$ for the most recent $<$30~Myr, which agrees with our \sfravg{30}{Myr}. 

The remaining discrepancy seen in Leo A, DDO 210, and SagDIG may be due to additional physical factors not fully accounted in the computation of the H$\alpha$ SFR conversion, such as: (1) a lack of ionizing stars due to both incomplete stellar IMF sampling and stochastic sampling of the cluster mass function in galaxies with low SFRs, (2) the absorption of ionizing photons by neutral hydrogen gas and dust, and (3) the escape of ionizing photons without interacting with the surrounding neutral ISM. While rigorous testing of each of these effects is beyond the scope of this paper, we will briefly discuss each effect in more depth. 

First, the upper MS of these three galaxies are very sparsely populated, suggesting a lack of ionizing stars. For example, \citet{Gull2022} analyzed two H\textsc{ii} regions in Leo A and confirmed that a single early O-type star powers each region. Notably, Leo A contains only a handful of H\textsc{ii} regions in total \citep{vanZee2006b}. The net result of stochastic sampling of the stellar and cluster IMFs is the formation of a lower number of massive stars in low-mass galaxies than would otherwise be expected \citep[e.g.,][]{Weisz2012}. Consequently, the SFR$_{\rm H\alpha}$ measurements in these low-mass galaxies are likely biased low, as the reduced number of ionizing stars leads to lower H$\alpha$ luminosity than expected from a fully sampled IMF. 

Second, these three galaxies each exhibit a broader upper MS than the other galaxies in our sample, while also exhibiting a well-defined and narrow RGB branch, resulting in higher \dAvy\,values in the CMD fits (see Table~\ref{tab:param_summary}). This suggests there is a negligible amount of dust surrounding the older stellar populations in these galaxies, but a significant amount of dust surrounding their young stellar populations. This result contrasts with the zero nebular reddening values measured from H\textsc{ii} regions in Leo A using the Balmer decrement method \citep{vanZee2006b}. However, non-zero stellar extinctions have been reported for individual O/B stars \citep{Gull2022} and from galaxy SED fits \citep{Dale2023}. Dust may be highly localized around young stars, causing features in hydrogen emission lines to be diluted when averaged over a larger area outside H\textsc{ii} regions within the long-slit spectroscopy's field of view \citep[e.g.,][]{MacKenty2000, MaizApellaniz2004}. Ultimately, this localized dust obscuration likely causes an underestimation of the Balmer decrement and thus the intrinsic H$\alpha$ luminosity, biasing the SFR$_{\rm H\alpha}$ measurements to lower values than the actual SF activity.

Third, the possibility of ionizing photons escaping from the galaxies' SF regions cannot be ruled out, given the relatively faint H$\alpha$ luminosity of these three galaxies despite being gas-rich. In metal-poor dwarf galaxies, a significant fraction of ionizing photons could escape from individual H\textsc{ii} regions and eventually into the intergalactic medium through low-density channels created by stellar feedback \citep[e.g.,][]{Lee2009, Choi2020}. If this escape fraction is not accounted for in the intrinsic H$\alpha$ luminosity, SFR$_{\rm H\alpha}$ estimates will be systematically underestimated relative to the true SFR. 

Finally, it is worth noting that these galaxies share common characteristics: they are isolated, relatively gas-rich, and exhibit delayed SFHs \citep{Cole2007, Cole2014, Higgs2016}, which may directly or indirectly contribute to the above factors, resulting in the more significant discrepancies between the H$\alpha$-based SFRs and CMD-based \sfravg{10}{Myr}.

\begin{figure}[ht]
 \centering
      \hspace{-2mm}
      \includegraphics[width=\linewidth]{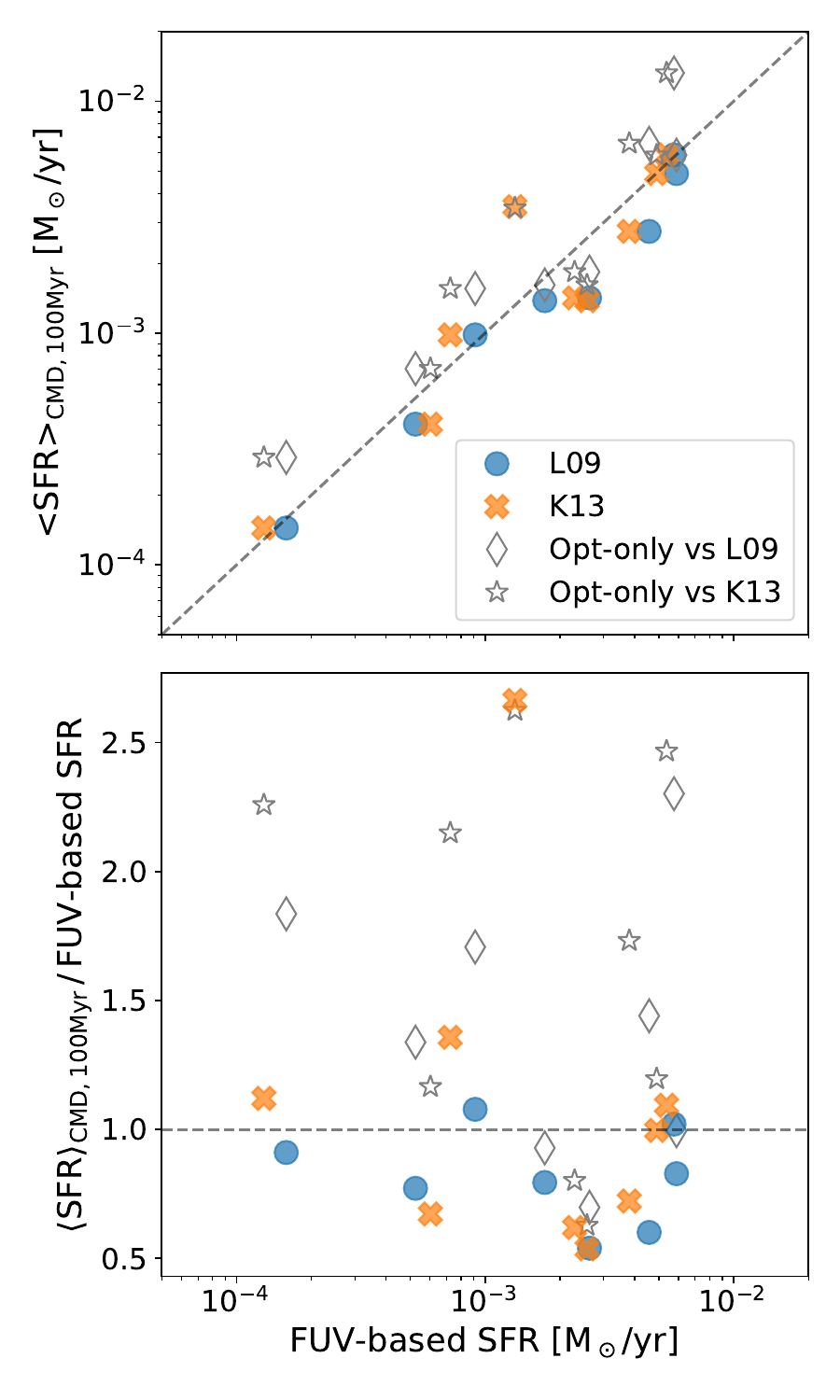}
      \hspace{-2mm}
       \caption{Comparison of the CMD-based \sfravg{100}{Myr} and the FUV-based SFRs. Symbols and lines are the same as in Figure~\ref{fig:HaSFRs}. In most galaxies, the CMD-based \sfravg{100}{Myr} values from the UVopt SFHs are lower than the FUV-based SFRs, whereas the \sfravg{100}{Myr} values from the Opt-only SFHs are higher. Despite the opposing trends between the two SFH cases, the overall deviations are moderate, with ratios ranging from approximately 0.5 to 2.
      \label{fig:fuvSFRs}}
\end{figure}

UGC 8508 and UGCA 292 are the two galaxies that show reasonable agreement between their UVopt \sfravg{10}{Myr} and at least one of the literature SFR$_{\rm H\alpha}$ measurements, indicating that the canonical assumptions underlying the H$\alpha$-SFR conversion might be valid for this particular galaxy. In fact, UGC 8508's differential SFH reveals multiple SF events within the last 10~Myr, with higher rates occurring in the earlier half of this period (i.e., 4--10~Myr ago). In addition, \citet{Berg2012} obtained an optical spectrum of one of the H$\alpha$ knots in UGC 8508 and reported the extinction at H$\beta$, resulting in $A_V \simeq$ 0.2, which is consistent with our best-fit \dAvy= 0.2.

For UGCA 292, its \sfravg{10}{Myr} aligns with the SFR$_{\rm H\alpha}$ reported both by \citetalias{Lee2009} and \citetalias{Karachentsev2013} within the uncertainty, although the \citetalias{Lee2009} value is $\sim$70\% lower than the \citetalias{Karachentsev2013} value. Unlike UGC 8508, UGCA 292 shows a pronounced SF event in the most recent time bin ($<$2~Myr), suggesting that the apparent agreement between \sfravg{10}{Myr} and the literature values may be coincidental rather than indicative of a true match in SFRs. Notably, this galaxy is one of the most gas-rich dwarfs in the local universe, with M(H\textsc{i})/L$_B$ = 7 \citep{Young2003}, and the farthest galaxy in our sample. These factors may result in the omission of embedded young massive stars from the stellar photometry catalog and thus SFH measurements, and/or the misidentification of unresolved clusters as multiple bright stars \citep[e.g.,][]{Choi2020}, making it more challenging to draw robust conclusions. A more detailed study is therefore necessary to investigate its SF properties, H$\alpha$ luminosity, and potential ionizing photon escape.

In summary, it is known that H$\alpha$ luminosity is a less robust SFR indicator for dwarf galaxies with low SFRs \citep[e.g.,][]{Lee2009, Weisz2012}, despite the ability to directly trace the ionizing radiation from massive stars. The significant discrepancy between \sfravg{10}{Myr} and SFR$_{\rm H\alpha}$ detected in the majority of our sample corroborates the lack of H$\alpha$'s robustness as a reliable SFR indicator in the low SFR regime \citepalias[below $\sim$0.003\Msun\,yr$^{-1}$;][]{Lee2009}.

\begin{deluxetable*}{lcccccc}
\tablecaption{Average SFRs derived from our CMD-based SFHs and those adopted from the literature using integrated flux measurements.
\label{tab:averageSFRs}}
\tablewidth{0pt}
\tablehead{
\colhead{Galaxy} & \colhead{\sfravg{10}{Myr}\tablenotemark{a}} & \colhead{\sfravg{100}{Myr}\tablenotemark{a}} & \colhead{SFR$_{\rm H\alpha,L09}$\tablenotemark{b}} & \colhead{SFR$_{\rm H\alpha,K13}$\tablenotemark{c}} & \colhead{SFR$_{\rm FUV,L09}$\tablenotemark{b}} & \colhead{SFR$_{\rm FUV,K13}$\tablenotemark{c}} \\
& (\Msun~yr$^{-1}$) & (\Msun~yr$^{-1}$) & (\Msun~yr$^{-1}$) & (\Msun~yr$^{-1}$) & (\Msun~yr$^{-1}$) & (\Msun~yr$^{-1}$)
}
\startdata
Leo A & \asymerr{0.001871}{0.000154}{0.000154} & \asymerr{0.000405}{0.000026}{0.000026} & 0.000073 & 0.000093 & 0.000525 & 0.000603  \\
DDO210 & \asymerr{0.000722}{0.000103}{0.000103} & \asymerr{0.000144}{0.000014}{0.000014} & 0.000005 & 0.000004 & 0.000158 & 0.000129 \\
SagDIG & \asymerr{0.002551}{0.000187}{0.000187} & \asymerr{0.000983}{0.000053}{0.000050} & 0.000137  & 0.000085 & 0.000912 & 0.000724 \\
UGC 8091 & \asymerr{0.005247}{0.000358}{0.000358} & \asymerr{0.001419}{0.000060}{0.000060} & 0.002131 & 0.002042 & 0.002630 & 0.002291 \\
UGC 8508 & \asymerr{0.002667}{0.000333}{0.000333} & \asymerr{0.001867}{0.000133}{0.000133} & \ldots & 0.002239 & \ldots & \ldots\\
UGC 9240 & \asymerr{0.004097}{0.000683}{0.000683} & \asymerr{0.005872}{0.000273}{0.000273} & 0.001439 & 0.002188 & 0.005754 & 0.005370 \\
NGC 4163 & \asymerr{0.005148}{0.000858}{0.000858} & \asymerr{0.002745}{0.000172}{0.000172} & 0.000960 & 0.001072 & 0.004571 & 0.003802 \\
NGC 3741 & \asymerr{0.009218}{0.002304}{0.002304} & \asymerr{0.004878}{0.000346}{0.000384} & 0.004063 & 0.003631 & 0.005888 & 0.004898 \\
DDO6 & \asymerr{0.000397}{0.000198}{0.000198} & \asymerr{0.003511}{0.000218}{0.000218} & \ldots & 0.000069 & \ldots &  0.001318 \\
UGCA 292 & \asymerr{0.002002}{0.000263}{0.000263} & \asymerr{0.001380}{0.000100}{0.000105} & 0.001773 & 0.002239 & 0.001738 & 0.002570 \\
\enddata
\tablecomments{GALEX FUV observation of UGC 8508 was prohibited \citep{Lee2011}.}
\tablenotetext{a}{Calculated from the UVopt SFHs scaled by a factor of 1.03493 to match the Salpeter IMF assumption with mass limits of 0.1--100~\Msun used in \citetalias{Lee2009} and \citetalias{Karachentsev2013}.}
\tablenotetext{b}{SFRs from \citetalias{Lee2009}. No corresponding errors were reported.}
\tablenotetext{c}{SFRs from \citetalias{Karachentsev2013}. No corresponding errors were reported.}
\end{deluxetable*}

\begin{figure*}[ht]
 \centering
      \includegraphics[width=\linewidth]{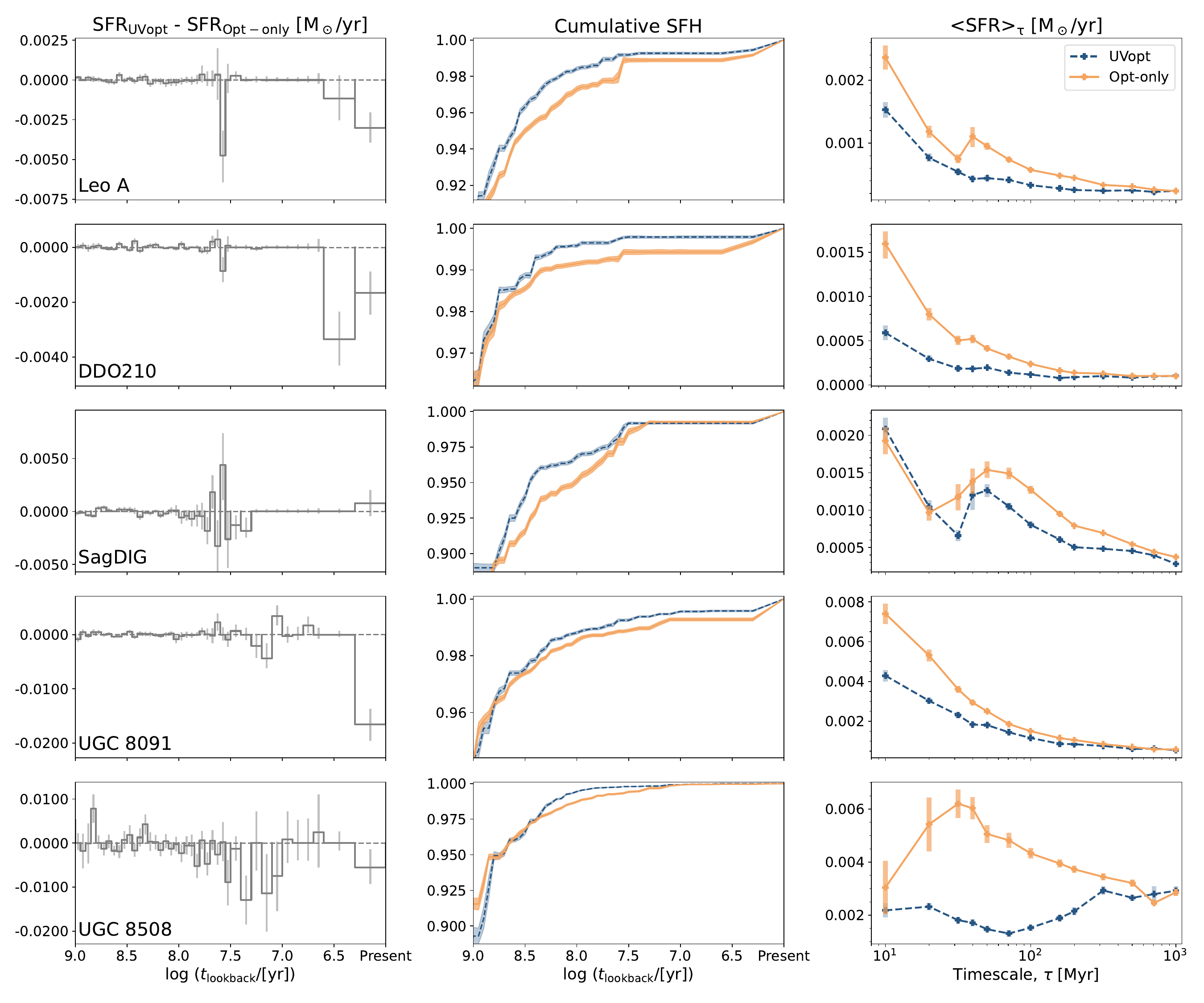}
       \caption{Each row compares the UVopt and Opt-only SFHs for Leo A, DDO 210, SagDIG, UGC 8091, and UGC 8508. \textit{Left:} Differences between the UVopt and Opt-only differential SFHs over the past 1~Gyr. The error bars represent the combined uncertainties from the UVopt and Opt-only SFHs, computed by adding their individual uncertainties in quadrature. Generally, the UVopt SFHs yield lower SFRs (i.e., negative values in the plot) and reduced uncertainties compared to the canonical Opt-only SFHs in most recent time bins. \textit{Middle:} Cumulative SFHs for both cases, shown with the 68\% confidence interval. The specific lookback time at which the two SFHs begin to diverge varies across galaxies. However, despite these variations in divergence timing, the overall impact on the total stellar mass formed remains small. Specifically, the discrepancies in the SFH affect at most the formation of the last few percent to 10 percent of the total stellar mass. \textit{Right:} Average SFRs as a function of $\tau$ for both cases. The calculation of average SFRs for each $\tau$ and their associated uncertainties is described at the beginning of Section~\ref{sec:avgSFR}. Notably, the UVopt SFHs show smaller uncertainties compared to the canonical Opt-only SFHs, particularly in age bins younger than 100~Myr, a trend also observed in the simulated SFHs. (Section~\ref{sec:simulations}). 
      \label{fig:compSFHs1}}
\end{figure*}

\begin{figure*}[ht]
 \centering
      \includegraphics[width=\linewidth]{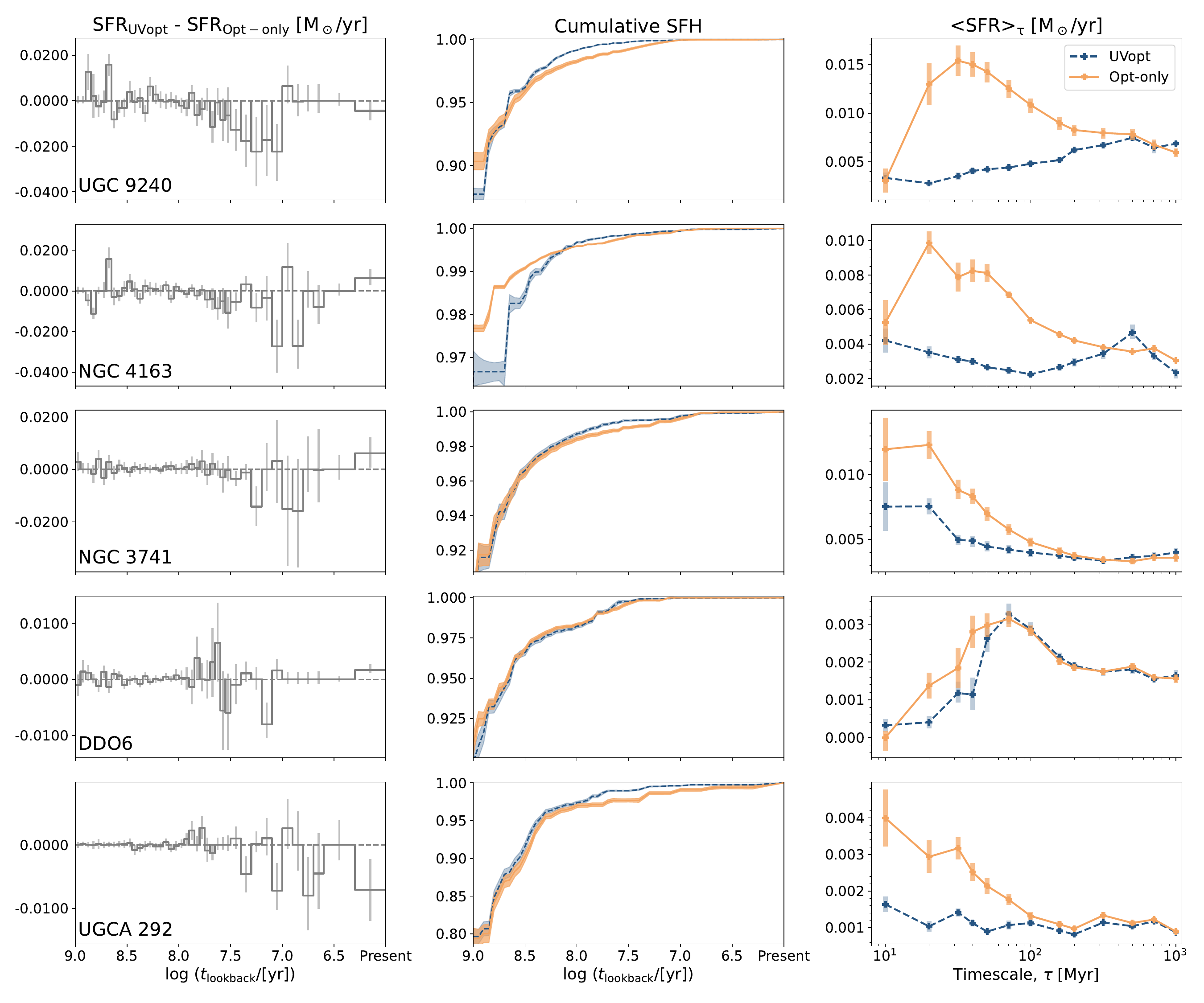}
       \caption{Same as Figure~\ref{fig:compSFHs1}, but for UGC 9240, NGC 4163, NGC 3741, DDO6, and UGCA 292. 
      \label{fig:compSFHs2}}
\end{figure*}

\subsubsection{CMD-based \sfravg{100}{Myr} vs. SFR$_{\rm FUV}$}\label{sec:comp_fuv}
Figure~\ref{fig:fuvSFRs} compares the CMD-based average SFRs over the FUV's nominal 100~Myr timescale, \sfravg{100}{Myr}, with SFR$_{\rm FUV}$ from the literature. Colored solid symbols represent our fiducial UVopt case, while gray open symbols correspond to the Opt-only case. The agreement between \sfravg{100}{Myr} and SFR$_{\rm FUV}$ is substantially better, with discrepancies limited to factors of less than $\sim$2.5--significantly smaller than the orders-of-magnitude deviations observed between \sfravg{10}{Myr} and SFR$_{\rm H\alpha}$ in Section~\ref{sec:comp_ha}. Focusing first on the UVopt case, the top panel shows a decent correlation between the two SFR estimates, whereas the bottom panel highlights their deviations. Most galaxies have lower \sfravg{100}{Myr} values compared to the SFR$_{\rm FUV}$, with a median ratio of 0.81 when compared to \citetalias{Lee2009} and 0.99 when compared to \citetalias{Karachentsev2013}. The HST footprints of our targets encompass (nearly) entire star-forming extents, ensuring the inclusion of the stellar populations responsible for the observed FUV fluxes, though the possibility of missing a small fraction of these stars outside our footprints cannot be entirely ruled out.  

The fact that \sfravg{100}{Myr} is generally lower than SFR$_{\rm FUV}$ is unsurprising. In Figure~\ref{fig:avgSFR}, most galaxies show at least one intersection between \sfravg{}{$\tau$} and SFR$_{\rm FUV}$ at $\tau \lesssim$ 100~Myr, where \sfravg{}{$\tau$} is higher than \sfravg{100}{Myr}. This implies that the observed FUV flux in most galaxies predominantly arises from SF events younger than 100~Myr. More specifically, the FUV fluxes predicted from our UVopt SFHs, following \citet{McQuinn2015}, indicate that over 90\% of the FUV flux arises from SF events younger than $\sim$2--3~Myr for Leo A, DDO 210, and UGC 8091; $\sim$20--50~Myr for SagDIG, NGC 4163, NGC 3741, and UGCA 292; $\sim$90~Myr for UGC 8508 and UGC 9240; and $\sim$180~Myr for DDO6. This finding is supported by \citet{Johnson2013}, who showed that the relative contributions of stellar populations to the FUV luminosity vary significantly with a galaxy's SFH. In their study of 50 nearby dwarf galaxies, some dwarf galaxies have up to $\sim$60\% of their total FUV luminosity arising from stars younger than 10~Myr, while others are dominated by older stars ($>$100~Myr). Similarly, \citet{FloresVelazquez2021} found in simulated dwarf galaxies from the FIRE project that the FUV timescale can span from $\sim$10~Myr to $\gtrsim$100~Myr, depending on the phase of the galaxy's fluctuating SFH. 

A lower \sfravg{100}{Myr} compared to SFR$_{\rm FUV}$ can arise from several factors. If a galaxy has recently experienced a burst of SF, the influx of young, massive stars significantly enhances the FUV luminosity. These stars, which dominate the FUV flux, have lifetimes of only a few Myr. Because the FUV-based SFR estimate is highly sensitive to these young stars and assumes a constant SFR over a given period, a recent burst can lead to an overestimated SFR$_{\rm FUV}$ relative to the longer-term average captured by \sfravg{100}{Myr}, which smooths over a short-lived burst. Additionally, the CMD-based method might miss some of the youngest or most heavily obscured massive stars, further contributing to a lower derived \sfravg{100}{Myr}. Furthermore, uncertainties in dust correction can disproportionately affect SFR$_{\rm FUV}$ estimates, as incorrect extinction corrections can either artificially inflate or suppress SFR$_{\rm FUV}$, introducing additional discrepancies between the two indicators.

Interestingly, this finding contrasts with previous studies that compared optical CMD-based and FUV-based SFRs across various types of dwarf galaxies. Those studies reported CMD-based SFRs to be systematically higher than FUV-based SFRs by up to factors of $\sim$4 for (post-)starburst dwarf galaxies with a median ratio of 1.54 \citep{McQuinn2015}, 2.3--2.5 for M31 \citep{Lewis2017}, and $\sim$5 for dwarf galaxies with higher stellar mass and metallicity with a median ratio of $\sim$2 \citep{Cignoni2019}. In fact, in our Opt-only cases, the CMD-based \sfravg{100}{Myr} are higher than SFR$_{\rm FUV}$ by a factor of up to $\sim$2.6 with a median ratio of $\sim$1.39 when compared to \citetalias{Lee2009} and $\sim$1.73 when compared to \citetalias{Karachentsev2013}. This aligns with traditional optical CMD modeling results but slightly to a lesser extent. In the next Section, we will explore the differences in the measured SFHs between the UVopt and Opt-only cases. 

Various explanations have been proposed in the literature to account for systematically higher optical CMD-based \sfravg{100}{Myr} values compared to SFR$_{\rm FUV}$. These include incomplete sampling of the stellar IMF in dwarf galaxies with low SFRs, variations in the shape and upper-mass cutoff of an assumed IMF, differences in dust correction methods, and choice of SFR scaling relations \citep[e.g.,][]{McQuinn2015, Lewis2017, Cignoni2019}. After a detailed investigation of starburst and post-starburst dwarf galaxies best approximating a constant SFH over $\sim$100~Myr, \citet{McQuinn2015} concluded that improvements in stellar evolutionary and/or atmospheric models at FUV wavelengths are likely needed to resolve this systematic discrepancy. Notably, the stellar atmospheric models used for FUV-based SFR calibrations were originally established with a limited sample of stars. Improvements can now be made using a significantly larger dataset of $\sim$220 stars across 10 star-forming regions in the Milky Way and nearby low-metallicity dwarf galaxies, provided by the Ultraviolet Legacy Library of Young Stars as Essential Standards (ULLYSES) program \citep{Roman-Duval2020}.

\begin{figure*}[ht]
 \centering
    \includegraphics[width=\linewidth]{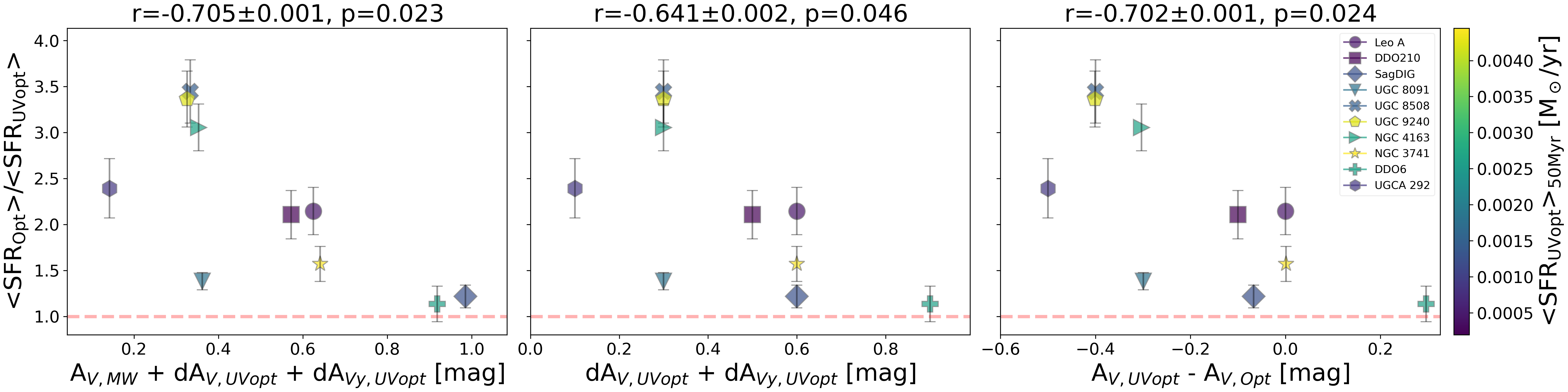}
    \caption{Ratio of \sfravg{}{Opt-only} to \sfravg{}{UVopt} at the $\tau$ = 50~Myr against the total dust extinction (left) and differential dust extinction (middle) derived from the UVopt solutions, as well as the difference in the total dust extinction between the UVopt and Opt-only solutions (right). These three cases are the only ones that show statistically significant correlations (p-value $<$ 0.05) among the various properties investigated, including dust content, H$\alpha$- and FUV-based SFRs, and oxygen abundance over 10~Myr $< \tau <$ 1~Gyr. The Pearson correlation coefficient (r) with its uncertainty and the p-value are noted at the top of each panel. The common $\tau$ = 50~Myr across all three correlations aligns with the 40~Myr threshold in \texttt{MATCH}, where d$A_{V_{\rm y}}$ is fully applied to younger stars and decreases linearly to zero by 100~Myr. This suggests that differences in dust extinction treatment between the canonical Opt-only and UVopt cases primarily drive the SFH discrepancies in recent time bins. \label{fig:SFRratio}}
\end{figure*}

\section{Comparison with the canonical optical CMD modeling}\label{sec:UVimpact}
Thus far, we have demonstrated the enhanced precision in recent SFH measurements achieved by incorporating a UV CMD into the CMD modeling process and highlighted the UV's capability to separate BHeB stars from MS stars, an area where current standard stellar models face challenges in the low-metallicity regime. 

In this section, we compare our fiducial UVopt SFHs with the canonical Opt-only SFHs, evaluate their consistencies and discrepancies across our sample of 10 galaxies, and investigate potential causes for the observed differences. 

\subsection{Deriving SFHs from Canonical Optical CMD Modeling}\label{sec:optSFHs}
Deriving a SFH via the modeling of a single optical CMD is a well established technique, for which \texttt{MATCH} has been used extensively. For our canonical Opt-only SFHs, we follow the procedure described in the literature for using \texttt{MATCH} to determine the SFH that best reproduces the observed CMD \citep[e.g.,][]{Dolphin02, Skillman2003, McQuinn2010a, Weisz2011, Williams2011, Choi2015, Lewis2015, Lazzarini2022}. 

One key distinction from the UVopt case is the way in which dust extinction is constrained. The canonical Opt-only case solves for a maximum of two dust parameters: $A_{V, \rm MW}$ and d$A_{V}$, while fixing d$A_{V_{\rm y}}$ to \texttt{MATCH}'s default value of 0.5~mag, which is based on CMD fitting of the Sextans-A upper MS \citep{Dolphin2003}. Therefore, for our Opt-only SFH fitting, we also fix d$A_{V_{\rm y}}$ to the default value of 0.5~mag to adhere as closely as possible to the canonical optical CMD modeling. In contrast, the UVopt case solves for all three dust parameters: $A_{V, \rm MW}$, d$A_{V}$, and d$A_{V_{\rm y}}$. Another key distinction from the UVopt case is the inclusion of BHeB stars in the CMD fits, also for consistency with previous studies. With a fixed \dAvy, \texttt{MATCH} has far less flexibility to misinterpret BHeB stars as highly reddened MS stars in the Opt-only case. This approach naturally prevents \texttt{MATCH} from fitting the CMD with unreasonably large d$A_{V_{\rm y}}$ values, though a fixed d$A_{V_{\rm y}}$ of 0.5~mag may not be ideal for any given galaxy. These two differences in the fitting procedure likely amplify the observed differences between the Opt-only and UVopt SFH results. All other assumptions and configurations for the Opt-only \texttt{MATCH} runs, including the distance modulus, remain consistent with those used in the UVopt case.

While we aim to adhere to canonical methods as closely as possible for our Opt-only case, there are several key differences between this study and previous work. These include an improved data reduction pipeline, updated stellar evolution models over time, a refined method for selecting and placing artificial stars, different strategies for determining the best-fit solutions, and slightly different spatial coverage (as this study is confined to the area with both UV and optical observations). Thus, we emphasize that the goal of the Opt-only case is not to precisely reproduce previous SFHs but to assess the impact of including a UV CMD on SFH measurements within a consistent framework. Consequently, the resulting Opt-only SFHs should not be considered as an attempt to reproduce, nor as directly representative of, the \texttt{MATCH} solutions reported in the literature.

\subsection{UVopt SFHs vs. Opt-only SFHs}\label{sec:comp_uvopt_optonly}
Figures~\ref{fig:compSFHs1} and \ref{fig:compSFHs2} compare the UVopt and Opt-only solutions by presenting the difference in their differential SFHs (left column), their cumulative SFHs (middle column), and their average SFRs (right column) for each of the 10 galaxies. In comparing the two SFHs for each galaxy, we report only random uncertainties of their best-fit SFHs. Systematic uncertainties, which are the largest source of error in SFH measurements due to systematics in the adopted stellar models \citep{Dolphin2012}, can be ignored in this comparison. This is because we use the same PARSEC stellar models \citep{Bressan2012} for all galaxies, aiming to understand the impact of including a UV CMD in deriving SFHs compared to the canonical Opt-only case. Experiments using the MIST stellar models \citep{Choi2016} yield trends consistent with those obtained using the PARSEC models presented in this paper. This demonstrates that the conclusions reported here are qualitatively robust to the choice of stellar models. 

From the left panels of Figures~\ref{fig:compSFHs1} and \ref{fig:compSFHs2}, we identify three common features in all galaxies over the past 1~Gyr. First, the Opt-only SFHs exhibit significant temporal fluctuations, similar to those observed in the UVopt SFHs (Figure~\ref{fig:uvoptSFH}). Second, discrepancies between the two solutions are more pronounced in recent time bins but diminish with increasing lookback time, with SFRs converging between approximately 100~Myr and 1~Gyr. The specific lookback time of convergence varies among galaxies. Third, the Opt-only solutions generally produce higher SFRs than the UVopt solutions, particularly within the past $\sim$100~Myr. This discrepancy primarily stems from differences in the treatment of dust and BHeB stars between the two cases, as detailed in the previous section. 

The middle panels of Figures~\ref{fig:compSFHs1} and \ref{fig:compSFHs2} compare the classical cumulative SFHs of both the UVopt and Opt-only cases, representing the cumulative stellar mass fraction normalized to 1 at the present day and 0 at the oldest age bin. The shaded region represents the 68\% confidence interval. The cumulative SFHs reveal that the point of divergence between the two solutions varies across galaxies, but the impact of these differences on the total stellar mass is limited. In most cases, the discrepancies only influence how the last few to up to 10\% of the stellar mass is formed. However, the difference between the two cumulative SFHs does not exceed a few percent in any time bin. This highlights the overall consistency between the two approaches on longer timescales while drawing attention to their divergence at younger ages.

However, NGC 4163 stands out as an exception, showing the largest discrepancy between the two solutions, with approximately 87\% more total stellar mass formed in the Opt-only case compared to the UVopt case. Notably, the total stellar mass formed in the UVopt case aligns well with the value estimated by \citet{Berg2012} using 4.5~$\mu$m luminosity and K - [4.5] and B - K colors following \citet{Lee2006}. Achieving satisfactory fits for the optical CMD is particularly challenging for this galaxy. Specifically, the CMD region predominantly occupied by HeB stars, located between the MS and RGB, proves difficult to reproduce and results in poor fit values. Moreover, as demonstrated in both \citetalias{Gilbert2025} and Section~\ref{sec:ASTs}, NGC 4163 experiences the most severe crowding among the 10 galaxies. However, the UV CMD is less affected by crowding due to the intrinsically sparse spatial distribution of young stars, which are the primary UV sources. Thus, incorporating the UV CMD might help mitigate the impact of crowding on the derived UVopt recent SFH for this galaxy.

\citet{Zinchenko2022} proposed recent pristine gas infall as a possible explanation for NGC 4163's unusual negative trend between the nitrogen-to-oxygen (N/O) ratio and oxygen abundance. When we relax the monotonically increasing metallicity assumption, \texttt{MATCH} yields significantly improved CMD fits, suggesting that NGC 4163 may have a more complex chemical enrichment history than other dwarf galaxies in our sample. However, allowing \texttt{MATCH} to have full flexibility in metallicity results in a highly variable (and perhaps nonphysical) metallicity history. Therefore, we report the best-fit SFH for NGC 4163 obtained using the \texttt{-zinc} flag to maintain consistency with the other galaxies in our sample. Further investigation into the chemical evolution for NGC 4163 is necessary to better constrain its SFH.

Although to a lesser degree than NGC 4163, we also note that UGC 9240, NGC 3741, and UGC 8508 exhibit particularly notable discrepancies between the two SFHs compared to the other galaxies. These galaxies are also subject to more significant crowding effects, as discussed in Section~\ref{sec:ASTs}.

The right panels of Figures~\ref{fig:compSFHs1} and \ref{fig:compSFHs2} compare the average SFRs of the UVopt and Opt-only cases as a function of $\tau$. The uncertainties are calculated using the same method described in Section~\ref{sec:avgSFR}. Here, the UVopt SFHs demonstrate significantly reduced uncertainties, particularly in age bins younger than 100~Myr in all galaxies. This improvement underscores the enhanced sensitivity of UV CMDs to young stellar populations in real galaxies as well, as in the case of simulated galaxies demonstrated in Section~\ref{sec:simulations}. As a result, the effective temporal resolution of the recovered SFHs can increase, aligning with suggestions by the LEGUS studies \citep{Cignoni2018, Cigan2021}. More specifically, for Leo A, the UVopt case achieves an overall precision improvement of $\sim$20\% over the past 100~Myr, consistent with results derived from synthetic CMDs based on a simplified version of Leo A's SFH. For galaxies with more complex SFHs than Leo A, the precision gains are even greater, ranging from $\sim$20--70\% at $\tau$ = 100~Myr. The reduced uncertainties could be particularly helpful when comparing with other recent SFR tracers by providing a stronger independent constraint. Another notable trend is that the Opt-only SFHs generally yield higher average SFRs than the UVopt solutions, particularly over the past few hundred Myr, as hinted at in the comparisons with the H$\alpha$-based and FUV-based SFRs in Section~\ref{sec:comp_ha} and Section~\ref{sec:comp_fuv}.

\subsection{What Drives the Differences?}
To explore the origin of the discrepancies between the canonical Opt-only SFHs and fiducial UVopt SFHs, as shown in Figures~\ref{fig:compSFHs1} and \ref{fig:compSFHs2}, we examine the ratio of \sfravgcol{\tau}{Myr}{Opt-only} to \sfravgcol{\tau}{Myr}{UVopt} (referred to as the SFR ratio) in relation to various parameters. We explore the SFR ratio as a function of derived dust-related properties, such as total dust extinction and differential dust extinction derived from the UVopt solutions, as well as the difference in total dust extinction derived from the UVopt and Opt-only solutions. We also explore the SFR ratio as a function of galactic properties like oxygen abundance, H$\alpha$- and FUV-based SFRs. The analysis spans the same range of $\tau$ explored in Section~\ref{sec:avgSFR}, from 10~Myr to 1~Gyr.

To quantify the strength of these correlations, we calculate the Pearson correlation coefficient, its associated p-value, and uncertainties. Uncertainties are derived through 2,000 Monte Carlo simulations applied to the given x and y data, incorporating errors in the y data. Here, the ratios of the average SFRs for the 10 galaxies serve as the y data. This process is repeated as a function of $\tau$ to assess whether any of these correlations strengthens over specific timescales.

Figure~\ref{fig:SFRratio} highlights the cases with statistically significant correlations ($p$-value $<$ 0.05). Below, we provide a brief discussion of the correlations shown in each panel, from the left to the right. 
\begin{itemize}
    \item SFR ratio vs.\ total dust extinction: We observe a statistically significant negative correlation at $\tau$ = 50~Myr (r $=$ -0.705$\pm$0.001), indicating that the discrepancy between the Opt-only and UVopt cases is pronounced for galaxies with small intrinsic amounts of dust. For instance, in the most metal-poor, and thus likely the least dusty galaxy, UGCA 292, the canonical Opt-only solution with a fixed \dAvy = 0.5 shows significantly enhanced SFRs in the young age bins. As discussed in Section~\ref{sec:bheb}, when \texttt{MATCH} is constrained to a fixed \dAvy = 0.5, it is compelled to produce a broader upper MS than what is actually observed in the galaxy. As a result, to compensate for the artificially broadened MS, the model must overproduce the total number of MS stars by enhancing the SFRs at young ages to match the number of stars along the observed narrower MS. This does not occur in the UVopt solutions with a more appropriate dust combination of \Avmw, \dAv\,and \dAvy.
    \item SFR ratio vs.\ differential dust extinction: There is a statistically significant negative correlation at $\tau$ = 50~Myr (r $=$ -0.641$\pm$0.002). This trend is likely due to the same factors as in the case of total extinction, since \Avmw\ is generally smaller than the combined contributions of \dAv\ and \dAvy\ for these galaxies.
    \item SFR ratio vs.\ the difference in total dust extinction ($A_{V, \rm UVopt}$ - $A_{V, \rm Opt-only}$): A statistically significant negative correlation is observed at $\tau$ = 50~Myr (r $=$ -0.702$\pm$0.001), suggesting that the discrepancy increases as the total extinction in the Opt-only SFHs (with a fixed \dAvy = 0.5) exceeds that in the UVopt SFHs, where three dust parameters are fitted. This trend is likewise driven by the same factors underlying the previous two correlations. 
\end{itemize}

The common timescale of 50~Myr observed across all three trends roughly coincides with the age of 40~Myr, where \dAvy\ applies its full value to stars younger than 40~Myr and then linearly decreases to zero between 40~Myr and 100~Myr in \texttt{MATCH} (see Section~\ref{sec:sfh}). Therefore, we conclude that the difference in the treatment of dust extinction within the CMD modeling process between the canonical Opt-only and UVopt cases is likely the primary cause of the discrepancy in their resulting SFHs in recent time bins.

\section{Discussion}\label{sec:discussion}
We have shown that incorporating UV CMDs into the CMD modeling process significantly enhances the precision of recent SFH measurements for our low-metallicity dwarf galaxies. However, there remains room for improvement in constraining their SFHs. Key areas for advancement include the refinement of standard stellar evolutionary models that are more accurate for HeB stars in these low-metallicity galaxies (see Section~\ref{sec:bheb} for detailed discussion), the adoption of more appropriate dust extinction relations tailored to metal-poor environments, and accounting for crowding effects in galaxies where confusion is significant, among others. Addressing the first factor is beyond the scope of this paper; however, an upcoming study will focus on calibrating poorly understood evolutionary phases, including the core HeB phase, using LUVIT data (G. Pastorelli et al., in prep). Here, we will discuss the potential impact of the remaining two factors.

The UV's sensitivity to variations in dust attenuation is not fully leveraged due to the current limitations of \texttt{MATCH}, which only supports the \citet{Cardelli1989} extinction relationship with $R_{V} = 3.1$, representing a Milky Way-type dust curve. However, dust attenuation in low-metallicity galaxies more closely resembles an SMC-type dust curve \citep[e.g.,][]{Gordon2003, Shivaei2020}. The distinction between these dust curves is more pronounced in UV filters compared to optical filters. For instance, with $A_{V} = 1$, the Milky Way-type curve produces extinctions of ($A_{F275W}$, $A_{F336W}$, $A_{F475W}$, $A_{F814W}$) = (2.023, 1.635, 1.202, 0.589), whereas the SMC-type curve yields ($A_{F275W}$, $A_{F336W}$, $A_{F475W}$, $A_{F814W}$) = (2.268, 1.827, 1.237, 0.572). The steeper slope of the SMC-type curve in shorter wavelengths predicts $\sim$12\% higher extinction in F275W and F336W and $\sim$14\% more reddening in F275W-F336W than the Milky Way-type curve, while showing minimal differences in extinction and reddening in the optical CMD for the same dust content.

Therefore, the SMC-type curve would require a smaller \dAvy\,value to reproduce the observed broadening in a given UV CMD compared to the Milky Way-type curve. With a reduced \dAvy, there would be no/less need to artificially inflate SFRs to match the number of stars on the bluer edge of the upper MS. This adjustment would lead to lower SFRs in the young age bins for the UVopt solutions, further increasing the discrepancy with the canonical Opt-only SFHs, particularly for galaxies with intrinsically low dust content. Incorporating an SMC-type dust model into \texttt{MATCH} would enable more accurate characterization of UV and optical CMDs simultaneously, likely improving its performance for low-metallicity galaxies.

As noted in Section~\ref{sec:ASTs}, four of the ten galaxies--NGC 4163, UGC 9240, NGC 3741, and UGC 8508--suffer from crowding, listed in descending order of crowding severity. Consequently, their SFHs are likely influenced by crowding effects. While ASTs are performed across different stellar density regions within each galaxy, deriving a global 50\% completeness per filter from combined ASTs across the entire galaxy inherently fails to accurately reflect variations in local stellar densities. Regions more crowded than the average will have shallower 50\% completeness limits, whereas less crowded regions will exhibit deeper limits. To address this issue, deriving SFHs in a spatially resolved manner is necessary. However, even spatially resolved SFHs cannot entirely eliminate the risk of missing stars or misclassifying small star clusters in the most crowded regions. This limitation arises from the practical and fundamental physical constraints on angular resolution, even with space-based telescopes.

When reconstructing the SFH of a metal-poor star-forming galaxy using only an optical CMD, we recommend closely examining the CMD residuals around the upper MS and BHeB regions to ensure that dust is not excessively applied to fit BHeB stars as highly reddened MS stars. Additionally, even for metal-poor dwarf galaxies, it is crucial to model both the global differential extinction for the entire stellar population (\dAv\,in \texttt{MATCH}) and an additional differential extinction specific to young stellar populations (\dAvy\,in \texttt{MATCH}). Incorporating these factors can significantly impact the derived recent SFHs.

\section{Summary and Conclusion}\label{sec:summary}
In this paper, we present SFHs of ten metal-poor, star-forming dwarf galaxies observed in multiple bands as part of the LUVIT survey. Our focus is on investigating and evaluating the benefits of incorporating UV data into the CMD modeling process for low-metallicity galaxies. 

We reconfirm that standard stellar evolutionary models for core HeB stars fail to accurately represent the observed HeB stars both in UV and optical CMDs in our low-metallicity dwarf galaxies. Specifically, isochrones at the metallicity of 10\%\Zsun, appropriate for young stellar populations in our galaxies, are too red to reproduce the observed BHeB stars. This discrepancy between the observations and models leads to artificially elevated SFRs in the most recent time bins. We mitigate this impact by leveraging the strong discriminatory power of the (F275W-F475W, F275W) CMDs, which effectively separate BHeB stars from upper MS stars and allow us to exclude them from the CMD modeling process. 

To quantitatively evaluate the impact of incorporating a UV CMD (UVopt case) on SFH measurements compared to the traditional approach of modeling a single optical CMD (Opt-only case), we first conduct rigorous tests using synthetic CMDs with various dust configurations--(\dAv, \dAvy) = (0.1, 0.2), (0.2, 0.1), (0.0, 0.5), and (0.2, 0.7)--and two distinct input SFHs: a realistic Leo A-like SFH and a constant SFH. These tests focus on quantifying improvements in precision and accuracy. 

Our results show that the UVopt case recovers input SFHs with higher precision and accuracy in recent time bins, extending up to $\sim$1~Gyr ago. While the degree of improvement varies with dust configuration and input SFH, the UVopt case consistently reduces overall uncertainties. For example, our results with Leo A-like synthetic CMDs achieve uncertainty reductions of $\sim$4–8\% over the past 10~Myr, $\sim$8–20\% over the past 100~Myr, and $\sim$8–14\% over the past 1~Gyr. The smallest improvements occur in scenarios with the highest total dust content (\dAv, \dAvy) = (0.2, 0.7) at $\tau$ = 10~Myr, while the most significant gains are observed at $\tau$ = 100~Myr across all dust combinations. 

We then derive SFHs for ten metal-poor, star-forming dwarf galaxies representative of the LUVIT sample, using both UV and optical CMDs, while excluding BHeB stars from the CMD modeling process. The resulting SFHs exhibit significant fluctuations, deviating from a constant SFH scenario, as commonly expected for dwarf galaxies. From the comparisons with other widely used SFR indicators, H$\alpha$ emission and FUV flux, we reaffirm the unreliability of H$\alpha$ emission as an SFR indicator over its nominal 10~Myr timescale in galaxies with low SFRs. We also find that although FUV flux provides a more robust tracer of recent SFR than H$\alpha$, the \sfravgcol{100}{Myr}{UVopt} values are systematically lower than the SFR$_{\rm FUV}$ values by a factor of up to $\sim$2, accompanied by significant scatter. This suggests that the nominal 100~Myr timescale used for calibrating FUV-based SFRs may not accurately apply to metal-poor dwarf galaxies with highly variable SFHs. 

Finally, we compare the UVopt SFHs with the Opt-only SFHs, derived in a manner consistent with the majority of prior single optical CMD modeling using \texttt{MATCH}, to identify differences and investigate the underlying causes of the observed discrepancies. Two key differences emerge from this comparison. First, the inclusion of UV CMD significantly reduces uncertainties in SFH measurements of these real galaxies compared to the Opt-only case, consistent with findings from the synthetic CMD analysis. This improvement is essential for accurately characterizing the fluctuating SF that is typical of dwarf galaxies, especially in low-metallicity environments. Second, the UVopt SFHs generally yield lower SFRs in recent time bins, primarily due to the implementation of more sophisticated and flexible dust modeling for young stellar populations, in contrast to the fixed \dAvy=0.5~mag assumption in the canonical Opt-only case. This reduced recent SFRs alleviate the overestimation of 100~Myr-averaged SFRs from CMD modeling, thereby mitigating discrepancy between CMD-based \sfravg{100}{Myr} and SFR$_{\rm FUV}$. 

In conclusion, our results highlight the importance of incorporating UV data in refining HeB-phase stellar models and in recent SFH analyses for metal-poor, star-forming dwarf galaxies. The improved precision of the resulting SFHs offered by the UVopt case over the Opt-only case eventually allow us to better understand the episodic nature of SF in metal-poor environments. These findings are crucial for interpreting SF in both local and high-redshift galaxies, providing a more detailed understanding of the evolutionary processes that shape galaxies over cosmic time.

\acknowledgements{
Support for this work was provided by NASA through grants GO-15275, GO-16162, GO-16292, and AR-16120 from the Space Telescope Science Institute. This research is based on observations made with the NASA/ESA Hubble Space Telescope obtained from the Space Telescope Science Institute, which is operated by the Association of Universities for Research in Astronomy, Inc., under NASA contract NAS 5–26555. The work of Y. Choi is supported by NOIRLab, which is managed by the Association of Universities for Research in Astronomy (AURA) under a cooperative agreement with the U.S. National Science Foundation. The Flatiron Institute is funded by the Simons Foundation.

\facility{HST(ACS/WFC), HST(WFC3/UVIS)}

\software{\texttt{Astropy} \citep{Astropy2013,Astropy2018,Astropy2022}, \texttt{BEAST} \citep{Gordon2016}, \texttt{DOLPHOT} \citep{Dolphin2000, Dolphin2016}, \texttt{MATCH} \citep{Dolphin02}, \texttt{Matplotlib} \citep{Hunter2007}, \texttt{NumPy} \citep{vanderwalt2011, harris2020}, \texttt{SciPy} \citep{Scipy2020}}}


\begin{thebibliography}{}
\expandafter\ifx\csname natexlab\endcsname\relax\def\natexlab#1{#1}\fi
\providecommand{\url}[1]{\href{#1}{#1}}
\providecommand{\dodoi}[1]{doi:~\href{http://doi.org/#1}{\nolinkurl{#1}}}
\providecommand{\doeprint}[1]{\href{http://ascl.net/#1}{\nolinkurl{http://ascl.net/#1}}}
\providecommand{\doarXiv}[1]{\href{https://arxiv.org/abs/#1}{\nolinkurl{https://arxiv.org/abs/#1}}}

\bibitem[{{Aparicio} {et~al.}(1996){Aparicio}, {Gallart}, {Chiosi}, \&
  {Bertelli}}]{Aparicio1996}
{Aparicio}, A., {Gallart}, C., {Chiosi}, C., \& {Bertelli}, G. 1996, \apjl,
  469, L97, \dodoi{10.1086/310279}

\bibitem[{{Astropy Collaboration} {et~al.}(2013){Astropy Collaboration},
  {Robitaille}, {Tollerud}, {Greenfield}, {Droettboom}, {Bray}, {Aldcroft},
  {Davis}, {Ginsburg}, {Price-Whelan}, {Kerzendorf}, {Conley}, {Crighton},
  {Barbary}, {Muna}, {Ferguson}, {Grollier}, {Parikh}, {Nair}, {Unther},
  {Deil}, {Woillez}, {Conseil}, {Kramer}, {Turner}, {Singer}, {Fox}, {Weaver},
  {Zabalza}, {Edwards}, {Azalee Bostroem}, {Burke}, {Casey}, {Crawford},
  {Dencheva}, {Ely}, {Jenness}, {Labrie}, {Lim}, {Pierfederici}, {Pontzen},
  {Ptak}, {Refsdal}, {Servillat}, \& {Streicher}}]{Astropy2013}
{Astropy Collaboration}, {Robitaille}, T.~P., {Tollerud}, E.~J., {et~al.} 2013,
  \aap, 558, A33, \dodoi{10.1051/0004-6361/201322068}

\bibitem[{{Astropy Collaboration} {et~al.}(2018){Astropy Collaboration},
  {Price-Whelan}, {Sip{\H{o}}cz}, {G{\"u}nther}, {Lim}, {Crawford}, {Conseil},
  {Shupe}, {Craig}, {Dencheva}, {Ginsburg}, {VanderPlas}, {Bradley},
  {P{\'e}rez-Su{\'a}rez}, {de Val-Borro}, {Aldcroft}, {Cruz}, {Robitaille},
  {Tollerud}, {Ardelean}, {Babej}, {Bach}, {Bachetti}, {Bakanov}, {Bamford},
  {Barentsen}, {Barmby}, {Baumbach}, {Berry}, {Biscani}, {Boquien}, {Bostroem},
  {Bouma}, {Brammer}, {Bray}, {Breytenbach}, {Buddelmeijer}, {Burke},
  {Calderone}, {Cano Rodr{\'\i}guez}, {Cara}, {Cardoso}, {Cheedella}, {Copin},
  {Corrales}, {Crichton}, {D'Avella}, {Deil}, {Depagne}, {Dietrich}, {Donath},
  {Droettboom}, {Earl}, {Erben}, {Fabbro}, {Ferreira}, {Finethy}, {Fox},
  {Garrison}, {Gibbons}, {Goldstein}, {Gommers}, {Greco}, {Greenfield},
  {Groener}, {Grollier}, {Hagen}, {Hirst}, {Homeier}, {Horton}, {Hosseinzadeh},
  {Hu}, {Hunkeler}, {Ivezi{\'c}}, {Jain}, {Jenness}, {Kanarek}, {Kendrew},
  {Kern}, {Kerzendorf}, {Khvalko}, {King}, {Kirkby}, {Kulkarni}, {Kumar},
  {Lee}, {Lenz}, {Littlefair}, {Ma}, {Macleod}, {Mastropietro}, {McCully},
  {Montagnac}, {Morris}, {Mueller}, {Mumford}, {Muna}, {Murphy}, {Nelson},
  {Nguyen}, {Ninan}, {N{\"o}the}, {Ogaz}, {Oh}, {Parejko}, {Parley}, {Pascual},
  {Patil}, {Patil}, {Plunkett}, {Prochaska}, {Rastogi}, {Reddy Janga},
  {Sabater}, {Sakurikar}, {Seifert}, {Sherbert}, {Sherwood-Taylor}, {Shih},
  {Sick}, {Silbiger}, {Singanamalla}, {Singer}, {Sladen}, {Sooley},
  {Sornarajah}, {Streicher}, {Teuben}, {Thomas}, {Tremblay}, {Turner},
  {Terr{\'o}n}, {van Kerkwijk}, {de la Vega}, {Watkins}, {Weaver}, {Whitmore},
  {Woillez}, {Zabalza}, \& {Astropy Contributors}}]{Astropy2018}
{Astropy Collaboration}, {Price-Whelan}, A.~M., {Sip{\H{o}}cz}, B.~M., {et~al.}
  2018, \aj, 156, 123, \dodoi{10.3847/1538-3881/aabc4f}

\bibitem[{{Astropy Collaboration} {et~al.}(2022){Astropy Collaboration},
  {Price-Whelan}, {Lim}, {Earl}, {Starkman}, {Bradley}, {Shupe}, {Patil},
  {Corrales}, {Brasseur}, {N{\"o}the}, {Donath}, {Tollerud}, {Morris},
  {Ginsburg}, {Vaher}, {Weaver}, {Tocknell}, {Jamieson}, {van Kerkwijk},
  {Robitaille}, {Merry}, {Bachetti}, {G{\"u}nther}, {Aldcroft},
  {Alvarado-Montes}, {Archibald}, {B{\'o}di}, {Bapat}, {Barentsen},
  {Baz{\'a}n}, {Biswas}, {Boquien}, {Burke}, {Cara}, {Cara}, {Conroy},
  {Conseil}, {Craig}, {Cross}, {Cruz}, {D'Eugenio}, {Dencheva}, {Devillepoix},
  {Dietrich}, {Eigenbrot}, {Erben}, {Ferreira}, {Foreman-Mackey}, {Fox},
  {Freij}, {Garg}, {Geda}, {Glattly}, {Gondhalekar}, {Gordon}, {Grant},
  {Greenfield}, {Groener}, {Guest}, {Gurovich}, {Handberg}, {Hart},
  {Hatfield-Dodds}, {Homeier}, {Hosseinzadeh}, {Jenness}, {Jones}, {Joseph},
  {Kalmbach}, {Karamehmetoglu}, {Ka{\l}uszy{\'n}ski}, {Kelley}, {Kern},
  {Kerzendorf}, {Koch}, {Kulumani}, {Lee}, {Ly}, {Ma}, {MacBride}, {Maljaars},
  {Muna}, {Murphy}, {Norman}, {O'Steen}, {Oman}, {Pacifici}, {Pascual},
  {Pascual-Granado}, {Patil}, {Perren}, {Pickering}, {Rastogi}, {Roulston},
  {Ryan}, {Rykoff}, {Sabater}, {Sakurikar}, {Salgado}, {Sanghi}, {Saunders},
  {Savchenko}, {Schwardt}, {Seifert-Eckert}, {Shih}, {Jain}, {Shukla}, {Sick},
  {Simpson}, {Singanamalla}, {Singer}, {Singhal}, {Sinha}, {Sip{\H{o}}cz},
  {Spitler}, {Stansby}, {Streicher}, {{\v{S}}umak}, {Swinbank}, {Taranu},
  {Tewary}, {Tremblay}, {de Val-Borro}, {Van Kooten}, {Vasovi{\'c}}, {Verma},
  {de Miranda Cardoso}, {Williams}, {Wilson}, {Winkel}, {Wood-Vasey}, {Xue},
  {Yoachim}, {Zhang}, {Zonca}, \& {Astropy Project Contributors}}]{Astropy2022}
{Astropy Collaboration}, {Price-Whelan}, A.~M., {Lim}, P.~L., {et~al.} 2022,
  \apj, 935, 167, \dodoi{10.3847/1538-4357/ac7c74}

\bibitem[{{Berg} {et~al.}(2012){Berg}, {Skillman}, {Marble}, {van Zee},
  {Engelbracht}, {Lee}, {Kennicutt}, {Calzetti}, {Dale}, \&
  {Johnson}}]{Berg2012}
{Berg}, D.~A., {Skillman}, E.~D., {Marble}, A.~R., {et~al.} 2012, \apj, 754,
  98, \dodoi{10.1088/0004-637X/754/2/98}

\bibitem[{{Bressan} {et~al.}(2012){Bressan}, {Marigo}, {Girardi}, {Salasnich},
  {Dal Cero}, {Rubele}, \& {Nanni}}]{Bressan2012}
{Bressan}, A., {Marigo}, P., {Girardi}, L., {et~al.} 2012, \mnras, 427, 127,
  \dodoi{10.1111/j.1365-2966.2012.21948.x}

\bibitem[{{Calzetti} {et~al.}(2015){Calzetti}, {Lee}, {Sabbi}, {Adamo},
  {Smith}, {Andrews}, {Ubeda}, {Bright}, {Thilker}, {Aloisi}, {Brown},
  {Chandar}, {Christian}, {Cignoni}, {Clayton}, {da Silva}, {de Mink}, {Dobbs},
  {Elmegreen}, {Elmegreen}, {Evans}, {Fumagalli}, {Gallagher}, {Gouliermis},
  {Grebel}, {Herrero}, {Hunter}, {Johnson}, {Kennicutt}, {Kim}, {Krumholz},
  {Lennon}, {Levay}, {Martin}, {Nair}, {Nota}, {{\"O}stlin}, {Pellerin},
  {Prieto}, {Regan}, {Ryon}, {Schaerer}, {Schiminovich}, {Tosi}, {Van Dyk},
  {Walterbos}, {Whitmore}, \& {Wofford}}]{Calzetti2015}
{Calzetti}, D., {Lee}, J.~C., {Sabbi}, E., {et~al.} 2015, \aj, 149, 51,
  \dodoi{10.1088/0004-6256/149/2/51}

\bibitem[{{Cardelli} {et~al.}(1989){Cardelli}, {Clayton}, \&
  {Mathis}}]{Cardelli1989}
{Cardelli}, J.~A., {Clayton}, G.~C., \& {Mathis}, J.~S. 1989, \apj, 345, 245,
  \dodoi{10.1086/167900}

\bibitem[{{Castelli} \& {Kurucz}(2003)}]{Castelli2003}
{Castelli}, F., \& {Kurucz}, R.~L. 2003, in Modelling of Stellar Atmospheres,
  ed. N.~{Piskunov}, W.~W. {Weiss}, \& D.~F. {Gray}, Vol. 210, A20.
\newblock \doarXiv{astro-ph/0405087}

\bibitem[{{Chen} {et~al.}(2015){Chen}, {Bressan}, {Girardi}, {Marigo}, {Kong},
  \& {Lanza}}]{Chen2015}
{Chen}, Y., {Bressan}, A., {Girardi}, L., {et~al.} 2015, \mnras, 452, 1068,
  \dodoi{10.1093/mnras/stv1281}

\bibitem[{{Choi} {et~al.}(2016){Choi}, {Dotter}, {Conroy}, {Cantiello},
  {Paxton}, \& {Johnson}}]{Choi2016}
{Choi}, J., {Dotter}, A., {Conroy}, C., {et~al.} 2016, \apj, 823, 102,
  \dodoi{10.3847/0004-637X/823/2/102}

\bibitem[{{Choi} {et~al.}(2015){Choi}, {Dalcanton}, {Williams}, {Weisz},
  {Skillman}, {Fouesneau}, \& {Dolphin}}]{Choi2015}
{Choi}, Y., {Dalcanton}, J.~J., {Williams}, B.~F., {et~al.} 2015, \apj, 810, 9,
  \dodoi{10.1088/0004-637X/810/1/9}

\bibitem[{{Choi} {et~al.}(2020){Choi}, {Dalcanton}, {Williams}, {Skillman},
  {Fouesneau}, {Gordon}, {Sandstrom}, {Weisz}, \& {Gilbert}}]{Choi2020}
---. 2020, \apj, 902, 54, \dodoi{10.3847/1538-4357/abb467}

\bibitem[{{Cigan} {et~al.}(2021){Cigan}, {Young}, {Gomez}, {Madden}, {De Vis},
  {Hunter}, {Elmegreen}, {Brinks}, \& {LITTLE THINGS Team}}]{Cigan2021}
{Cigan}, P., {Young}, L.~M., {Gomez}, H.~L., {et~al.} 2021, \aj, 162, 83,
  \dodoi{10.3847/1538-3881/abfd2e}

\bibitem[{{Cignoni} {et~al.}(2018){Cignoni}, {Sacchi}, {Aloisi}, {Tosi},
  {Calzetti}, {Lee}, {Sabbi}, {Adamo}, {Cook}, {Dale}, {Elmegreen},
  {Gallagher}, {Gouliermis}, {Grasha}, {Grebel}, {Hunter}, {Johnson}, {Messa},
  {Smith}, {Thilker}, {Ubeda}, \& {Whitmore}}]{Cignoni2018}
{Cignoni}, M., {Sacchi}, E., {Aloisi}, A., {et~al.} 2018, \apj, 856, 62,
  \dodoi{10.3847/1538-4357/aab041}

\bibitem[{{Cignoni} {et~al.}(2019){Cignoni}, {Sacchi}, {Tosi}, {Aloisi},
  {Cook}, {Calzetti}, {Lee}, {Sabbi}, {Thilker}, {Adamo}, {Dale}, {Elmegreen},
  {Gallagher}, {Grebel}, {Johnson}, {Messa}, {Smith}, \& {Ubeda}}]{Cignoni2019}
{Cignoni}, M., {Sacchi}, E., {Tosi}, M., {et~al.} 2019, \apj, 887, 112,
  \dodoi{10.3847/1538-4357/ab53d5}

\bibitem[{{Cohen} {et~al.}(2024){Cohen}, {McQuinn}, {Murray}, {Williams},
  {Choi}, {Lindberg}, {Burhenne}, {Gordon}, {Yanchulova Merica-Jones},
  {Gilbert}, {Boyer}, {Goldman}, {Dolphin}, \& {Telford}}]{Cohen2024}
{Cohen}, R.~E., {McQuinn}, K. B.~W., {Murray}, C.~E., {et~al.} 2024, \apj, 975,
  42, \dodoi{10.3847/1538-4357/ad6cd5}

\bibitem[{{Cole} {et~al.}(2014){Cole}, {Weisz}, {Dolphin}, {Skillman},
  {McConnachie}, {Brooks}, \& {Leaman}}]{Cole2014}
{Cole}, A.~A., {Weisz}, D.~R., {Dolphin}, A.~E., {et~al.} 2014, \apj, 795, 54,
  \dodoi{10.1088/0004-637X/795/1/54}

\bibitem[{{Cole} {et~al.}(2007){Cole}, {Skillman}, {Tolstoy}, {Gallagher},
  {Aparicio}, {Dolphin}, {Gallart}, {Hidalgo}, {Saha}, {Stetson}, \&
  {Weisz}}]{Cole2007}
{Cole}, A.~A., {Skillman}, E.~D., {Tolstoy}, E., {et~al.} 2007, \apjl, 659,
  L17, \dodoi{10.1086/516711}

\bibitem[{{Dalcanton} {et~al.}(2012){Dalcanton}, {Williams}, {Lang}, {Lauer},
  {Kalirai}, {Seth}, {Dolphin}, {Rosenfield}, {Weisz}, {Bell}, {Bianchi},
  {Boyer}, {Caldwell}, {Dong}, {Dorman}, {Gilbert}, {Girardi}, {Gogarten},
  {Gordon}, {Guhathakurta}, {Hodge}, {Holtzman}, {Johnson}, {Larsen}, {Lewis},
  {Melbourne}, {Olsen}, {Rix}, {Rosema}, {Saha}, {Sarajedini}, {Skillman}, \&
  {Stanek}}]{Dalcanton2012}
{Dalcanton}, J.~J., {Williams}, B.~F., {Lang}, D., {et~al.} 2012, \apjs, 200,
  18, \dodoi{10.1088/0067-0049/200/2/18}

\bibitem[{{Dale} {et~al.}(2023){Dale}, {Boquien}, {Turner}, {Calzetti},
  {Kennicutt}, \& {Lee}}]{Dale2023}
{Dale}, D.~A., {Boquien}, M., {Turner}, J.~A., {et~al.} 2023, \aj, 165, 260,
  \dodoi{10.3847/1538-3881/accffe}

\bibitem[{{Dohm-Palmer} {et~al.}(2002){Dohm-Palmer}, {Skillman}, {Mateo},
  {Saha}, {Dolphin}, {Tolstoy}, {Gallagher}, \& {Cole}}]{DohmPalmer2002}
{Dohm-Palmer}, R.~C., {Skillman}, E.~D., {Mateo}, M., {et~al.} 2002, \aj, 123,
  813, \dodoi{10.1086/324635}

\bibitem[{{Dohm-Palmer} {et~al.}(1997){Dohm-Palmer}, {Skillman}, {Saha},
  {Tolstoy}, {Mateo}, {Gallagher}, {Hoessel}, {Chiosi}, \&
  {Dufour}}]{Dohm-Palmer1997}
{Dohm-Palmer}, R.~C., {Skillman}, E.~D., {Saha}, A., {et~al.} 1997, \aj, 114,
  2527, \dodoi{10.1086/118665}

\bibitem[{{Dohm-Palmer} {et~al.}(1998){Dohm-Palmer}, {Skillman}, {Gallagher},
  {Tolstoy}, {Mateo}, {Dufour}, {Saha}, {Hoessel}, \&
  {Chiosi}}]{DohmPalmer1998}
{Dohm-Palmer}, R.~C., {Skillman}, E.~D., {Gallagher}, J., {et~al.} 1998, \aj,
  116, 1227, \dodoi{10.1086/300514}

\bibitem[{{Dolphin}(2016)}]{Dolphin2016}
{Dolphin}, A. 2016, {DOLPHOT: Stellar photometry}, Astrophysics Source Code
  Library, record ascl:1608.013.
\newblock \doeprint{1608.013}

\bibitem[{{Dolphin}(2000)}]{Dolphin2000}
{Dolphin}, A.~E. 2000, \pasp, 112, 1383, \dodoi{10.1086/316630}

\bibitem[{{Dolphin}(2002)}]{Dolphin02}
---. 2002, \mnras, 332, 91, \dodoi{10.1046/j.1365-8711.2002.05271.x}

\bibitem[{{Dolphin}(2012)}]{Dolphin2012}
---. 2012, \apj, 751, 60, \dodoi{10.1088/0004-637X/751/1/60}

\bibitem[{{Dolphin}(2013)}]{Dolphin13}
---. 2013, \apj, 775, 76, \dodoi{10.1088/0004-637X/775/1/76}

\bibitem[{{Dolphin} {et~al.}(2003){Dolphin}, {Saha}, {Skillman}, {Dohm-Palmer},
  {Tolstoy}, {Cole}, {Gallagher}, {Hoessel}, \& {Mateo}}]{Dolphin2003}
{Dolphin}, A.~E., {Saha}, A., {Skillman}, E.~D., {et~al.} 2003, \aj, 126, 187,
  \dodoi{10.1086/375761}

\bibitem[{{Dotter}(2016)}]{Dotter2016}
{Dotter}, A. 2016, \apjs, 222, 8, \dodoi{10.3847/0067-0049/222/1/8}

\bibitem[{{Duane} {et~al.}(1987){Duane}, {Kennedy}, {Pendleton}, \&
  {Roweth}}]{Duane1987}
{Duane}, S., {Kennedy}, A.~D., {Pendleton}, B.~J., \& {Roweth}, D. 1987,
  Physics Letters B, 195, 216, \dodoi{10.1016/0370-2693(87)91197-X}

\bibitem[{{Durbin} {et~al.}(2020){Durbin}, {Beaton}, {Dalcanton}, {Williams},
  \& {Boyer}}]{Durbin2020}
{Durbin}, M.~J., {Beaton}, R.~L., {Dalcanton}, J.~J., {Williams}, B.~F., \&
  {Boyer}, M.~L. 2020, \apj, 898, 57, \dodoi{10.3847/1538-4357/ab9cbb}

\bibitem[{{Flores Vel{\'a}zquez} {et~al.}(2021){Flores Vel{\'a}zquez},
  {Gurvich}, {Faucher-Gigu{\`e}re}, {Bullock}, {Starkenburg}, {Moreno},
  {Lazar}, {Mercado}, {Stern}, {Sparre}, {Hayward}, {Wetzel}, \&
  {El-Badry}}]{FloresVelazquez2021}
{Flores Vel{\'a}zquez}, J.~A., {Gurvich}, A.~B., {Faucher-Gigu{\`e}re}, C.-A.,
  {et~al.} 2021, \mnras, 501, 4812, \dodoi{10.1093/mnras/staa3893}

\bibitem[{{Gallart} {et~al.}(2005){Gallart}, {Zoccali}, \&
  {Aparicio}}]{Gallart2005}
{Gallart}, C., {Zoccali}, M., \& {Aparicio}, A. 2005, \araa, 43, 387,
  \dodoi{10.1146/annurev.astro.43.072103.150608}

\bibitem[{{Gilbert} {et~al.}(2025){Gilbert}, {Choi}, {Boyer}, {Williams},
  {Weisz}, {Bell}, {Dalcanton}, {McQuinn}, {Skillman}, {Costa}, {Dolphin},
  {Fouesneau}, {Girardi}, {Goldman}, {Gordon}, {Guhathakurta}, {Gull}, {Hagen},
  {Huynh}, {Lindberg}, {Marigo}, {Murray}, {Pastorelli}, \& {Yanchulova
  Merica-Jones}}]{Gilbert2025}
{Gilbert}, K.~M., {Choi}, Y., {Boyer}, M.~L., {et~al.} 2025, \apjs, 276, 8,
  \dodoi{10.3847/1538-4365/ad76af}

\bibitem[{{Gordon} {et~al.}(2003){Gordon}, {Clayton}, {Misselt}, {Landolt}, \&
  {Wolff}}]{Gordon2003}
{Gordon}, K.~D., {Clayton}, G.~C., {Misselt}, K.~A., {Landolt}, A.~U., \&
  {Wolff}, M.~J. 2003, \apj, 594, 279, \dodoi{10.1086/376774}

\bibitem[{{Gordon} {et~al.}(2016){Gordon}, {Fouesneau}, {Arab}, {Tchernyshyov},
  {Weisz}, {Dalcanton}, {Williams}, {Bell}, {Bianchi}, {Boyer}, {Choi},
  {Dolphin}, {Girardi}, {Hogg}, {Kalirai}, {Kapala}, {Lewis}, {Rix},
  {Sandstrom}, \& {Skillman}}]{Gordon2016}
{Gordon}, K.~D., {Fouesneau}, M., {Arab}, H., {et~al.} 2016, \apj, 826, 104,
  \dodoi{10.3847/0004-637X/826/2/104}

\bibitem[{{Gull} {et~al.}(2022){Gull}, {Weisz}, {Senchyna}, {Sandford}, {Choi},
  {McLeod}, {El-Badry}, {G{\"o}tberg}, {Gilbert}, {Boyer}, {Dalcanton},
  {GuhaThakurta}, {Goldman}, {Marigo}, {McQuinn}, {Pastorelli}, {Stark},
  {Skillman}, {Ting}, \& {Williams}}]{Gull2022}
{Gull}, M., {Weisz}, D.~R., {Senchyna}, P., {et~al.} 2022, \apj, 941, 206,
  \dodoi{10.3847/1538-4357/aca295}

\bibitem[{{Harris} {et~al.}(2020){Harris}, {Millman}, {van der Walt},
  {Gommers}, {Virtanen}, {Cournapeau}, {Wieser}, {Taylor}, {Berg}, {Smith},
  {Kern}, {Picus}, {Hoyer}, {van Kerkwijk}, {Brett}, {Haldane}, {del R{\'\i}o},
  {Wiebe}, {Peterson}, {G{\'e}rard-Marchant}, {Sheppard}, {Reddy}, {Weckesser},
  {Abbasi}, {Gohlke}, \& {Oliphant}}]{harris2020}
{Harris}, C.~R., {Millman}, K.~J., {van der Walt}, S.~J., {et~al.} 2020, \nat,
  585, 357, \dodoi{10.1038/s41586-020-2649-2}

\bibitem[{{Harris} \& {Zaritsky}(2001)}]{Harris2001}
{Harris}, J., \& {Zaritsky}, D. 2001, \apjs, 136, 25, \dodoi{10.1086/321792}

\bibitem[{{Higgs} {et~al.}(2016){Higgs}, {McConnachie}, {Irwin}, {Bate},
  {Lewis}, {Walker}, {C{\^o}t{\'e}}, {Venn}, \& {Battaglia}}]{Higgs2016}
{Higgs}, C.~R., {McConnachie}, A.~W., {Irwin}, M., {et~al.} 2016, \mnras, 458,
  1678, \dodoi{10.1093/mnras/stw257}

\bibitem[{{Holtzman} {et~al.}(1999){Holtzman}, {Gallagher}, {Cole}, {Mould},
  {Grillmair}, {Ballester}, {Burrows}, {Clarke}, {Crisp}, {Evans}, {Griffiths},
  {Hester}, {Hoessel}, {Scowen}, {Stapelfeldt}, {Trauger}, \&
  {Watson}}]{Holtzman1999}
{Holtzman}, J.~A., {Gallagher}, III, J.~S., {Cole}, A.~A., {et~al.} 1999, \aj,
  118, 2262, \dodoi{10.1086/301097}

\bibitem[{{Hunter}(2007)}]{Hunter2007}
{Hunter}, J.~D. 2007, Computing in Science and Engineering, 9, 90,
  \dodoi{10.1109/MCSE.2007.55}

\bibitem[{{Jacobs} {et~al.}(2009){Jacobs}, {Rizzi}, {Tully}, {Shaya},
  {Makarov}, \& {Makarova}}]{Jacobs2009}
{Jacobs}, B.~A., {Rizzi}, L., {Tully}, R.~B., {et~al.} 2009, \aj, 138, 332,
  \dodoi{10.1088/0004-6256/138/2/332}

\bibitem[{{Johnson} {et~al.}(2013){Johnson}, {Weisz}, {Dalcanton}, {Johnson},
  {Dale}, {Dolphin}, {Gil de Paz}, {Kennicutt}, {Lee}, {Skillman}, {Boquien},
  \& {Williams}}]{Johnson2013}
{Johnson}, B.~D., {Weisz}, D.~R., {Dalcanton}, J.~J., {et~al.} 2013, \apj, 772,
  8, \dodoi{10.1088/0004-637X/772/1/8}

\bibitem[{{Karachentsev} {et~al.}(2013){Karachentsev}, {Makarov}, \&
  {Kaisina}}]{Karachentsev2013}
{Karachentsev}, I.~D., {Makarov}, D.~I., \& {Kaisina}, E.~I. 2013, \aj, 145,
  101, \dodoi{10.1088/0004-6256/145/4/101}

\bibitem[{{Karakas} {et~al.}(2018){Karakas}, {Lugaro}, {Carlos}, {Cseh},
  {Kamath}, \& {Garc{\'\i}a-Hern{\'a}ndez}}]{Karakas2018}
{Karakas}, A.~I., {Lugaro}, M., {Carlos}, M., {et~al.} 2018, \mnras, 477, 421,
  \dodoi{10.1093/mnras/sty625}

\bibitem[{{Kennicutt}(1998)}]{Kennicutt1998}
{Kennicutt}, Robert~C., J. 1998, \araa, 36, 189,
  \dodoi{10.1146/annurev.astro.36.1.189}

\bibitem[{{Kennicutt} {et~al.}(2008){Kennicutt}, {Lee}, {Funes}, {J.}, {Sakai},
  \& {Akiyama}}]{Kennicutt2008}
{Kennicutt}, Robert~C., J., {Lee}, J.~C., {Funes}, J.~G., {et~al.} 2008, \apjs,
  178, 247, \dodoi{10.1086/590058}

\bibitem[{{Kennicutt} {et~al.}(1994){Kennicutt}, {Tamblyn}, \&
  {Congdon}}]{Kennicutt1994}
{Kennicutt}, Robert~C., J., {Tamblyn}, P., \& {Congdon}, C.~E. 1994, \apj, 435,
  22, \dodoi{10.1086/174790}

\bibitem[{{Kennicutt} \& {Evans}(2012)}]{Kennicutt2012}
{Kennicutt}, R.~C., \& {Evans}, N.~J. 2012, \araa, 50, 531,
  \dodoi{10.1146/annurev-astro-081811-125610}

\bibitem[{{Kroupa}(2001)}]{Kroupa2001}
{Kroupa}, P. 2001, \mnras, 322, 231, \dodoi{10.1046/j.1365-8711.2001.04022.x}

\bibitem[{{Lanz} \& {Hubeny}(2003)}]{Lanz2003}
{Lanz}, T., \& {Hubeny}, I. 2003, \apjs, 146, 417, \dodoi{10.1086/374373}

\bibitem[{{Lanz} \& {Hubeny}(2007)}]{Lanz2007}
---. 2007, \apjs, 169, 83, \dodoi{10.1086/511270}

\bibitem[{{Lazzarini} {et~al.}(2022){Lazzarini}, {Williams}, {Durbin},
  {Dalcanton}, {Smercina}, {Bell}, {Choi}, {Dolphin}, {Gilbert},
  {Guhathakurta}, {Rosolowsky}, {Skillman}, {Telford}, \&
  {Weisz}}]{Lazzarini2022}
{Lazzarini}, M., {Williams}, B.~F., {Durbin}, M.~J., {et~al.} 2022, \apj, 934,
  76, \dodoi{10.3847/1538-4357/ac7568}

\bibitem[{{Lee} {et~al.}(2006){Lee}, {Skillman}, {Cannon}, {Jackson}, {Gehrz},
  {Polomski}, \& {Woodward}}]{Lee2006}
{Lee}, H., {Skillman}, E.~D., {Cannon}, J.~M., {et~al.} 2006, \apj, 647, 970,
  \dodoi{10.1086/505573}

\bibitem[{{Lee} {et~al.}(2009){Lee}, {Gil de Paz}, {Tremonti}, {Kennicutt},
  {Salim}, {Bothwell}, {Calzetti}, {Dalcanton}, {Dale}, {Engelbracht}, {Funes},
  {Johnson}, {Sakai}, {Skillman}, {van Zee}, {Walter}, \& {Weisz}}]{Lee2009}
{Lee}, J.~C., {Gil de Paz}, A., {Tremonti}, C., {et~al.} 2009, \apj, 706, 599,
  \dodoi{10.1088/0004-637X/706/1/599}

\bibitem[{{Lee} {et~al.}(2011){Lee}, {Gil de Paz}, {Kennicutt}, {Bothwell},
  {Dalcanton}, {Jos{\'e} G. Funes S.}, {Johnson}, {Sakai}, {Skillman},
  {Tremonti}, \& {van Zee}}]{Lee2011}
{Lee}, J.~C., {Gil de Paz}, A., {Kennicutt}, Robert~C., J., {et~al.} 2011,
  \apjs, 192, 6, \dodoi{10.1088/0067-0049/192/1/6}

\bibitem[{{Leitherer} {et~al.}(1999){Leitherer}, {Schaerer}, {Goldader},
  {Delgado}, {Robert}, {Kune}, {de Mello}, {Devost}, \&
  {Heckman}}]{Leitherer99}
{Leitherer}, C., {Schaerer}, D., {Goldader}, J.~D., {et~al.} 1999, \apjs, 123,
  3, \dodoi{10.1086/313233}

\bibitem[{{Le{\v{s}}{\v{c}}inskait{\.{e}}}
  {et~al.}(2022){Le{\v{s}}{\v{c}}inskait{\.{e}}}, {Stonkut{\.{e}}}, \&
  {Vansevi{\v{c}}ius}}]{Lescinskaite2022}
{Le{\v{s}}{\v{c}}inskait{\.{e}}}, A., {Stonkut{\.{e}}}, R., \&
  {Vansevi{\v{c}}ius}, V. 2022, \aap, 660, A79,
  \dodoi{10.1051/0004-6361/202142743}

\bibitem[{{Lewis} {et~al.}(2015){Lewis}, {Dolphin}, {Dalcanton}, {Weisz},
  {Williams}, {Bell}, {Seth}, {Simones}, {Skillman}, {Choi}, {Fouesneau},
  {Guhathakurta}, {Johnson}, {Kalirai}, {Leroy}, {Monachesi}, {Rix}, \&
  {Schruba}}]{Lewis2015}
{Lewis}, A.~R., {Dolphin}, A.~E., {Dalcanton}, J.~J., {et~al.} 2015, \apj, 805,
  183, \dodoi{10.1088/0004-637X/805/2/183}

\bibitem[{{Lewis} {et~al.}(2017){Lewis}, {Simones}, {Johnson}, {Dalcanton},
  {Skillman}, {Weisz}, {Dolphin}, {Williams}, {Bell}, {Fouesneau}, {Kapala},
  {Rosenfield}, \& {Schruba}}]{Lewis2017}
{Lewis}, A.~R., {Simones}, J.~E., {Johnson}, B.~D., {et~al.} 2017, \apj, 834,
  70, \dodoi{10.3847/1538-4357/834/1/70}

\bibitem[{{Lindberg} {et~al.}(2024){Lindberg}, {Murray}, {Yanchulova
  Merica-Jones}, {Bot}, {Burhenne}, {Choi}, {Clark}, {Cohen}, {Gilbert},
  {Goldman}, {Gordon}, {Hirschauer}, {McQuinn}, {Roman-Duval}, {Sandstrom},
  {Tarantino}, \& {Williams}}]{Lindberg2024}
{Lindberg}, C.~W., {Murray}, C.~E., {Yanchulova Merica-Jones}, P., {et~al.}
  2024, arXiv e-prints, arXiv:2410.19910, \dodoi{10.48550/arXiv.2410.19910}

\bibitem[{{MacKenty} {et~al.}(2000){MacKenty}, {Ma{\'\i}z-Apell{\'a}niz},
  {Pickens}, {Norman}, \& {Walborn}}]{MacKenty2000}
{MacKenty}, J.~W., {Ma{\'\i}z-Apell{\'a}niz}, J., {Pickens}, C.~E., {Norman},
  C.~A., \& {Walborn}, N.~R. 2000, \aj, 120, 3007, \dodoi{10.1086/316841}

\bibitem[{{Ma{\'\i}z-Apell{\'a}niz} {et~al.}(2004){Ma{\'\i}z-Apell{\'a}niz},
  {P{\'e}rez}, \& {Mas-Hesse}}]{MaizApellaniz2004}
{Ma{\'\i}z-Apell{\'a}niz}, J., {P{\'e}rez}, E., \& {Mas-Hesse}, J.~M. 2004,
  \aj, 128, 1196, \dodoi{10.1086/422925}

\bibitem[{{Massana} {et~al.}(2022){Massana}, {Ruiz-Lara}, {No{\"e}l},
  {Gallart}, {Nidever}, {Choi}, {Sakowska}, {Besla}, {Olsen}, {Monelli},
  {Dorta}, {Stringfellow}, {Cassisi}, {Bernard}, {Zaritsky}, {Cioni},
  {Monachesi}, {van der Marel}, {de Boer}, \& {Walker}}]{Massana2022}
{Massana}, P., {Ruiz-Lara}, T., {No{\"e}l}, N.~E.~D., {et~al.} 2022, \mnras,
  513, L40, \dodoi{10.1093/mnrasl/slac030}

\bibitem[{{Mazzi} {et~al.}(2021){Mazzi}, {Girardi}, {Zaggia}, {Pastorelli},
  {Rubele}, {Bressan}, {Cioni}, {Clementini}, {Cusano}, {Rocha}, {Gullieuszik},
  {Kerber}, {Marigo}, {Ripepi}, {Bekki}, {Bell}, {de Grijs}, {Groenewegen},
  {Ivanov}, {Oliveira}, {Sun}, \& {van Loon}}]{Mazzi2021}
{Mazzi}, A., {Girardi}, L., {Zaggia}, S., {et~al.} 2021, \mnras, 508, 245,
  \dodoi{10.1093/mnras/stab2399}

\bibitem[{{McQuinn} {et~al.}(2009){McQuinn}, {Skillman}, {Cannon}, {Dalcanton},
  {Dolphin}, {Stark}, \& {Weisz}}]{McQuinn2009}
{McQuinn}, K. B.~W., {Skillman}, E.~D., {Cannon}, J.~M., {et~al.} 2009, \apj,
  695, 561, \dodoi{10.1088/0004-637X/695/1/561}

\bibitem[{{McQuinn} {et~al.}(2011){McQuinn}, {Skillman}, {Dalcanton},
  {Dolphin}, {Holtzman}, {Weisz}, \& {Williams}}]{McQuinn2011}
{McQuinn}, K. B.~W., {Skillman}, E.~D., {Dalcanton}, J.~J., {et~al.} 2011,
  \apj, 740, 48, \dodoi{10.1088/0004-637X/740/1/48}

\bibitem[{{McQuinn} {et~al.}(2015){McQuinn}, {Skillman}, {Dolphin}, \&
  {Mitchell}}]{McQuinn2015}
{McQuinn}, K. B.~W., {Skillman}, E.~D., {Dolphin}, A.~E., \& {Mitchell}, N.~P.
  2015, \apj, 808, 109, \dodoi{10.1088/0004-637X/808/2/109}

\bibitem[{{McQuinn} {et~al.}(2010{\natexlab{a}}){McQuinn}, {Skillman},
  {Cannon}, {Dalcanton}, {Dolphin}, {Hidalgo-Rodr{\'\i}guez}, {Holtzman},
  {Stark}, {Weisz}, \& {Williams}}]{McQuinn2010a}
{McQuinn}, K. B.~W., {Skillman}, E.~D., {Cannon}, J.~M., {et~al.}
  2010{\natexlab{a}}, \apj, 721, 297, \dodoi{10.1088/0004-637X/721/1/297}

\bibitem[{{McQuinn} {et~al.}(2010{\natexlab{b}}){McQuinn}, {Skillman},
  {Cannon}, {Dalcanton}, {Dolphin}, {Hidalgo-Rodr{\'\i}guez}, {Holtzman},
  {Stark}, {Weisz}, \& {Williams}}]{McQuinn2010b}
---. 2010{\natexlab{b}}, \apj, 724, 49, \dodoi{10.1088/0004-637X/724/1/49}

\bibitem[{{Murray} {et~al.}(2024){Murray}, {Lindberg}, {Yanchulova
  Merica-Jones}, {Williams}, {Cohen}, {Gordon}, {McQuinn}, {Choi}, {Burhenne},
  {Sandstrom}, {Bot}, {Johnson}, {Goldman}, {Clark}, {Roman-Duval}, {Gilbert},
  {Peek}, {Hirschauer}, {Boyer}, \& {Dolphin}}]{Murray2024}
{Murray}, C.~E., {Lindberg}, C.~W., {Yanchulova Merica-Jones}, P., {et~al.}
  2024, \apjs, 275, 5, \dodoi{10.3847/1538-4365/ad6de2}

\bibitem[{{Olsen}(1999)}]{Olsen1999}
{Olsen}, K. A.~G. 1999, \aj, 117, 2244, \dodoi{10.1086/300854}

\bibitem[{{Osterbrock}(1989)}]{Osterbrock1989}
{Osterbrock}, D.~E. 1989, {Astrophysics of gaseous nebulae and active galactic
  nuclei}

\bibitem[{{Roman-Duval} {et~al.}(2020){Roman-Duval}, {Proffitt}, {Taylor},
  {Monroe}, {Fischer}, {Fischer}, {Fullerton}, {Aloisi}, {Britt}, {Busko},
  {Carlberg}, {De Rosa}, {Jedrzejewski}, {Lockwood}, {Frazer}, {Hernandez},
  {James}, {Oliveira}, {Plesha}, {Riedel}, {Riley}, {Sahnow}, {Sankrit},
  {Shaw}, {Smith}, {Sohn}, {Som}, {Ubeda}, \& {Welty}}]{Roman-Duval2020}
{Roman-Duval}, J., {Proffitt}, C.~R., {Taylor}, J.~M., {et~al.} 2020, Research
  Notes of the American Astronomical Society, 4, 205,
  \dodoi{10.3847/2515-5172/abca2f}

\bibitem[{{Rubele} {et~al.}(2012){Rubele}, {Kerber}, {Girardi}, {Cioni},
  {Marigo}, {Zaggia}, {Bekki}, {de Grijs}, {Emerson}, {Groenewegen},
  {Gullieuszik}, {Ivanov}, {Miszalski}, {Oliveira}, {Tatton}, \& {van
  Loon}}]{Rubele2012}
{Rubele}, S., {Kerber}, L., {Girardi}, L., {et~al.} 2012, \aap, 537, A106,
  \dodoi{10.1051/0004-6361/201117863}

\bibitem[{{Rubele} {et~al.}(2015){Rubele}, {Girardi}, {Kerber}, {Cioni},
  {Piatti}, {Zaggia}, {Bekki}, {Bressan}, {Clementini}, {de Grijs}, {Emerson},
  {Groenewegen}, {Ivanov}, {Marconi}, {Marigo}, {Moretti}, {Ripepi},
  {Subramanian}, {Tatton}, \& {van Loon}}]{Rubele2015}
{Rubele}, S., {Girardi}, L., {Kerber}, L., {et~al.} 2015, \mnras, 449, 639,
  \dodoi{10.1093/mnras/stv141}

\bibitem[{{Rubele} {et~al.}(2018){Rubele}, {Pastorelli}, {Girardi}, {Cioni},
  {Zaggia}, {Marigo}, {Bekki}, {Bressan}, {Clementini}, {de Grijs}, {Emerson},
  {Groenewegen}, {Ivanov}, {Muraveva}, {Nanni}, {Oliveira}, {Ripepi}, {Sun}, \&
  {van Loon}}]{Rubelel2018}
{Rubele}, S., {Pastorelli}, G., {Girardi}, L., {et~al.} 2018, \mnras, 478,
  5017, \dodoi{10.1093/mnras/sty1279}

\bibitem[{{Saviane} {et~al.}(2002){Saviane}, {Rizzi}, {Held}, {Bresolin}, \&
  {Momany}}]{Saviane2002}
{Saviane}, I., {Rizzi}, L., {Held}, E.~V., {Bresolin}, F., \& {Momany}, Y.
  2002, \aap, 390, 59, \dodoi{10.1051/0004-6361:20020750}

\bibitem[{{Savino} {et~al.}(2023){Savino}, {Weisz}, {Skillman}, {Dolphin},
  {Cole}, {Kallivayalil}, {Wetzel}, {Anderson}, {Besla}, {Boylan-Kolchin},
  {Brown}, {Bullock}, {Collins}, {Cooper}, {Deason}, {Dotter}, {Fardal},
  {Ferguson}, {Fritz}, {Geha}, {Gilbert}, {Guhathakurta}, {Ibata}, {Irwin},
  {Jeon}, {Kirby}, {Lewis}, {Mackey}, {Majewski}, {Martin}, {McConnachie},
  {Patel}, {Rich}, {Simon}, {Sohn}, {Tollerud}, \& {van der
  Marel}}]{Savino2023}
{Savino}, A., {Weisz}, D.~R., {Skillman}, E.~D., {et~al.} 2023, \apj, 956, 86,
  \dodoi{10.3847/1538-4357/acf46f}

\bibitem[{{Schlafly} \& {Finkbeiner}(2011)}]{Schlafly2011}
{Schlafly}, E.~F., \& {Finkbeiner}, D.~P. 2011, \apj, 737, 103,
  \dodoi{10.1088/0004-637X/737/2/103}

\bibitem[{{Schlegel} {et~al.}(1998){Schlegel}, {Finkbeiner}, \&
  {Davis}}]{Schlegel1998}
{Schlegel}, D.~J., {Finkbeiner}, D.~P., \& {Davis}, M. 1998, \apj, 500, 525,
  \dodoi{10.1086/305772}

\bibitem[{{Shivaei} {et~al.}(2020){Shivaei}, {Reddy}, {Rieke}, {Shapley},
  {Kriek}, {Battisti}, {Mobasher}, {Sanders}, {Fetherolf}, {Azadi}, {Coil},
  {Freeman}, {de Groot}, {Leung}, {Price}, {Siana}, \& {Zick}}]{Shivaei2020}
{Shivaei}, I., {Reddy}, N., {Rieke}, G., {et~al.} 2020, \apj, 899, 117,
  \dodoi{10.3847/1538-4357/aba35e}

\bibitem[{{Skillman} {et~al.}(2003){Skillman}, {Tolstoy}, {Cole}, {Dolphin},
  {Saha}, {Gallagher}, {Dohm-Palmer}, \& {Mateo}}]{Skillman2003}
{Skillman}, E.~D., {Tolstoy}, E., {Cole}, A.~A., {et~al.} 2003, \apj, 596, 253,
  \dodoi{10.1086/377635}

\bibitem[{{Smartt}(2009)}]{Smartt2009}
{Smartt}, S.~J. 2009, \araa, 47, 63,
  \dodoi{10.1146/annurev-astro-082708-101737}

\bibitem[{{Tang} {et~al.}(2014){Tang}, {Bressan}, {Rosenfield}, {Slemer},
  {Marigo}, {Girardi}, \& {Bianchi}}]{Tang2014}
{Tang}, J., {Bressan}, A., {Rosenfield}, P., {et~al.} 2014, \mnras, 445, 4287,
  \dodoi{10.1093/mnras/stu2029}

\bibitem[{{Tang} {et~al.}(2016){Tang}, {Bressan}, {Slemer}, {Marigo},
  {Girardi}, {Bianchi}, {Rosenfield}, \& {Momany}}]{Tang2016}
{Tang}, J., {Bressan}, A., {Slemer}, A., {et~al.} 2016, \mnras, 455, 3393,
  \dodoi{10.1093/mnras/stv2491}

\bibitem[{{Tolstoy} {et~al.}(2009){Tolstoy}, {Hill}, \& {Tosi}}]{Tolstoy2009}
{Tolstoy}, E., {Hill}, V., \& {Tosi}, M. 2009, \araa, 47, 371,
  \dodoi{10.1146/annurev-astro-082708-101650}

\bibitem[{{Tolstoy} \& {Saha}(1996)}]{Tolstoy1996}
{Tolstoy}, E., \& {Saha}, A. 1996, \apj, 462, 672, \dodoi{10.1086/177181}

\bibitem[{{Tosi} {et~al.}(1991){Tosi}, {Greggio}, {Marconi}, \&
  {Focardi}}]{Tosi1991}
{Tosi}, M., {Greggio}, L., {Marconi}, G., \& {Focardi}, P. 1991, \aj, 102, 951,
  \dodoi{10.1086/115925}

\bibitem[{{van der Walt} {et~al.}(2011){van der Walt}, {Colbert}, \&
  {Varoquaux}}]{vanderwalt2011}
{van der Walt}, S., {Colbert}, S.~C., \& {Varoquaux}, G. 2011, Computing in
  Science and Engineering, 13, 22, \dodoi{10.1109/MCSE.2011.37}

\bibitem[{{van Zee} \& {Haynes}(2006)}]{vanZee2006}
{van Zee}, L., \& {Haynes}, M.~P. 2006, \apj, 636, 214, \dodoi{10.1086/498017}

\bibitem[{{van Zee} {et~al.}(2006){van Zee}, {Skillman}, \&
  {Haynes}}]{vanZee2006b}
{van Zee}, L., {Skillman}, E.~D., \& {Haynes}, M.~P. 2006, \apj, 637, 269,
  \dodoi{10.1086/498298}

\bibitem[{Virtanen {et~al.}(2020)Virtanen, Gommers, Oliphant, Haberland, Reddy,
  Cournapeau, Burovski, Peterson, Weckesser, Bright, {van der Walt}, Brett,
  Wilson, Millman, Mayorov, Nelson, Jones, Kern, Larson, Carey, Polat, Feng,
  Moore, {VanderPlas}, Laxalde, Perktold, Cimrman, Henriksen, Quintero, Harris,
  Archibald, Ribeiro, Pedregosa, {van Mulbregt}, \& {SciPy 1.0
  Contributors}}]{Scipy2020}
Virtanen, P., Gommers, R., Oliphant, T.~E., {et~al.} 2020, Nature Methods, 17,
  261, \dodoi{10.1038/s41592-019-0686-2}

\bibitem[{{Walmswell} {et~al.}(2015){Walmswell}, {Tout}, \&
  {Eldridge}}]{Walmswell2015}
{Walmswell}, J.~J., {Tout}, C.~A., \& {Eldridge}, J.~J. 2015, \mnras, 447,
  2951, \dodoi{10.1093/mnras/stu2666}

\bibitem[{{Weisz} {et~al.}(2014){Weisz}, {Dolphin}, {Skillman}, {Holtzman},
  {Gilbert}, {Dalcanton}, \& {Williams}}]{Weisz2014}
{Weisz}, D.~R., {Dolphin}, A.~E., {Skillman}, E.~D., {et~al.} 2014, \apj, 789,
  147, \dodoi{10.1088/0004-637X/789/2/147}

\bibitem[{{Weisz} {et~al.}(2011){Weisz}, {Dalcanton}, {Williams}, {Gilbert},
  {Skillman}, {Seth}, {Dolphin}, {McQuinn}, {Gogarten}, {Holtzman}, {Rosema},
  {Cole}, {Karachentsev}, \& {Zaritsky}}]{Weisz2011}
{Weisz}, D.~R., {Dalcanton}, J.~J., {Williams}, B.~F., {et~al.} 2011, \apj,
  739, 5, \dodoi{10.1088/0004-637X/739/1/5}

\bibitem[{{Weisz} {et~al.}(2012){Weisz}, {Johnson}, {Johnson}, {Skillman},
  {Lee}, {Kennicutt}, {Calzetti}, {van Zee}, {Bothwell}, {Dalcanton}, {Dale},
  \& {Williams}}]{Weisz2012}
{Weisz}, D.~R., {Johnson}, B.~D., {Johnson}, L.~C., {et~al.} 2012, \apj, 744,
  44, \dodoi{10.1088/0004-637X/744/1/44}

\bibitem[{{Williams} {et~al.}(2011){Williams}, {Dalcanton}, {Gilbert}, {Seth},
  {Weisz}, {Skillman}, \& {Dolphin}}]{Williams2011}
{Williams}, B.~F., {Dalcanton}, J.~J., {Gilbert}, K.~M., {et~al.} 2011, \apj,
  735, 22, \dodoi{10.1088/0004-637X/735/1/22}

\bibitem[{{Williams} {et~al.}(2009){Williams}, {Dalcanton}, {Seth}, {Weisz},
  {Dolphin}, {Skillman}, {Harris}, {Holtzman}, {Girardi}, {de Jong}, {Olsen},
  {Cole}, {Gallart}, {Gogarten}, {Hidalgo}, {Mateo}, {Rosema}, {Stetson}, \&
  {Quinn}}]{Williams2009}
{Williams}, B.~F., {Dalcanton}, J.~J., {Seth}, A.~C., {et~al.} 2009, \aj, 137,
  419, \dodoi{10.1088/0004-6256/137/1/419}

\bibitem[{{Williams} {et~al.}(2014){Williams}, {Lang}, {Dalcanton}, {Dolphin},
  {Weisz}, {Bell}, {Bianchi}, {Byler}, {Gilbert}, {Girardi}, {Gordon},
  {Gregersen}, {Johnson}, {Kalirai}, {Lauer}, {Monachesi}, {Rosenfield},
  {Seth}, \& {Skillman}}]{Williams2014}
{Williams}, B.~F., {Lang}, D., {Dalcanton}, J.~J., {et~al.} 2014, \apjs, 215,
  9, \dodoi{10.1088/0067-0049/215/1/9}

\bibitem[{{Williams} {et~al.}(2021){Williams}, {Durbin}, {Dalcanton}, {Lang},
  {Girardi}, {Smercina}, {Dolphin}, {Weisz}, {Choi}, {Bell}, {Rosolowsky},
  {Skillman}, {Koch}, {Lindberg}, {Hagen}, {Gordon}, {Seth}, {Gilbert},
  {Guhathakurta}, {Lauer}, \& {Bianchi}}]{Williams2021}
{Williams}, B.~F., {Durbin}, M.~J., {Dalcanton}, J.~J., {et~al.} 2021, \apjs,
  253, 53, \dodoi{10.3847/1538-4365/abdf4e}

\bibitem[{{Young} {et~al.}(2003){Young}, {van Zee}, {Lo}, {Dohm-Palmer}, \&
  {Beierle}}]{Young2003}
{Young}, L.~M., {van Zee}, L., {Lo}, K.~Y., {Dohm-Palmer}, R.~C., \& {Beierle},
  M.~E. 2003, \apj, 592, 111, \dodoi{10.1086/375581}

\bibitem[{{Zinchenko} \& {Pilyugin}(2022)}]{Zinchenko2022}
{Zinchenko}, I.~A., \& {Pilyugin}, L.~S. 2022, Astronomische Nachrichten, 343,
  e20220048, \dodoi{10.1002/asna.20220048}

\end{thebibliography}

\end{document}